\documentclass[reqno,oneside,final]{amsart}
\usepackage[margin=3cm]{geometry}
\usepackage{amssymb,amsmath}
\usepackage{amsfonts}
\usepackage{rotating}
\usepackage{upgreek}
\usepackage{graphicx}
\usepackage{tikz}
\usepackage{tikz-uml}
\tikzset{
 every overlay node/.style={
   draw=white,fill=white,rounded corners,anchor=north west,
 },
}
\usetikzlibrary[patterns,arrows,positioning,patterns,calc]

\usepackage{graphics,caption}
\usepackage[labelformat=simple]{subcaption}

\usepackage[utf8]{inputenc} 
\usepackage[T1]{fontenc} 
\usepackage{hyperref}
\usepackage[numbers,sort&compress]{natbib}
\usepackage{multirow}
\usepackage{hhline}
\usepackage{colortbl}
\newcolumntype{L}{>{\centering\arraybackslash}m{0.46\textwidth}}
\usepackage{arydshln}
\usepackage{comment}

\usepackage{changes}
\usepackage{multirow}

\usepackage[colorinlistoftodos,prependcaption,textsize=small,linecolor=blue,backgroundcolor=blue!25,bordercolor=blue,textwidth=2.5cm]{todonotes}

\usepackage[ruled, lined, longend, linesnumbered]{algorithm2e}
\usepackage{xcolor}

\SetAlFnt{\footnotesize}
\SetArgSty{textnormal}

\SetAlCapSty{xAlCapSty}

\SetCommentSty{xCommentSty}

\SetNlSty{mynlfont}{}{} 

\LinesNumbered

\SetSideCommentRight

\DontPrintSemicolon

\RestyleAlgo{algoruled}

\SetInd{0.5em}{0.5em}

\usepackage{acronym}
\acrodef{fe}[FE]{Finite Element}
\acrodef{dd}[DD]{Domain Decomposition}
\acrodef{bddc}[BDDC]{Balancing Domain Decomposition by Constraints}
\acrodef{dof}[DOF]{Degree Of Freedom}
\acrodefplural{dof}[DOFs]{Degrees of Freedom}
\acrodef{am}[AM]{Additive Manufacturing}
\acrodef{amr}[AMR]{Adaptive Mesh Refinement}
\acrodef{mpi}[MPI]{Message Passing Interface}
\acrodef{ebm}[EBM]{Element-Birth Method}
\acrodef{hst}[HST]{Hyperplane Separation Theorem}
\acrodef{pcg}[PCG]{Preconditioned Conjugate Gradient}
\acrodef{hav}[HAV]{Heat Affected Volume}
\acrodef{amg}[AMG]{Algebraic MultiGrid}
\acrodef{pde}[PDE]{Partial Differential Equation}
\acrodefplural{pde}[PDEs]{Partial Differential Equations}

\graphicspath{{Figures/}{Figures/results/}{Figures/results/logPlots/}{Figures/results/wiggle/}}

\def\ENrew#1{{\color{blue} \textbf{REWRITE:}~#1}}

\begin{document}

\title[Parallel FE framework for growing geometries in metal AM]{A scalable 
parallel finite element framework for growing geometries. Application to metal 
additive manufacturing.}

\author{Eric Neiva \and Santiago Badia \and Alberto F. Martín \and Michele 
Chiumenti}

\date{\today}

\renewcommand{\thefootnote}{\arabic{footnote}}

\maketitle

\begin{center}
\small{Universitat Polit\`ecnica de Catalunya (UPC), Jordi Girona 1-3, Edifici 
C1, 08034 Barcelona, Spain. \\
\vspace{0.15cm}
Centre Internacional de M\`etodes Num\`erics en Enginyeria (CIMNE), Building 
C1, \\
Campus Nord UPC, Gran Capitán S/N, 08034 Barcelona, Spain.\\
\vspace{0.15cm}
\texttt{\{eneiva,sbadia,amartin,michele\}@cimne.upc.edu}.}
\end{center}

\begin{abstract}
\textit{This work introduces an innovative parallel, fully-distributed finite 
element framework for growing geometries and its application to metal additive 
manufacturing. It is well-known that virtual part design and qualification in 
additive manufacturing requires highly-accurate multiscale and multiphysics 
analyses. Only high performance computing tools are able to handle such 
complexity in time frames compatible with time-to-market. However, efficiency, 
without loss of accuracy, has rarely held the centre stage in the numerical 
community. Here, in contrast, the framework is designed to adequately exploit 
the resources of high-end distributed-memory machines. It is grounded on three 
building blocks: (1) Hierarchical adaptive mesh refinement with octree-based 
meshes; (2) a parallel strategy to model the growth of the geometry; (3) 
\replaced{state-of-the-art parallel iterative linear solvers.}{the 
customization of a parallel iterative linear solver, which leverages the 
so-called balancing domain decomposition by constraints preconditioning 
approach for fast convergence and high parallel scalability.} Computational 
experiments consider the heat transfer analysis at the part scale of the 
printing process by powder-bed technologies. After verification against a 3D 
benchmark, a strong-scaling analysis \replaced{assesses performance and 
identifies major sources of parallel overhead. A third numerical example 
examines the efficiency and robustness of (2) in a curved 3D shape. 
Unprecedented parallelism and scalability were achieved in this work.}{is 
carried out for a simulation of 48 layers printed in a cuboid. The cuboid is 
adaptively meshed to model a layer-by-layer metal deposition process and the 
average global problem size amounts to 10.3 million unknowns. An unprecedented 
scalability for problems with growing domains is achieved, with the capability 
of simulating the printing and recoat of a single layer in 8 seconds average on 
3,072 processors.} Hence, this framework contributes to take on higher 
complexity and/or accuracy, not only of part-scale simulations of metal or 
polymer additive manufacturing, but also in welding, sedimentation, 
atherosclerosis, or any other physical problem where the physical domain of 
interest grows in time.}
\end{abstract}

\noindent{\bf Keywords:} 
Parallel computing, domain decomposition, finite elements, adaptive mesh 
refinement, additive manufacturing, powder-bed fusion.


\pagestyle{myheadings}
\thispagestyle{plain}


\section{Introduction}
\label{sec:introduction}

\ac{am}, broadly known as 3D Printing, is introducing a disruptive design 
paradigm in the manufacturing landscape. The key potential of \ac{am} is the 
ability to cost-effectively create \emph{on-demand} objects with complex shapes 
and enhanced properties, that are near impossible or impractical to produce 
with conventional technologies, such as casting or forging. Adoption of \ac{am} 
is undergoing an exponential growth lead by the aerospace, defence, medical and 
dental industries and the prospect is a stronger and wider presence as a 
manufacturing technology~\citep{wohlers2017wohlers}.

Nowadays, one of the main showstoppers in the \ac{am} industry, especially for 
metals, is the lack of a software ecosystem supporting fast and reliable 
product and process design. Part qualification is chiefly based on slow and 
expensive trial-and-error physical experimentation and the understanding of the 
process-structure-performance link is still very obscure. This situation 
precludes further implementation of \ac{am} and it is a call to action to shift 
to a virtual-based design model, based on predictive computer simulation tools. 
Only then will it be possible to 
fully leverage the geometrical freedom, cost efficiency and immediacy of this 
technology.

This work addresses the numerical simulation of metal \ac{am} processes through 
\emph{High Performance Computing (HPC) tools}. The mathematical modelling of 
the process involves dealing with multiple scales in space (e.g. part, melt 
pool, microstructure), multiple scales in time (e.g. microseconds, hours), 
coupled multiphysics~\citep{de1999formulation,khairallah2016laser} (e.g. 
thermomechanics, phase-change, melt pool flow) and arbitrarily complex 
geometries that grow in time. As a result, high-fidelity analyses, which are 
vital for part qualification, can be extremely expensive and require vast 
computational resources. In this sense, HPC tools capable to run these highly 
accurate simulations in a time scale compatible with the time-to-market of 
\ac{am} are of great importance. By efficiently exploiting HPC resources with 
scalable methods, one can drastically reduce CPU time, allowing the 
optimization of the \ac{am} building process and the virtual certification of 
the final component in reasonable time.

Experience acquired in modelling traditional processes, such as casting or 
welding~\citep{cervera1999thermo,chiumenti2008numerical,lindgren2014computational},
has been the cornerstone of the first models for metal \ac{am} 
processes~\citep{bugeda1999numerical,roberts_three-dimensional_2009,anca_computational_2011,lundback_modelling_2011,kolossov_3d_2004}.
At the part scale, \ac{fe} modelling has proved to be useful to assess the 
influence of process parameters~\citep{parry_understanding_2016}, compute 
temperature 
distributions~\citep{denlinger_thermal_2016,chiumenti_numerical_2017}, or 
evaluate distortions and residual 
stresses~\citep{dunbar_experimental_2016,prabhakar_computational_2015,lu2018finite}.
Recent contributions have introduced microstructure simulations of grain 
growth~\citep{Rodgers2017,Lian2018} and crystal 
plasticity~\citep{kergassner2018modeling}, melt-pool-scale 
models~\citep{YAN2018210,khairallah2016laser} and even multiscale and 
multiphysics 
solvers~\citep{Yan2018,keller2017application,Yang2018,lindgren2016simulation,salsi2018modeling}.
Furthermore, advanced frameworks (e.g. grounded on multi-level $hp$-\ac{fe} 
methods combined with implicit boundary 
methods~\citep{KOLLMANNSBERGER20181483}) or applications to topology 
optimization~\citep{Steuben2017} have also been considered.

However, in spite of the active scientific progress in the field, the authors 
have detected that very little effort has turned to the design of \emph{large 
scale} \ac{fe} methods for metal \ac{am}. Even if computational efficiency has 
been taken into consideration in several works, all approaches have been 
limited to 
\ac{amr}~\citep{patil_generalized_2015,denlinger_thermomechanical_2017} or  
simplifications~\citep{hodge_experimental_2016,irwin_line_2016,chiumenti_neiva_2017,setien2018empirical}
that sacrifice the accuracy of the model. Parallelism and scalability has been 
generally disregarded (with few 
exceptions~\citep{Lian2018,mozaffar2019acceleration}), even if it is 
fundamental to face more complexity and/or provide increased accuracy at 
acceptable CPU times. For instance, for the high-fidelity melt-pool solver in 
~\citep{YAN2018210}, a simulation of 16 ms of physical time with 7 million 
cells requires 700 h of CPU time on a common desktop with an Intel Core i7-2600.

The purpose of this work is to design a novel scalable parallel \ac{fe} 
framework for metal \ac{am} \emph{at the part scale}. Our approach considers 
three main building blocks:

\begin{enumerate}
	
	\item{\emph{Hierarchical \ac{amr} with octree meshes} (see 
	Sect.~\ref{sec:octrees}). The dimensions of a part are in the order of 
	$[\mathrm{mm}]$ or $[\mathrm{cm}]$, but relevant physical phenomena tend to 
	concentrate around the melt pool $[\upmu \mathrm{m}]$. Likewise, in 
	powder-bed fusion, the layer size is also $[\upmu \mathrm{m}]$, i.e. the 
	scale of growth is much smaller than the scale of the part. Hence, adaptive 
	meshing can be suitably exploited for the highly-localized nature of the 
	problem. Here, a parallel octree-based $h$-adaptive \ac{fe} 
	framework~\citep{badia2018on} is established.}
	
	\item{\emph{Modelling the growth of the geometry} (see 
	Sect.~\ref{sec:growth}). In welding and \ac{am} processes, the addition of 
	material into the part has been typically modelled by adding new elements 
	into the computational mesh. To this effect, the simulation starts with the 
	generation of a \emph{static} background mesh comprising the substrate and 
	the filling, i.e. the final part. \replaced{Common practice in the literature 
	essentially considers two different techniques 
	\citep{michaleris_modeling_2014,chiumenti_numerical_2016}: 
	\emph{quiet}-element method and \ac{ebm}. The only difference among them is 
	how they treat the elements in the filling at the start of the simulation. 
	While the former assigns to them penalized material properties, perturbing 
	the original problem; in the latter, they have no \acp{dof}. At each time 
	step, the computational mesh is updated: elements inside the incremental 
	growth region are found with a search operation and assigned the usual 
	material properties or new degrees of freedom, respectively.}{At 
	each time step, the elements in the region that has not yet been filled are 
	either assigned penalized properties (\emph{quiet}-element method, which 
	perturbs the original problem) or excluded from the computational domain in 
	the \ac{ebm}, both detailed in.} This work \replaced{extends}{adapts} the 
	\replaced{\ac{ebm}}{latter} to \replaced{(1).}{the parallel hierarchical 
	\ac{amr} framework above (Sect.~\ref{subsec:ebm}).} \added{The 
	parallelization approach is designed such that it can be accommodated in a 
	general-purpose \ac{fe} code. It only requires two interprocessor 
	communications and it is completed with an efficient and embarrassingly 
	parallel intersection test \emph{for rectangular bodies} to drive the update 
	of the computational mesh (Sect.~\ref{subsec:search}).} 
	\replaced{Finally,}{In addition,} a strategy is devised to balance the 
	computational load among processors, during the growth of the geometry 
	(Sect.~\ref{subsec:dynamic}). \added{The parallel and adaptive \ac{ebm} is 
	central to a parallel \ac{fe} framework for growing domains and constitutes 
	the main novelty of this work.}}
	
	\item{\emph{\replaced{State-of-the-art parallel}{Adapting a highly scalable 
	Parallel} iterative linear solver\added{s}}. Compared to sparse direct 
	solvers, iterative solvers can be efficiently implemented in parallel 
	computer codes for distributed-memory machines. However, they must also be 
	equipped with efficient preconditioning schemes\replaced{, i.e. 
	preconditioners able to keep a bound of the condition number, independent of 
	mesh resolution, while still exposing a high degree of parallelism. Examples 
	of preconditioners that the framework is able to use include 
	\ac{amg}~\citep{chow2006survey} or \ac{bddc}~\citep{badia2014highly}, 
	although they are not actually exploited in this work, for reasons made clear 
	in Sect.~\ref{sec:formulation}, related to the application problem.}{to 
	accomplish parallel/algorithmic scalability. To this end, this work relies on 
	nonoverlapping, by considering a customization to the problem at hand of the 
	\ac{bddc} method.}}
	
\end{enumerate}

As the originality of the framework centres upon the computational aspects to 
efficiently deal with growing geometries, \replaced{(1) and (2)}{these blocks} 
are presented in an abstract way, i.e. without considering a reference physical 
problem. Afterwards, Sect.~\ref{sec:formulation} considers an application to 
the heat transfer analysis of metal \ac{am} processes by powder-bed fusion. 
Nonetheless, the authors believe that the framework can readily be extended to 
a coupled thermomechanical analysis, \added{as long as proper treatment of 
history variables within the \ac{amr} framework can be guaranteed. Likewise, it 
may also be} adapted to other metal or polymer technologies, or even be useful 
to other domains of study, such as the simulation of sedimentation processes.

Computer implementation was supported by
\texttt{FEMPAR}~\citep{fempar-web-page,badia-fempar}, a general-purpose 
object-oriented multi-threaded/message-passing scientific software for the fast 
solution of multiphysics problems governed by \acp{pde}. \texttt{FEMPAR} adapts 
to a range of computing environments, from desktop and laptop computers to the 
most advanced HPC clusters and supercomputers. \deleted{On the other hand, it 
contains an implementation of \ac{bddc} that scales up to 458,672 cores on the 
JUQUEEN supercomputer (JSC, Germany)~\citep{badia2016multilevel}.} 
\replaced{The \texttt{FEMPAR-AM} module for \ac{fe} analyses of metal \ac{am} 
processes has been developed on top of this high-end infrastructure. Its main 
software abstractions are described in Sect.~\ref{sec:implementation}. The 
exposition is intended to help in the customization of any general-purpose 
\ac{fe} code for growing domains.}{The computer implementation of the 
framework, using this high-end infrastructure, has given rise to the 
\texttt{FEMPAR-AM} module for \ac{fe} analyses of metal \ac{am} processes.}

The numerical study of the framework in Sect.~\ref{sec:results} starts with a 
verification of the thermal \ac{fe} model against a well-known 3D benchmark. 
Validation of this heat transfer formulation has already been object of 
previous works~\citep{chiumenti_numerical_2017,chiumenti_neiva_2017} and it is 
not covered here. A strong-scaling analysis follows in 
\added{Sect.}~\ref{sec:ss_example} \replaced{to analyse the performance of the 
computer implementation and expose sources of load imbalance}{, with a special 
focus on sources of load imbalance}, identified as a major parallel overhead 
threatening the efficiency of the implementation.\deleted{, and corresponding 
corrective measures.} \replaced{The simulation considers the printing of 48 
layers in a cuboid, one layer printed per time step (followed by an interlayer 
cooling step). A relevant outcome is the capability of simulating the printing 
and cooling of a single layer (two linearized time steps) in a 10 million 
unknown problem in merely 2.5 seconds average with 6,144 processors.}{A 
relevant outcome is the capability of simulating the printing of a single layer 
in a 10 million unknown problem in merely 8 seconds average with 3,072 
processors.} \added{The last example in Sect.~\ref{subsec:curved} considers a 
curved 3D shape and follows the actual laser path point-to-point. A second 
order adaptive mesh with no geometrical error is transformed during the 
simulation to accommodate the laser path. Regardless of having nonrectangular 
cells, the parallel \ac{ebm} is capable of tracking the laser path, as long as 
some quasi-rectangularity conditions hold.} In conclusion (see 
Sect.~\ref{sec:conclusions}), the fully-distributed, parallel framework 
presented in this work is set to contribute to the efficient simulation of 
\ac{am} processes, a critical aspect long identified, though mostly neglected 
by the numerical community.

\section{Mesh generation by hierarchical AMR}
\label{sec:octrees}

Physical phenomena are often characterized by multiple scales in both space and 
time. When the smallest ones are highly-localized in the physical domain of 
analysis, uniformly refined meshes tend to be impractical from the 
computational viewpoint, even for the largest available supercomputers.

The purpose of \ac{amr} is to reach a compromise between the high-accuracy 
requirements in the regions of interest and the computational effort of solving 
for the whole system. To this end, the mesh is refined in the regions of the 
domain that present a complex behaviour of the solution, while larger mesh 
sizes are prescribed in other areas. 

In this work, the areas of interest are known \emph{a priori} and correspond to 
the growing regions. It is assumed that the geometrical scale of growth is much 
smaller than the domain of study, as it is the case of welding or \ac{am} 
processes. \deleted{However, the most usual approach in \ac{fe} analysis is to 
drive the \ac{amr} with \textit{a posteriori} error 
estimates~\citep{ainsworth2011posteriori}. These quantities are cell-wise 
indicators of approximation error generated from the \ac{fe} solution to the 
PDE.} Besides, the framework is restricted to $h$-adaptivity, i.e. only the 
mesh size changes among cells, in contrast to $hp$-adaptivity, where the 
polynomial order $p$ of the \acp{fe} may also vary among cells. This 
computational framework is briefly outlined in this section; the reader is 
referred to~\citep{badia2018on} for a thorough exposition.

\subsection{Hierarchical AMR with octree meshes}
\label{subsec:octrees}

Let us suppose that $\Omega \subset \mathbb{R}^d$ is an open bounded polyhedral 
domain, being $d = 2,3$ the dimension of the physical space. Let 
$\mathcal{T}_h^0$ be a conforming and quasi-uniform partition of $\Omega$ into 
quadrilaterals ($d = 2$) or hexahedra ($d = 3$), where every $K \in 
\mathcal{T}_h^0$ is the image of a reference element $\hat{K}$ through a smooth 
bijective mapping $F_K$. If not stated otherwise, these hypotheses are common 
to all sections of this document.

Hierarchical \ac{amr} is a multi-step process. The mesh generation consists in 
the transformation of $\mathcal{T}_h^0$, typically as simple as a single 
quadrilateral or hexahedron, into an objective mesh $\mathcal{T}_h$ via a 
finite number of refinement/coarsening steps; in other words, the \ac{amr} 
process generates a sequence $\mathcal{T}_h^0, \mathcal{T}_h^1, \ldots, 
\mathcal{T}_h^m \equiv \mathcal{T}_h$ such that $\mathcal{T}_h^i = 
\mathcal{R}(\mathcal{T}_h^{i-1},\theta^i), \ i = 1, \ldots, m < \infty$, where 
$\mathcal{R}$ applies the refinement/coarsening procedure over 
$\mathcal{T}_h^{i-1}$ and $\theta^i : \mathcal{T}_h^{i-1} \rightarrow 
\{-1,0,1\}$ is an array establishing the action to be taken at each cell: -1 
for coarsening, 0 for "do nothing" and 1 for refinement. 

\begin{figure}[!h]
	\centering
	\scalebox{.60}{

\begin{tikzpicture}[font=\LARGE]
   
	\node[fill=white          ,draw=black,inner sep=0pt,minimum size=1.2cm] (0)      at (-8,8) {\Huge{$\mathcal{T}_h^0$}};
	\node[fill=white          ,draw=black,inner sep=0pt,minimum size=0.6cm] (00)     at (-11,6)   {};
	\node[fill=cyan!30!white  ,draw=black,inner sep=0pt,minimum size=0.6cm] (01)     at (-9,6)    {};
	\node[fill=cyan!30!white  ,draw=black,inner sep=0pt,minimum size=0.6cm] (10)     at (-7,6)    {};
	\node[fill=white          ,draw=black,inner sep=0pt,minimum size=0.6cm] (11)     at (-5,6)    {};
	\node[fill=olive!30!white ,draw=black,inner sep=0pt,minimum size=0.6cm] (0000)   at (-12.5,4) {};
	\node[fill=olive!30!white ,draw=black,inner sep=0pt,minimum size=0.6cm] (0001)   at (-11.5,4) {};
	\node[fill=olive!30!white ,draw=black,inner sep=0pt,minimum size=0.6cm] (0010)   at (-10.5,4) {};
	\node[fill=olive!30!white ,draw=black,inner sep=0pt,minimum size=0.6cm] (0011)   at (-9.5,4)  {};
	\node[fill=cyan!30!white  ,draw=black,inner sep=0pt,minimum size=0.6cm] (1100)   at (-6.5,4)  {};
	\node[fill=cyan!30!white  ,draw=black,inner sep=0pt,minimum size=0.6cm] (1101)   at (-5.5,4)  {};
	\node[fill=white          ,draw=black,inner sep=0pt,minimum size=0.6cm] (1110)   at (-4.5,4)  {};
	\node[fill=purple!30!white,draw=black,inner sep=0pt,minimum size=0.6cm] (1111)   at (-3.5,4)  {};
	\node[fill=purple!30!white,draw=black,inner sep=0pt,minimum size=0.6cm] (111000) at (-6,2)    {};
	\node[fill=purple!30!white,draw=black,inner sep=0pt,minimum size=0.6cm] (111001) at (-5,2)    {};
	\node[fill=purple!30!white,draw=black,inner sep=0pt,minimum size=0.6cm] (111010) at (-4,2)    {};
	\node[fill=purple!30!white,draw=black,inner sep=0pt,minimum size=0.6cm] (111011) at (-3,2)    {};
    
    \draw[black] (0) -- (00); \draw[black] (0) -- (01); \draw[black] (0) -- (10); \draw[black] (0) -- (11);
    \draw[black] (00) -- (0000); \draw[black] (00) -- (0001); \draw[black] (00) -- (0010); \draw[black] (00) -- (0011);
    \draw[black] (11) -- (1100); \draw[black] (11) -- (1101); \draw[black] (11) -- (1110); \draw[black] (11) -- (1111);
    \draw[black] (1110) -- (111000); \draw[black] (1110) -- (111001); \draw[black] (1110) -- (111010); \draw[black] (1110) -- (111011);
    
    \draw[black, line width = 0.25 mm, ->, double] (-12.5,4) node[anchor=east,xshift=-0.5cm] {\Huge{$\mathcal{T}_h$}} -- (-11.5,4) -- (-10.5,4) -- (-9.5,4) -- (-9,6) -- (-7,6) -- (-6.5,4) -- (-5.5,4) -- (-6,2) -- (-5,2) -- (-4,2) -- (-3,2) -- (-3.5,4);
    
    \draw[darkgray, line width = 0.5 mm, dashed, dash pattern = on 10pt off 4pt] (0) -- (-10,6) -- (-10,5.5) -- (-8.5,4.5) -- (-8.5,0);
    \draw[darkgray, line width = 0.5 mm, dashed, dash pattern = on 10pt off 4pt] (0) -- (-5,5) -- (-5,3) -- (-6.5,3) -- (-6.5,0);
    
    \node[black] at (-11,0.5)  {$P_1$};
    \node[black] at (-7.5,0.5) {$P_2$};
    \node[black] at (-4.5,0.5) {$P_3$};
    
    
    \node[black] at (-15,8) {0};
    \node[black] at (-15,6) {1};
    \node[black] at (-15,4) {2};
    \node[black] at (-15,2) {3};
    \node[black] at (-16,5) {$l$};
    \draw[black] (-15.5,8.5) -- (-15.5,1.5);
    
    
    \fill[olive!30!white]  (0,0) rectangle (4,4);
	\fill[cyan!30!white]   (4,0) rectangle (8,6);
	\fill[cyan!30!white]   (0,4) rectangle (4,8);
	\fill[purple!30!white] (4,6) rectangle (8,8);

    \draw[darkgray, scale =  2] (0,0) grid (2,2);
    \draw[darkgray, scale =  2] (2,2) grid (4,4);
    \draw[darkgray, scale =  1] (4,6) grid (6,8);
    \node[black] at (1,7) {\Huge{$\mathcal{T}_h$}};
    
    \draw[black, line width = 0.5 mm, dashed, dash pattern = on 10pt off 4pt] (4,0) -- (4,4) -- (0,4);
    \draw[black, line width = 0.5 mm, dashed, dash pattern = on 10pt off 4pt] (4,8) -- (4,6) -- (8,6);
    
    \draw[black, line width = 0.5 mm]    (0,0) rectangle (8,8);
    \draw[black, line width = 0.5 mm,->] (0,0) -- (1,0) node[anchor=north,yshift=-0.2cm] {x};
    \draw[black, line width = 0.5 mm,->] (0,0) -- (0,1) node[anchor=east ,xshift=-0.2cm] {y};
    
    \draw[black, line width = 0.25 mm, ->, double] (1,1) -- (3,1) -- (1,3) -- (3,3) -- (6,2) -- (2,6) -- (5,5) -- (7,5) -- (4.5,6.5) -- (5.5,6.5) -- (4.5,7.5) -- (5.5,7.5) -- (7,7);

	\fill[blue]  (2,4) circle [radius=0.15] node[xshift=0.3cm,yshift=0.3cm] {$\bf{b}$};
	\fill[red]   (4,7) circle [radius=0.15] node[xshift=0.3cm,yshift=0.3cm] {$\bf{a}$};	
	\fill[black] (0,4) circle [radius=0.15] node[xshift=0.3cm,yshift=0.3cm] {$\bf{1}$};
	\fill[black] (4,4) circle [radius=0.15] node[xshift=0.3cm,yshift=0.3cm] {$\bf{2}$};

\end{tikzpicture}}
	\caption{The hierarchical construction of $\mathcal{T}_h$ gives rise to a 
	one-to-one correspondence between the cells of $\mathcal{T}_h$ and the leaves 
	of a quadtree rooted in $\mathcal{T}_h^0$. $\mathcal{T}_h$ does not satisfy 
	the \emph{2:1 balance}, e.g. the red hanging vertex $a$ is not permitted. 
	Assuming a discretization with conforming Lagrangian linear \acp{fe}, the 
	value of the \ac{dof} at hanging vertex $b$ is subject to the constraint $V_b 
	= 0.5 V_1 + 0.5 V_2$. $\mathcal{T}_h$ is partitioned into three subdomains, 
	$P1$, $P2$ and $P3$, using the $z$-order curve obtained by traversing the 
	leaves of the octree in increasing Morton index. Adapted 
	from~\citep{BursteddeWilcoxGhattas11}.}
	\label{fig:octree}
\end{figure}
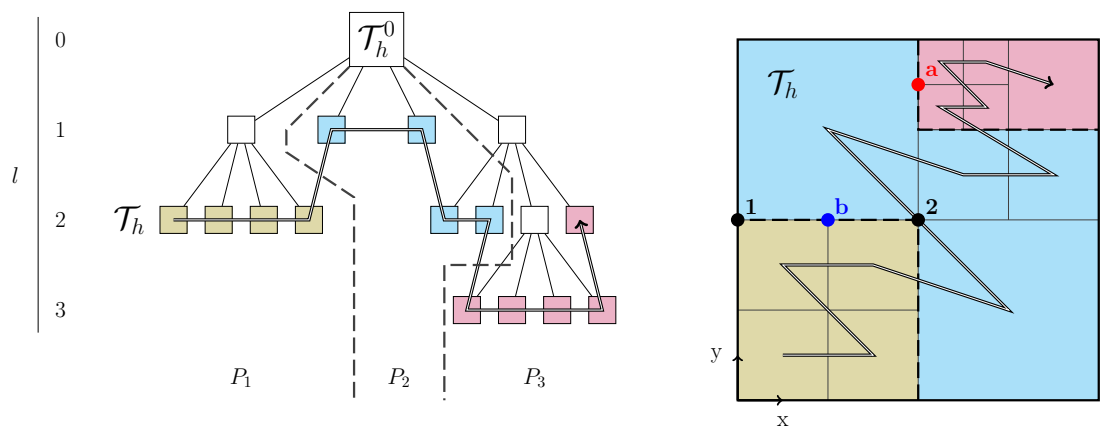

A cell marked for refinement is partitioned into four (2D) or eight (3D) 
children cells by bisecting all cell edges. As a result, $\mathcal{T}_h$ can be 
interpreted as a collection of quadtrees (2D) or octrees (3D), where the cells 
of $\mathcal{T}_h^0$ are the roots of these trees, and children cells (a.k.a. 
quadrants or octants) branch off their parent cells. The leaf cells in this 
hierarchy form the mesh in the usual meaning, i.e. $\mathcal{T}_h$. 
Furthermore, for any cell $K \in \mathcal{T}_h$, $l(K)$ is the \emph{level} of 
refinement of $K$ in the aforementioned hierarchy. In particular, $l(K) = 0$ 
for the root cells and $l(K) = l(\rm{parent}(K)) + 1$, otherwise. The level can 
also be defined for lower dimensional mesh entities (vertices, edges, faces) as 
in~\citep[Def. 2.4]{isaac2015recursive}. Fig.~\ref{fig:octree} illustrates this 
recursive tree structure with cells at different levels of refinement stemming 
from a single root.

Octree meshes admit a very compact computer representation, based on Morton 
encoding~\citep{morton1966computer} by bit interleaving, which enables 
efficient manipulation in high-end distributed-memory 
computers~\citep{BursteddeWilcoxGhattas11}. Moreover, they provide 
multi-resolution capability by local adaptation, as the leaves in the hierarchy 
can be at different levels of refinement. However, they are potentially 
nonconforming by construction, e.g. there can be \emph{hanging} vertices in the 
middle of an edge or face or \emph{hanging} edges or faces in touch with a 
coarser geometrical entity. 

Nonconformity introduces additional complexity in the implementation of 
conforming \acp{fe}, especially in parallel codes for distributed-memory 
computers. This degree of complexity is nevertheless significantly reduced by 
enforcing the so-called \emph{2:1 balance} relation, i.e. adjacent cells may 
differ at most by a single level of refinement. In this sense, the mesh in 
Fig.~\ref{fig:octree} violates the 2:1 balance, because hanging vertex $a$ is 
surrounded by cells that differ by two levels.

\added{In order to preserve the continuity of a conforming \ac{fe} 
approximation, \acp{dof} sitting on \emph{hanging} geometric entities cannot 
have an arbitrary value, they are constrained by neighbouring nonhanging 
\acp{dof}. The approach advocated in this work to handle the hanging node 
constraints~\citep{kuus2017coupling} consists in eliminating them from the 
local matrices, before assembling the global matrix. In this case, hanging 
\acp{dof} are not associated to global \acp{dof} and, thus, a row/column in the 
global linear system of equations.}

%

\deleted{On the other hand, as with constrained \acp{dof} of Dirichlet type, 
there are two main strategies to handle hanging \acp{dof}. One option is to 
keep them in the global system and impose additional constraints to ensure 
continuity. Alternatively, the approach advocated in this 
work~\citep{kuus2017coupling} eliminates the hanging node constraints of the 
local matrices, before assembling the global matrix. In this case, hanging 
\acp{dof} are not associated to global \acp{dof} and, thus, a row/column in the 
global linear system of equations.}

\subsection{Partitioning the octree mesh}

Efficient and scalable parallel partitioning schemes for adaptively refined 
meshes are still an active research topic. Our computational 
framework relies on the \texttt{p4est} 
library~\citep{BursteddeWilcoxGhattas11}. \texttt{p4est} is a \ac{mpi} library 
for efficiently handling (forests of) octrees that has scaled up to hundreds of 
thousands of cores~\citep{bangerth2011algorithms}. Using the properties of the 
Morton encoding, \texttt{p4est} offers, among others, parallel subroutines to 
refine (or coarsen) the octants of an octree, enforce the 2:1 balance ratio and 
partition/redistribute the octants among the available processors to keep the 
computational load 
balanced~\citep{BursteddeWilcoxGhattas11,isaac2015recursive}. \added{Data 
structures and algorithms involved in the interface between \texttt{p4est} and 
\texttt{FEMPAR} are detailed in~\citep{badia2018on}. They are configured 
according to a two-layered meshing approach. The first or \emph{inner} layer is 
a light-weight encoding of the forest-of-trees, handled by \texttt{p4est}. The 
second or \emph{outer} layer is a richer mesh representation, suitable for 
generic finite elements.}

The principle underlying the mesh partitioner is the use of space-filling 
curves. Octants within an octree can be naturally assigned an ordering by a 
traversal across all leaves, e.g. increasing Morton index, as shown in 
Fig.~\ref{fig:octree}. Application of the one-to-one correspondence between 
tree nodes and octants reveals that this one-dimensional sequence corresponds 
exactly to a $z$-order space-filling curve of the triangulation 
$\mathcal{T}_h$. Hence, the problem of partitioning $\mathcal{T}_h$ can be 
reformulated into the much simpler problem of subdividing a one-dimensional 
curve. This circumvents the parallel scaling bottleneck that commonly arises 
from {\em dynamic load balancing} via graph partitioning 
algorithms~\citep{Karypis1998,Karypis1999}.

However, the simplicity comes at a price related to the fact that, among 
space-filling curves, $z$-curves have unbounded locality and low bounding-box 
quality~\citep{haverkort2008locality}. In our context, this leads to the 
emergence of poorly-shaped and, possibly, disconnected subdomains (with at most 
two components~\citep{Burstedde2018} for single-octree meshes). Bad quality of 
subdomains affects the performance of nonoverlapping \ac{dd} 
methods~\citep{kuus2017coupling}.

\section{Modelling the growth of the geometry}
\label{sec:growth}

\subsection{\replaced{Parallel}{The} element-birth method}
\label{subsec:ebm}

As mentioned in Sect.~\ref{sec:introduction}, the growth of the geometry is 
modelled in a \emph{background} \ac{fe} mesh that, even if it is refined or 
coarsened, always covers the same  domain.

Let $\Omega(t)$ be a \emph{growing-in-time} domain.\deleted{ with the initial 
hypotheses stated in Sect.~\ref{subsec:octrees}.} During the time interval 
$[t_{\rm i},t_{\rm f}]$, $\Omega(t)$ transforms from an initial domain 
$\Omega_{\rm i}$ to a final one $\Omega_{\rm f}$. For the sake of simplicity, 
\ac{amr} is not considered for now, only later in the exposition, and it is 
assumed that there exists (1) a \emph{background} conforming partition 
$\mathcal{T}_h \equiv \mathcal{T}_h^0 = \{K\}$ of $\Omega_{\rm f}$\added{, 
where $\Omega_{\rm f}$ and $\mathcal{T}_h^0$ satisfy the hypotheses stated in 
the first paragraph of Sect.~\ref{subsec:octrees}}, and (2) a time 
discretization $t_{\rm i} = t_0 < t_1 < \ldots < t_{N_t} = t_{\rm f}$ such 
that, for all $j = 0, \ldots, N_t$, a partition $\mathcal{T}_{h,j}$ of 
$\Omega(t_j)$ can be obtained as a subset of cells of $\mathcal{T}_{h}$. In 
other words, a body-fitted mesh of $\Omega_{\rm f}$ can be built so that 
subsets of this mesh can be taken as body-fitted meshes of $\Omega(t_j),$ $j = 
0, \ldots, N_t$. As $\Omega(t)$ grows in time, the relation $\mathcal{T}_{h,\rm 
i} = \mathcal{T}_{h,0} \subseteq \mathcal{T}_{h,1} \subseteq \ldots \subseteq 
\mathcal{T}_{h,N_t} = \mathcal{T}_{h,\rm f}$ holds.

This setting is typical in welding or \ac{am} simulations. In \ac{am}, for 
instance, it is frequently required that the mesh of the component conforms to 
the layers. On the other hand, the method presented below can be adapted with 
little effort to a more general setting, where the growth-fitting requirement 
is dismissed by resorting to unfitted or immersed boundary methods 
\cite{badia2018aggregated}. In this case, $\mathcal{T}_{h}$ is a triangulation 
of an artificial domain $\Omega_{\rm art}$, such that it includes the final 
physical domain, i.e. $\Omega_{\rm f} \subset \Omega_{\rm art}$, but it is also 
characterized by a simple geometry, easy to mesh with Cartesian grids.

Consider now partitions of $\mathcal{T}_{h}$ of the form 
$\{\mathcal{T}_{h,j},\mathcal{T}_{h} \setminus \mathcal{T}_{h,j}\}, \ j = 0, 
\ldots, N_t$. In this classification, the cells in $\mathcal{T}_{h,j}$ are 
referred to as the \emph{active} cells $K_{\rm ac}$, while the ones in 
$\mathcal{T}_{h} \setminus \mathcal{T}_{h,j}$ as the \emph{inactive} cells 
$K_{\rm in}$. The key point of the \ac{ebm} is to assign 
degrees of freedom only in active cells, that is, the computational domain, at 
any $j = 0, \ldots, N_t$, is defined by $\mathcal{T}_{h,j} = \{K_{\rm ac}\}$. 
\added{In this way, inactive cells do not play any role in the numerical 
approximation; they have no contribution to the global linear system, in 
contrast with the quiet-element method~\citep{michaleris_modeling_2014}. 
Besides,} note the similarities of this approach to the one employed in 
unfitted \ac{fe} methods~\citep{badia2017robust,badia2018aggregated}, 
distinguishing interior and cut (active) cells from exterior (inactive) cells.

This representation of a growing domain is completed with a procedure to update 
the computational mesh during a time increment. The most usual approach is to 
use a search algorithm to find the set of cells in $\mathcal{T}_{h,j+1} 
\setminus \mathcal{T}_{h,j}, \ j = 0, \ldots, N_t - 1$, referred to as the 
\emph{activated} cells $K_{\rm acd}$. Then, $\mathcal{T}_{h,j+1} = 
\mathcal{T}_{h,j} \cup \{K_{\rm acd}\}$ defines the next computational mesh, as 
illustrated in Fig.~\ref{fig:feactivation}. This means that 
$\mathcal{T}_{h,j+1}$ receives the old \ac{dof} global identifiers (and FE 
function \ac{dof} values) from $\mathcal{T}_{h,j}$ and has new degrees of 
freedom assigned in the activated cells. The initial value of these new 
\acp{dof} is set with a criterion that depends on the application problem, as 
seen in Sect.~\ref{sec:formulation}.


\added{Therefore, in a parallel distributed-memory environment, there are two 
different partitions of $\mathcal{T}_{h,j}$ playing a role in the simulation: 
(1) into subdomains $\mathcal{T}_{h,j}^i, \ i = 1, \ldots, n^{\rm sbd}$ and (2) 
into active $K_{\rm ac}$ and inactive $K_{\rm in}$ cells. Assuming a one-to-one 
mapping among subdomains and CPU cores, in our approach~\citep{badia2018on}, 
each processor stores the geometrical information corresponding to the 
subdomain portion $\mathcal{T}_{h,j}^i$ of the global mesh and one layer of 
off-processor cells surrounding the local subdomain mesh, the so-called 
\emph{ghost} cells. With regards to (2), each cell is associated a 
\emph{status}, e.g. an integer value, that expresses whether it is an active or 
inactive cell. It follows that $\mathcal{T}_{h,j}^i$ can be composed of both 
\emph{active} and \emph{inactive} cells.} 

\added{As explained, the update of (2) follows from finding the subset $K_{\rm 
acd}$. \emph{Local} activated cells, i.e. $\mathcal{T}_{h,j}^i \cap \{K_{\rm 
acd}\}$, are found with the search algorithm. Afterwards, a nearest-neighbour 
communication is carried out to update the status at the ghost cells. The 
second step is necessary to know whether a face sitting on the interprocessor 
contour is in the interior or at the boundary of the domain 
$\mathcal{T}_{h,j+1}$.}

\added{Bringing now briefly \ac{amr} into the discussion, some considerations 
with regards to the \ac{ebm} must be taken into account when applying mesh 
transformations. First, when refining a cell, its status is inherited by the 
child cells. A cell can only be coarsened, when all siblings have the same 
status, otherwise the computational domain is perturbed. Finally, after a 
partition/redistribution of cells across processors, the status of the 
redistributed cells must be migrated with interprocessor communication. Adding 
this to the update of the status at ghost cells, the \ac{ebm} in a parallel 
\ac{amr} framework only demands two extra interprocessor data transfers with 
respect to a standard \ac{fe} simulation pipeline.}

\subsection{\added{Parallel search algorithm.}}
\label{subsec:search}

\added{The update of the computational mesh consists in finding the cells in 
the mesh that are inside the \emph{known} growing region of the current time 
increment. In this sense, the problem is a standard and well known collision 
detection that can be tackled with any of the many existing algorithms. Our 
goal is to derive from them a strategy that is both computationally inexpensive 
and highly-parallelizable for octree-based meshes.}

\added{The approach adopted in this work is founded on the 
\ac{hst}~\citep{boyd2004convex}. It states that given $A$ and $B$ two disjoint, 
nonempty and \emph{convex} subsets of $\mathbb{R}^n$, there exists a nonzero 
vector $v$ and a real number $c$, such that $\langle x,v \rangle \geq c$ and 
$\langle y,v \rangle \leq c$, for any $x \in A, \ y \in B$. In other words, the 
hyperplane $\langle \cdot,v \rangle = c$, with $v$ its normal vector, separates 
$A$ and $B$. A corollary of this theorem is that, if $A$ and $B$ are convex 
\emph{polyhedra}, possible separating planes are either parallel to a face of 
one of the polyhedra or contain an edge from each of the polyhedra.}

\added{In our context, assuming the \ac{fe} mesh is formed by rectangular 
hexahedra and the search volume is a cuboid, as in Sect.~\ref{sec:formulation}, 
the purpose is to test the intersection between two cuboids (any mesh cell vs 
search volume). In this case, the \ac{hst} narrows down the number of potential 
separating planes to fifteen~\citep{eberly2002dynamic,gottschalk1996obbtree}: 
three for the independent faces of the cell, three for the independent faces of 
the search cuboid and nine generated by all possible independent pairs formed 
by an edge from the cell and an edge from the search cuboid. It follows that 
two cuboids intersect, if and only if, none of the fifteen possible separating 
planes exists.}

\begin{figure}[!h]
 \centering
 \resizebox{0.5\textwidth}{!}{%

\begin{tikzpicture}
	
	\filldraw[fill=orange!80!white] (2.0,4.0) circle [radius=1.75];
	\filldraw[fill=olive!80!white] (6.0,5.0) -- (8.0,7.0) -- (6.0,9.0) -- 
	(4.0,8.0) -- (4.0,6.0) -- cycle;
	
	\draw[ultra thick] (0.5,9.0) -- node[above,near start,sloped] {separating 
	plane} (7.5,2.0);
	\draw[ultra thick] (2.0,0.0) -- node[above,very near end,sloped] {separating 
	axis} (11.0,9.0);
	
	\draw[dashed] (0.763,2.763) -- (2.763,0.763);
	\draw[dashed] (3.237,5.237) -- (5.237,3.237);
	
	\draw[dashed] (4.0,6.0) -- (6.0,4.0);
	\draw[dashed] (8.0,7.0) -- (8.5,6.5);
	
	\draw[thin] (5.5,4.0) -- (5.25,3.75) -- (5.5,3.5);
	
	\draw[line width = 1.5mm, orange!80!white] (2.763,0.763) -- (5.237,3.237);
	\draw[line width = 1.5mm, olive!80!white]  (6.0,4.0) -- (8.5,6.5);
	
\end{tikzpicture}%
 }%
 \caption{\added{Illustration of the \ac{hst}. A separating plane can be tested 
 by examining whether the projections of the two convex bodies onto a line 
 perpendicular to the plane intersect or not.}}
 \label{fig:separating-axis}
\end{figure}
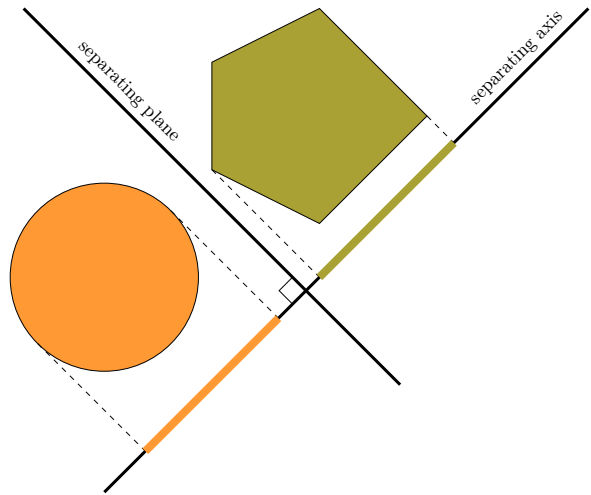

\added{A separating plane can be tested by comparing the projections of the 
cuboids onto a line perpendicular to the plane, referred to as the separating 
axis (Fig.~\ref{fig:separating-axis}). If the intervals of the projections do 
not intersect, then the cuboids do not intersect themselves. In 
\citep{eberly2002dynamic}, it is shown how this test amounts to compare two 
real quantities that depend on the dimensions and unit directions of the cell, 
the dimensions and unit directions of the search cuboid and the vector joining 
the centroids of the two cuboids. For each separating axis, the nonintersection 
test in terms of these quantities is given in~\citep[Tab. 1]{eberly2002dynamic} 
or~\citep[Sect. 4.6]{gottschalk2000collision}.} 

\added{At this point, it remains to see how this test can be exploited for the 
parallel search algorithm with adaptive octree-based meshes. Dropping the 
assumption in  Sect.~\ref{subsec:ebm} of sequential mesh inclusion, i.e. 
$\mathcal{T}_{h,j} \neq \mathcal{T}_{h,j+1}, \forall j = 0, \ldots, N_t$, now 
$\mathcal{T}_{h,j}$ is transformed into $\mathcal{T}_{h,j+1}$, by applying 
several refinement (coarsening) operations to the octants intersecting 
(nonintersecting) the search cuboid of the $j \rightarrow j+1$ time increment. 
The transformation finishes when all octants intersecting the search cuboid 
have a given maximum level of refinement. In fact, these octants form precisely 
the subset $K_{\rm acd}$. The number of transformations required is 
problem-dependent, but it is upper-bounded by the difference between the 
user-prescribed maximum and minimum levels of refinement.}

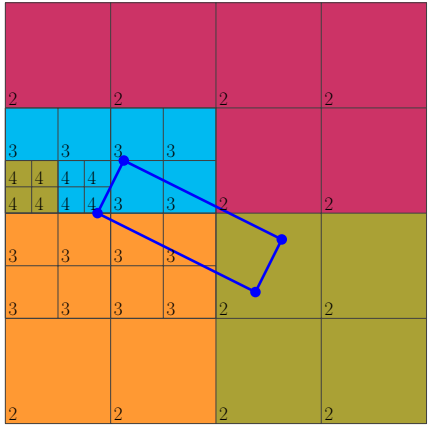
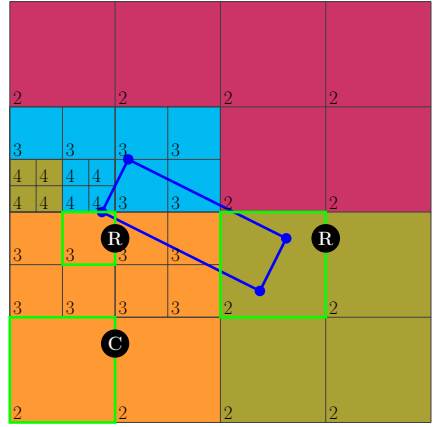
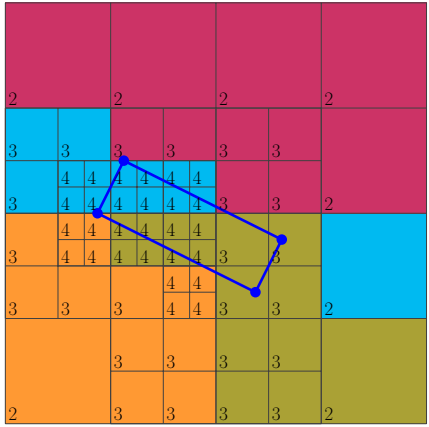
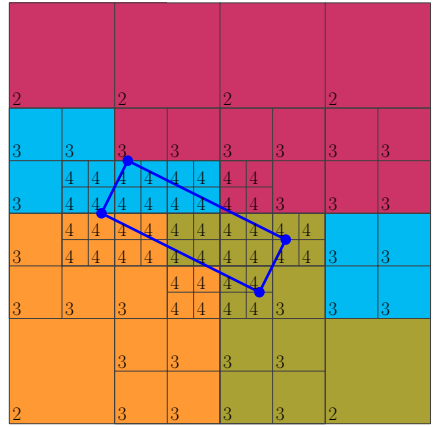
\begin{figure}[!h]
 \centering
 \begin{subfigure}[t]{0.45\textwidth}
  	\centering
   \scalebox{0.35}{

\begin{tikzpicture}[font=\Huge]
	
	\fill[orange!80!white] (0,0) rectangle (8,8);
	\fill[olive!80!white] (0,8) rectangle (2,10);
	\fill[cyan!80!white] (2,8) rectangle (4,10);
	
	\fill[olive!80!white] (8,0) rectangle (16,8);
	\fill[cyan!80!white] (0,10) rectangle (4,12);
	\fill[cyan!80!white] (4,8) rectangle (8,12);
	\fill[purple!80!white] (0,12) rectangle (8,16);
	\fill[purple!80!white] (8,8) rectangle (16,16);
	
	\draw[darkgray, scale =  4] (0,0) grid (4,4);
	\draw[darkgray, scale =  2] (0,2) grid (4,6);
	\draw[darkgray, scale =  1] (0,8) grid (4,10);
	
	\draw[blue, line width = 1mm] (3.5,8) -- (4.5,10) -- (10.5,7) -- (9.5,5) -- 
	cycle;
	
	\fill[blue] (3.5,8) circle [radius=0.2];
	\fill[blue] (4.5,10) circle [radius=0.2];
	\fill[blue] (10.5,7) circle [radius=0.2];
	\fill[blue] (9.5,5) circle [radius=0.2];
	
	
	\node[anchor=south west] at (0,0) {2};
	\node[anchor=south west] at (4,0) {2};
	\node[anchor=south west] at (8,0) {2};
	\node[anchor=south west] at (12,0) {2};
	
	\node[anchor=south west] at (0,4) {3};
	\node[anchor=south west] at (2,4) {3};
	\node[anchor=south west] at (4,4) {3};
	\node[anchor=south west] at (6,4) {3};
	
	\node[anchor=south west] at (8,4) {2};
	\node[anchor=south west] at (12,4) {2};
	
	\node[anchor=south west] at (0,6) {3};
	\node[anchor=south west] at (2,6) {3};
	\node[anchor=south west] at (4,6) {3};
	\node[anchor=south west] at (6,6) {3};
	
	\node[anchor=south west] at (0,8) {4};
	\node[anchor=south west] at (1,8) {4};
	\node[anchor=south west] at (2,8) {4};
	\node[anchor=south west] at (3,8) {4};
	
	\node[anchor=south west] at (0,9) {4};
	\node[anchor=south west] at (1,9) {4};
	\node[anchor=south west] at (2,9) {4};
	\node[anchor=south west] at (3,9) {4};
	
	\node[anchor=south west] at (4,8) {3};
	\node[anchor=south west] at (6,8) {3};
	\node[anchor=south west] at (0,10) {3};
	\node[anchor=south west] at (2,10) {3};
	\node[anchor=south west] at (4,10) {3};
	\node[anchor=south west] at (6,10) {3};
	
	\node[anchor=south west] at (8,8) {2};
	\node[anchor=south west] at (12,8) {2};
	\node[anchor=south west] at (0,12) {2};
	\node[anchor=south west] at (4,12) {2};
	\node[anchor=south west] at (8,12) {2};
	\node[anchor=south west] at (12,12) {2};
	
\end{tikzpicture}}
   \caption{Given $\mathcal{T}_{h,j}, j = 0, \ldots, N_t$, compute the 
   search volume for the time increment $\Delta t^{j \rightarrow j+1}$.}
   \label{fig:search-a}
 \end{subfigure}
 \quad
 \begin{subfigure}[t]{0.45\textwidth}
   \centering
   \scalebox{0.35}{

\begin{tikzpicture}[font=\Huge]
	
	\fill[orange!80!white] (0,0) rectangle (8,8);
	\fill[olive!80!white] (0,8) rectangle (2,10);
	\fill[cyan!80!white] (2,8) rectangle (4,10);
	
	\fill[olive!80!white] (8,0) rectangle (16,8);
	\fill[cyan!80!white] (0,10) rectangle (4,12);
	\fill[cyan!80!white] (4,8) rectangle (8,12);
	\fill[purple!80!white] (0,12) rectangle (8,16);
	\fill[purple!80!white] (8,8) rectangle (16,16);
	
	\draw[darkgray, scale =  4] (0,0) grid (4,4);
	\draw[darkgray, scale =  2] (0,2) grid (4,6);
	\draw[darkgray, scale =  1] (0,8) grid (4,10);
	
	\draw[blue, line width = 1mm] (3.5,8) -- (4.5,10) -- (10.5,7) -- (9.5,5) -- 
	cycle;
	
	\fill[blue] (3.5,8) circle [radius=0.2];
	\fill[blue] (4.5,10) circle [radius=0.2];
	\fill[blue] (10.5,7) circle [radius=0.2];
	\fill[blue] (9.5,5) circle [radius=0.2];
	
	\draw[green, line width = 1mm] (0.0,0.0) rectangle (4.0,4.0) 
	node[white,circle,fill=black,minimum size=1cm,yshift=-1cm] {\textbf{C}};
	
	\draw[green, line width = 1mm] (2.0,6.0) rectangle (4.0,8.0) 
	node[white,circle,fill=black,minimum size=1cm,yshift=-1cm] {\textbf{R}};
	
	\draw[green, line width = 1mm] (8.0,4.0) rectangle (12.0,8.0) 
	node[white,circle,fill=black,minimum size=1cm,yshift=-1cm] {\textbf{R}};
	
	
	\node[anchor=south west] at (0,0) {2};
	\node[anchor=south west] at (4,0) {2};
	\node[anchor=south west] at (8,0) {2};
	\node[anchor=south west] at (12,0) {2};
	
	\node[anchor=south west] at (0,4) {3};
	\node[anchor=south west] at (2,4) {3};
	\node[anchor=south west] at (4,4) {3};
	\node[anchor=south west] at (6,4) {3};
	
	\node[anchor=south west] at (8,4) {2};
	\node[anchor=south west] at (12,4) {2};
	
	\node[anchor=south west] at (0,6) {3};
	\node[anchor=south west] at (2,6) {3};
	\node[anchor=south west] at (4,6) {3};
	\node[anchor=south west] at (6,6) {3};
	
	\node[anchor=south west] at (0,8) {4};
	\node[anchor=south west] at (1,8) {4};
	\node[anchor=south west] at (2,8) {4};
	\node[anchor=south west] at (3,8) {4};
	
	\node[anchor=south west] at (0,9) {4};
	\node[anchor=south west] at (1,9) {4};
	\node[anchor=south west] at (2,9) {4};
	\node[anchor=south west] at (3,9) {4};
	
	\node[anchor=south west] at (4,8) {3};
	\node[anchor=south west] at (6,8) {3};
	\node[anchor=south west] at (0,10) {3};
	\node[anchor=south west] at (2,10) {3};
	\node[anchor=south west] at (4,10) {3};
	\node[anchor=south west] at (6,10) {3};
	
	\node[anchor=south west] at (8,8) {2};
	\node[anchor=south west] at (12,8) {2};
	\node[anchor=south west] at (0,12) {2};
	\node[anchor=south west] at (4,12) {2};
	\node[anchor=south west] at (8,12) {2};
	\node[anchor=south west] at (12,12) {2};

\end{tikzpicture}}
   \caption{Loop over cells in $\mathcal{T}_{h,j}$. If nonintersection test 
   fails (passes), then mark cell for refinement (coarsening). Some examples 
   are highlighted (R = refine, C = coarsen).}
   \label{fig:search-b}
 \end{subfigure} \\ \vspace{0.2cm}
 \begin{subfigure}[t]{0.45\textwidth}
  	\centering
   \scalebox{0.35}{

\begin{tikzpicture}[font=\Huge]
	
	\fill[orange!80!white] (0,0)  rectangle (8,6);
	\fill[orange!80!white] (0,6)  rectangle (4,8);
	
	\fill[olive!80!white]  (8,0)  rectangle (16,4);
	\fill[olive!80!white]  (8,4)  rectangle (12,6);
	\fill[olive!80!white]  (4,6)  rectangle (12,8);
	
	\fill[cyan!80!white]   (12,4) rectangle (16,8);
	\fill[cyan!80!white]   (0,8)  rectangle (8,12);
	\fill[cyan!80!white]   (0,10) rectangle (4,12);
	
	\fill[purple!80!white] (4,10) rectangle (8,12);
	\fill[purple!80!white] (8,8)  rectangle (16,12);
	\fill[purple!80!white] (0,12) rectangle (16,16);
	
	\draw[darkgray, scale =  4] (0,0) grid (4,4);
	\draw[darkgray, scale =  2] (0,2) grid (6,6);
	\draw[darkgray, scale =  2] (2,0) grid (6,2);
	\draw[darkgray, scale =  1] (2,6) grid (8,10);
	\draw[darkgray, scale =  1] (6,4) grid (8,6);
	
	\draw[blue, line width = 1mm] (3.5,8) -- (4.5,10) -- (10.5,7) -- (9.5,5) -- 
	cycle;
	
	\fill[blue] (3.5,8) circle [radius=0.2];
	\fill[blue] (4.5,10) circle [radius=0.2];
	\fill[blue] (10.5,7) circle [radius=0.2];
	\fill[blue] (9.5,5) circle [radius=0.2];
	
	
	\node[anchor=south west] at (0,0) {2};
	\node[anchor=south west] at (12,0) {2};
	\node[anchor=south west] at (0,12) {2};
	\node[anchor=south west] at (4,12) {2};
	\node[anchor=south west] at (8,12) {2};
	\node[anchor=south west] at (12,4) {2};
	\node[anchor=south west] at (12,8) {2};
	\node[anchor=south west] at (12,12) {2};
	
	\node[anchor=south west] at (4,0) {3};
	\node[anchor=south west] at (6,0) {3};
	\node[anchor=south west] at (4,2) {3};
	\node[anchor=south west] at (6,2) {3};
	
	\node[anchor=south west] at (8,0) {3};
	\node[anchor=south west] at (10,0) {3};
	\node[anchor=south west] at (8,2) {3};
	\node[anchor=south west] at (10,2) {3};
	
	\node[anchor=south west] at (8,4) {3};
	\node[anchor=south west] at (10,4) {3};
	\node[anchor=south west] at (8,6) {3};
	\node[anchor=south west] at (10,6) {3};
	
	\node[anchor=south west] at (8,8) {3};
	\node[anchor=south west] at (10,8) {3};
	\node[anchor=south west] at (8,10) {3};
	\node[anchor=south west] at (10,10) {3};
	
	\node[anchor=south west] at (4,4) {3};
	\node[anchor=south west] at (2,4) {3};
	\node[anchor=south west] at (0,4) {3};
	\node[anchor=south west] at (0,6) {3};
	\node[anchor=south west] at (0,8) {3};
	\node[anchor=south west] at (0,10) {3};
	\node[anchor=south west] at (2,10) {3};
	\node[anchor=south west] at (4,10) {3};
	\node[anchor=south west] at (6,10) {3};
	
	\node[anchor=south west] at (2,6) {4};
	\node[anchor=south west] at (3,6) {4};
	\node[anchor=south west] at (2,7) {4};
	\node[anchor=south west] at (3,7) {4};
	\node[anchor=south west] at (2,8) {4};
	\node[anchor=south west] at (3,8) {4};
	\node[anchor=south west] at (2,9) {4};
	\node[anchor=south west] at (3,9) {4};

	\node[anchor=south west] at (4,6) {4};
	\node[anchor=south west] at (5,6) {4};
	\node[anchor=south west] at (4,7) {4};
	\node[anchor=south west] at (5,7) {4};
	\node[anchor=south west] at (4,8) {4};
	\node[anchor=south west] at (5,8) {4};
	\node[anchor=south west] at (4,9) {4};
	\node[anchor=south west] at (5,9) {4};
	
	\node[anchor=south west] at (6,4) {4};
	\node[anchor=south west] at (7,4) {4};
	\node[anchor=south west] at (6,5) {4};
	\node[anchor=south west] at (7,5) {4};
	\node[anchor=south west] at (6,6) {4};
	\node[anchor=south west] at (7,6) {4};
	\node[anchor=south west] at (6,7) {4};
	\node[anchor=south west] at (7,7) {4};
	\node[anchor=south west] at (6,8) {4};
	\node[anchor=south west] at (7,8) {4};
	\node[anchor=south west] at (6,9) {4};
	\node[anchor=south west] at (7,9) {4};

\end{tikzpicture}}
   \caption{Refine, coarsen and redistribute cells.}
   \label{fig:search-c}
 \end{subfigure}
 \quad
 \begin{subfigure}[t]{0.45\textwidth}
   \centering
   \scalebox{0.35}{

\begin{tikzpicture}[font=\Huge]
	
	\fill[orange!80!white] (0,0)  rectangle (8,6);
	\fill[orange!80!white] (0,6)  rectangle (6,8);
	
	\fill[olive!80!white]  (6,6)  rectangle (12,8);
	\fill[olive!80!white]  (8,0)  rectangle (12,6);
	\fill[olive!80!white]  (12,0) rectangle (16,4);
	
	\fill[cyan!80!white]   (12,4) rectangle (16,8);
	\fill[cyan!80!white]   (0,8)  rectangle (4,12);
	\fill[cyan!80!white]   (4,8)  rectangle (6,10);
	
	\fill[purple!80!white] (6,8)  rectangle (16,16);
	\fill[purple!80!white] (0,12) rectangle (6,16);
	\fill[purple!80!white] (4,10) rectangle (6,12);
	
	\fill[cyan!80!white]   (4,8)  rectangle (8,10);
	
	\draw[darkgray, scale =  4] (0,0) grid (4,4);
	\draw[darkgray, scale =  2] (0,2) grid (6,6);
	\draw[darkgray, scale =  2] (2,0) grid (4,2);
	\draw[darkgray, scale =  2] (4,0) grid (6,2);
	\draw[darkgray, scale =  2] (6,2) grid (8,6);
	\draw[darkgray, scale =  1] (2,6) grid (10,10);
	\draw[darkgray, scale =  1] (6,4) grid (10,6);
	\draw[darkgray, scale =  1] (10,6) grid (12,8);
	
	\draw[blue, line width = 1mm] (3.5,8) -- (4.5,10) -- (10.5,7) -- (9.5,5) -- 
	cycle;
	
	\fill[blue] (3.5,8) circle [radius=0.2];
	\fill[blue] (4.5,10) circle [radius=0.2];
	\fill[blue] (10.5,7) circle [radius=0.2];
	\fill[blue] (9.5,5) circle [radius=0.2];

	
	\node[anchor=south west] at (0,0) {2};
	\node[anchor=south west] at (12,0) {2};
	\node[anchor=south west] at (0,12) {2};
	\node[anchor=south west] at (4,12) {2};
	\node[anchor=south west] at (8,12) {2};
	\node[anchor=south west] at (12,12) {2};
	
	\node[anchor=south west] at (4,0) {3};
	\node[anchor=south west] at (6,0) {3};
	\node[anchor=south west] at (4,2) {3};
	\node[anchor=south west] at (6,2) {3};
	
	\node[anchor=south west] at (8,0) {3};
	\node[anchor=south west] at (10,0) {3};
	\node[anchor=south west] at (8,2) {3};
	\node[anchor=south west] at (10,2) {3};
	
	\node[anchor=south west] at (10,4) {3};
	\node[anchor=south west] at (12,4) {3};
	\node[anchor=south west] at (14,4) {3};
	\node[anchor=south west] at (12,6) {3};
	\node[anchor=south west] at (14,6) {3};
	\node[anchor=south west] at (12,8) {3};
	\node[anchor=south west] at (14,8) {3};
	\node[anchor=south west] at (12,10) {3};
	\node[anchor=south west] at (14,10) {3};
	
	\node[anchor=south west] at (10,8) {3};
	\node[anchor=south west] at (8,10) {3};
	\node[anchor=south west] at (10,10) {3};
	\node[anchor=south west] at (4,4) {3};
	\node[anchor=south west] at (2,4) {3};
	\node[anchor=south west] at (0,4) {3};
	\node[anchor=south west] at (0,6) {3};
	\node[anchor=south west] at (0,8) {3};
	\node[anchor=south west] at (0,10) {3};
	\node[anchor=south west] at (2,10) {3};
	\node[anchor=south west] at (4,10) {3};
	\node[anchor=south west] at (6,10) {3};
	
	\node[anchor=south west] at (2,6) {4};
	\node[anchor=south west] at (3,6) {4};
	\node[anchor=south west] at (2,7) {4};
	\node[anchor=south west] at (3,7) {4};
	\node[anchor=south west] at (2,8) {4};
	\node[anchor=south west] at (3,8) {4};
	\node[anchor=south west] at (2,9) {4};
	\node[anchor=south west] at (3,9) {4};

	\node[anchor=south west] at (4,6) {4};
	\node[anchor=south west] at (5,6) {4};
	\node[anchor=south west] at (4,7) {4};
	\node[anchor=south west] at (5,7) {4};
	\node[anchor=south west] at (4,8) {4};
	\node[anchor=south west] at (5,8) {4};
	\node[anchor=south west] at (4,9) {4};
	\node[anchor=south west] at (5,9) {4};
	
	\node[anchor=south west] at (6,4) {4};
	\node[anchor=south west] at (7,4) {4};
	\node[anchor=south west] at (6,5) {4};
	\node[anchor=south west] at (7,5) {4};
	\node[anchor=south west] at (6,6) {4};
	\node[anchor=south west] at (7,6) {4};
	\node[anchor=south west] at (6,7) {4};
	\node[anchor=south west] at (7,7) {4};
	\node[anchor=south west] at (6,8) {4};
	\node[anchor=south west] at (7,8) {4};
	\node[anchor=south west] at (6,9) {4};
	\node[anchor=south west] at (7,9) {4};
	
	\node[anchor=south west] at (8,4) {4};
	\node[anchor=south west] at (9,4) {4};
	\node[anchor=south west] at (8,5) {4};
	\node[anchor=south west] at (9,5) {4};
	\node[anchor=south west] at (8,6) {4};
	\node[anchor=south west] at (9,6) {4};
	\node[anchor=south west] at (8,7) {4};
	\node[anchor=south west] at (9,7) {4};
	\node[anchor=south west] at (8,8) {4};
	\node[anchor=south west] at (9,8) {4};
	\node[anchor=south west] at (8,9) {4};
	\node[anchor=south west] at (9,9) {4};
	
	\node[anchor=south west] at (10,6) {4};
	\node[anchor=south west] at (11,6) {4};
	\node[anchor=south west] at (10,7) {4};
	\node[anchor=south west] at (11,7) {4};
	
\end{tikzpicture}}
   \caption{Repeat (B)-(C) until all cells with maximum level of refinement 
   that intersect the cuboid have been found.}
   \label{fig:search-d}
 \end{subfigure}
 \caption{\added{2D example illustrating the iterative procedure to transform 
 $\mathcal{T}_{h,j}$ into $\mathcal{T}_{h,j+1}$. The maximum and minimum levels 
 of refinement are 4 and 2. The level of each cell is written at their lower 
 left corner. Each colour represents a different subdomain. Note that, from one 
 step to the next one, some cells that do not intersect the search volume have 
 to be refined to keep the 2:1 balance.}}
 \label{fig:search}
\end{figure}

\added{Therefore, the mesh transformation from $\mathcal{T}_{h,j}$ into 
$\mathcal{T}_{h,j+1}$ is carried out in a finite number of 
refinement/coarsening steps, each one determined by a cell-wise search. 
Specifically, the criterion to decide whether an octant is refined or coarsened 
is to perform the nonintersection test against the search cuboid. If it 
passes (fails), the octant is coarsened (refined). An example of this procedure 
is shown in Fig.~\ref{fig:search}. As observed, if all processors know the 
dimensions of the search cuboid, the algorithm is embarrassingly parallel, in 
particular, it does not require interprocessor communication. By construction 
of \texttt{p4est} a subdomain can only prescribe refinement/coarsening 
operations to its own local cells. It follows that the nonintersection test on 
ghost cells is redundant; the status of ghost cells must be updated at the end 
of the mesh transformation with a nearest-neighbour communication.}

\added{The search algorithm can be further accelerated by intersecting 
beforehand the search cuboid (or a bounding box of it) against the subdomain 
limits. In this case, the procedure can be skipped on those subdomains that do 
not intersect the cuboid. On the other hand, faster tests can be designed for 
uniformly refined meshes, such as checking whether the centroid of the cell is 
inside the search cuboid. Note that this test is not equivalent to the method 
of separating axes. Besides, it is not suitable for octree meshes. For 
instance, it may not transform the mesh at all if the search cuboid sits on top 
of heavily coarsened cells.}

\added{Apart from that, the algorithm is limited to rectangular meshes and 
search volumes. In more general cases, e.g. high-order meshes, the hypothesis 
of convexity may not hold; i.e. the hyperplane separation theorem cannot be the 
starting point; a more general method must be adopted instead. However, as 
shown in the example of Sect.~\ref{subsec:curved}, a high order mesh can be 
configured such that the search cuboid always overlaps a region of the 
mesh, with enough resolution to assume its cells are quasi-rectangular.}

\subsection{Dynamic load balancing}
\label{subsec:dynamic}

\deleted{Assuming from Sect.~\ref{sec:octrees} a single-octree mesh, 
$\mathcal{T}_{h}$ is now a nonconforming triangulation of a voxel domain 
$\Omega_{\rm oct}$, such that $\Omega_{\rm f} \subseteq \Omega_{\rm oct}$. 
Adding to this the element-birth method, there are two different partitions of 
$\mathcal{T}_{h}$ playing a role in the simulation: (1) into subdomains and (2) 
into active $K_{\rm ac}$ and inactive $K_{\rm in}$ cells.}

When designing a scalable application, \replaced{the partition into 
subdomains}{(1)} must be defined such that it evenly distributes among 
processors the total computational load. However, to this goal, \replaced{the 
\ac{ebm}}{(2)} adds two mutually excluding constraints; indeed, while the size 
of data structures and the complexity of procedures that manipulate the mesh 
grow with the total number of cells, those concerning the FE space, the FE 
system and the linear solver depend on the number of active cells (i.e. number 
of \acp{dof}).

The distribution of computational work can be tuned by allowing for a 
user-specified \emph{weight function} $w$ that assigns a non-negative integer 
value to each octant. The partition can then be constructed by equally 
distributing the accumulated weights of the octants that each processor owns, 
instead of the number of owned octants per processor. As \texttt{p4est} 
provides such capability (see~\citep[Sect. 3.3]{BursteddeWilcoxGhattas11}), the 
remaining question is to decide how to define $w$, taking into account the 
constraints above.

The answer depends on how the computational time is distributed among the 
different stages of the FE simulation pipeline. In the context of growing 
domains, \deleted{the }FE analysis is \added{a long} transient and it may be 
often desirable to reuse the same mesh for several time steps, seeking to 
minimize the \ac{amr} events and, thus, reduce simulation times. In this 
scenario, the number of time steps (linear system solutions) is greater than 
the number of mesh transformations and $w$ should favour the balance of active 
cells. With this idea in mind, the weight function can be defined as
\begin{equation}
 w_K =
 \begin{cases}
	w_a & \text{if } K \in \{ K_{\rm ac} \} \\
	w_i & \text{if } K \in \{ K_{\rm in} \} \\
 \end{cases}
 \label{eq:weight}
\end{equation}
\noindent{where $w_a \in \mathbb{N}$ and $w_i \in \mathbb{N}^0$. }

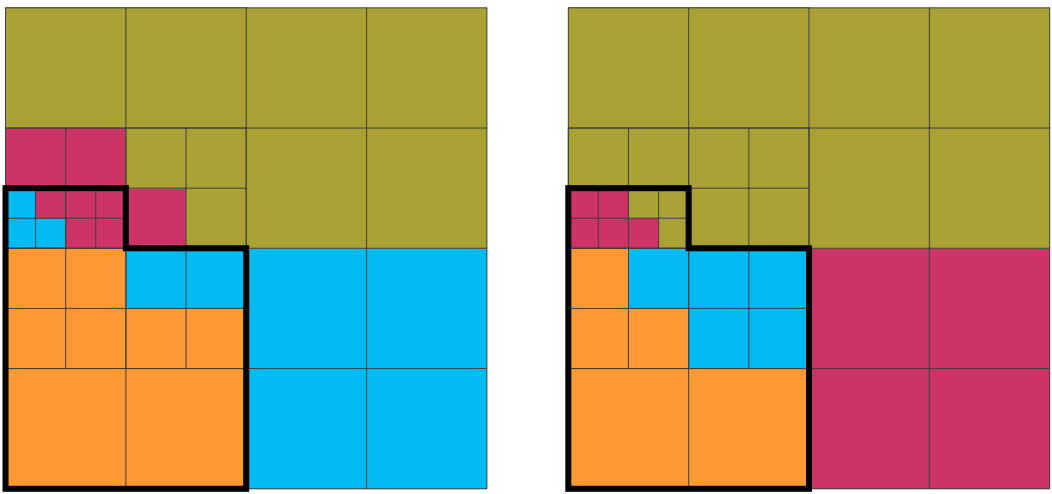
\begin{figure}[!h]
 \centering
 \begin{subfigure}[t]{0.45\textwidth}
  	\centering
   \scalebox{0.4}{

\begin{tikzpicture}[font=\Huge]
  
    \fill[orange!80!white] (0,0) -- (8,0) -- (8,6) -- (4,6) -- (4,8) -- (0,8) 
    -- cycle;
	\fill[cyan!80!white] (4,6) rectangle (8,8);
	\fill[cyan!80!white] (8,0) rectangle (16,8);
	\fill[cyan!80!white] (0,8) -- (2,8) -- (2,9) -- (1,9) -- (1,10) -- (0,10);
	\fill[purple!80!white] (2,8) rectangle (4,10);
	\fill[purple!80!white] (1,9) rectangle (2,10);
	\fill[purple!80!white] (0,10) rectangle (4,12);
	\fill[purple!80!white] (2,8) rectangle (6,10);
	\fill[olive!80!white] (4,10) rectangle (6,12);
	\fill[olive!80!white] (0,12) rectangle (8,16);
	\fill[olive!80!white] (6,8) rectangle (16,16);
    
    \draw[darkgray, scale =  4] (0,0) grid (4,4);
    \draw[darkgray, scale =  2] (0,2) grid (4,6);
    \draw[darkgray, scale =  1] (0,8) grid (4,10);
    
    \draw[black, line width = 2 mm] (0,0) -- (8,0) -- (8,8) -- (4,8) -- (4,10) -- (0,10) -- cycle;

\end{tikzpicture}}
   \caption{A default partition, that is, $(w_a,w_i) = (1,1)$, can result in a  
   poor balancing of \acp{dof} and even fully inactive parts (e.g. green 
   subdomain).}
   \label{fig:not_balanced}
 \end{subfigure}
 \quad
 \begin{subfigure}[t]{0.45\textwidth}
   \centering
   \scalebox{0.4}{

\begin{tikzpicture}[font=\Huge]
  
    \fill[orange!80!white] (0,0) -- (8,0) -- (8,4) -- (4,4) -- (4,6) -- (2,6) 
    -- (2,8) -- (0,8) -- cycle;
	\fill[cyan!80!white] (4,4) rectangle (8,8);
	\fill[cyan!80!white] (2,6) rectangle (4,8);
	\fill[purple!80!white] (0,8) rectangle (2,10);
	\fill[purple!80!white] (8,0) rectangle (16,8);
	\fill[olive!80!white] (2,8) rectangle (4,10);
	\fill[purple!80!white] (2,8) rectangle (3,9);
	\fill[olive!80!white] (0,10) rectangle (8,16);
	\fill[olive!80!white] (4,8) rectangle (8,10);
	\fill[olive!80!white] (8,8) rectangle (16,16);
    
    \draw[darkgray, scale =  4] (0,0) grid (4,4);
    \draw[darkgray, scale =  2] (0,2) grid (4,6);
    \draw[darkgray, scale =  1] (0,8) grid (4,10);
    
    \draw[black, line width = 2 mm] (0,0) -- (8,0) -- (8,8) -- (4,8) -- (4,10) -- (0,10) -- cycle;

\end{tikzpicture}}
   \caption{By setting partition weights to, e.g. $(w_a,w_i) = (10,1)$, the 
   active cells can be balanced, leading to a more equilibrated parallel 
   distribution of the \acp{dof}.}
   \label{fig:balanced}
 \end{subfigure}
 \caption{2D example illustrating how partition weights can be used to balance 
 dynamically the \acp{dof} across the processors. Each colour represents a 
 different subdomain. Active cells are enclosed by a thick contour polygon, 
 representing the computational domain.}
 \label{fig:weights}
\end{figure}

The effect of this weight function is illustrated in Fig.~\ref{fig:weights}. A 
uniform distribution of the octree octants, $(w_a,w_i) = (1,1)$, may lead to 
high load imbalance in the number of \acp{dof} per subdomain. There can even be 
fully inactive parts, as shown in Fig.~\ref{fig:not_balanced}, leaving the 
processors in charge of them mostly idling during local integration, assembly 
and solve phases. In contrast, $(w_a,w_i) = (1,0)$ gives the most uniform 
distribution of $\{K_{\rm ac}\}$, but it can also lead to extreme imbalance of 
$\{K_{\rm in}\}$ and, thus, the whole set of triangulation cells. 
Alternatively, pairs $(w_a,w_i)$ satisfying $w_a \gg w_i$ offer good compromise 
partitions. 

\section{Application to metal AM}
\label{sec:formulation}

\subsection{Heat transfer analysis}

After introducing the ingredients of the parallel \ac{fe} framework for growing 
geometries, the purpose now is to apply it to the thermal analysis of an 
additive manufacturing process by powder-bed fusion, such as Direct Metal Laser 
Sintering (DMLS). This manufacturing technology is illustrated in~\citep[Fig. 
1]{chiumenti_neiva_2017}. This will be the reference problem for the 
subsequent analysis with numerical experiments.

Let $\Omega(t)$ be a \emph{growing} domain in $\mathbb{R}^3$ as in 
Sect.~\ref{sec:growth}. Here, $\Omega(t)$ represents the component to be 
printed. The governing equation to find the temperature distribution $u$ in 
time is the balance of energy equation, expressed as
\begin{equation}
   C(u) \partial_{t}{u} - \boldsymbol{\nabla} \cdot ( k(u) \ \nabla u ) = f, 
   \quad \text{in} \enspace \Omega(t), \quad t \in [t_{\rm i},t_{\rm f}],    
\label{eq:balanceofenergy}
\end{equation}

\noindent{where $C(u)$ is the heat capacity coefficient, $k(u) \geq 0$ is the 
thermal conductivity and $f$ is the rate of energy supplied to the system per 
unit volume by the moving laser. $C(u)$ is given by the product of the density 
of the material $\rho(u)$ and the specific heat $c(u)$, but one may consider a 
modified heat capacity coefficient to also account for phase change 
effects~\citep{chiumenti_numerical_2017} or compute $C(u)$ with the CALPHAD 
approach~\citep{keller2017application,smith2016thermodynamically,kaufman1970computer}.}

Eq.~\eqref{eq:balanceofenergy} is subject to the initial condition
\begin{equation}
	u(\boldsymbol{x},t_{\rm i}) = u^0(\boldsymbol{x})
\label{eq:initial condition}
\end{equation}
\noindent{and the boundary conditions~\citep[Fig. 2]{chiumenti_neiva_2017} are 
(1) heat conduction through the building platform, (2) heat conduction through 
the powder-bed and (3) heat convection and radiation through the free surface. 
After linearising the Stefan-Boltzmann's law for heat 
radiation~\citep{chiumenti_numerical_2017}, all heat loss boundary conditions 
admit a unified expression in terms of Newton's law of cooling:}
\begin{equation}
   q_{\rm loss}(u,t) = h_{\rm loss}(u) (u - u_{\rm loss}(t)), \enspace 
   \text{in} \enspace \partial \Omega^{\rm loss}(t), \enspace t \in [t_{\rm 
   i},t_{\rm f}],
\label{eq:bondary_conditions}
\end{equation}

\noindent{where $loss$ refers to the kind of heat loss mechanism (conduction 
through solid, conduction through powder or convection and radiation) and the 
boundary region where it applies (contact with building platform, interface 
solid-powder or free surface).}

The weak form of the problem defined by 
Eqs.~(\ref{eq:balanceofenergy})-(\ref{eq:bondary_conditions}) can be stated as: 
\textit{Find $u(t) \in \mathcal{V}_t = \mathrm{H}^1(\Omega(t))$, almost 
everywhere in $(t_{\rm i},t_{\rm f}]$, such that}
\begin{equation}
   ( C(u) \partial_{t} u, v ) - ( k(u) \nabla u , \nabla v ) + 
   \langle h_{\rm loss}(u) u, v \rangle_{\partial \Omega^{\rm loss}} = \langle 
   f, v \rangle + \langle h_{\rm loss}(u) u_{\rm loss}, v \rangle_{\partial 
   \Omega^{\rm loss}}, \enspace \forall v \in \mathcal{V}_t.
\label{eq:weakform}
\end{equation}

Considering now $\mathcal{T}_h$ the triangulation of $\Omega_{\rm f}$, 
$\mathrm{V}_h \subset \mathcal{V}_{t_{\rm f}}$ a conforming \ac{fe} space for 
the temperature field and $\{ \varphi_j (\boldsymbol{x}) \}_{j=1}^{N_h}$ a 
\ac{fe} basis of the space $\mathrm{V}_h$, the semi-discrete form of 
Eq.~\eqref{eq:weakform}, after discretization in space with the Galerkin method 
and integration in time, e.g. with the semi-implicit backward Euler method, 
reads
\begin{equation}
	\begin{aligned}
		\left[ \frac{\mathbf{M}_C^n}{\Delta t^{n+1}} +  \mathbf{A}^n + 
		\mathbf{M}_{\rm loss}^n \right] \boldsymbol{\mathrm{U}}^{n+1} &= 
		\mathbf{b}_{f}^{n+1} + \frac{\mathbf{M}_C^n}{\Delta t^{n+1}} 
		\boldsymbol{\mathrm{U}}^n + \mathbf{b}_{\rm loss}^{n}, \\
		\boldsymbol{\mathrm{U}} (0) &= \boldsymbol{\mathrm{U}}^0,
	\end{aligned} 
\label{eq:semi-implicit}
\end{equation}
\noindent{where the time interval of interest $[t_{\rm i},t_{\rm f}]$ has been 
divided in subintervals $t_{\rm i} = t_0 < t_1 < \ldots < t_{N_t} = t_{\rm f}$ 
with $\Delta t^{n+1} = t_{n+1} - t_n$ variable for $n = 0, \ldots, N_t-1$. As a 
result of using the \ac{ebm} (Sect.~\ref{sec:growth}), 
Eq.~\eqref{eq:semi-implicit} is only constructed in $\Omega(t_n), \ n = 0, 
\ldots, N_t-1$. In other words, local integration and assembly of 
Eq.~\eqref{eq:weakform} is only carried out at the subset of active elements 
$\{K_{\rm ac}\} = \mathcal{T}_{h,n}$.}

Moreover, $\boldsymbol{\mathrm{U}} (t) = (\mathrm{U}_j(t))_{j=1}^{N_h}$ and 
$\boldsymbol{\mathrm{U}}^0 = (\mathrm{U}_j^0)_{j=1}^{N_h}$ are the components 
of $\mathrm{U}_h (t)$ and $\mathrm{U}_h^i$ with respect to the basis $\{ 
\varphi_j (\boldsymbol{x}) \}_{j=1}^{N_h}$, and the coefficients of 
$\mathbf{M}_\Box$, $\mathbf{A}$ and $\mathbf{b}_\Box$ are given by:
\begin{align*}
 & \mathrm{M}_{C}^{n,ij} = ( C (\boldsymbol{\mathrm{U}}^n) \varphi_i, \varphi_j 
 )_{\Omega(t_n)}, \enskip
   \mathrm{M}_{\rm loss}^{n,ij} = \langle h_{\rm loss} 
   (\boldsymbol{\mathrm{U}}^n) \varphi_i, \varphi_j \rangle_{\partial 
   \Omega^{\rm loss} \cap \partial \Omega(t_n)}, \enskip
   \mathrm{A}^{n,ij} =  ( k (\boldsymbol{\mathrm{U}}^n) \nabla \varphi_i, 
   \nabla \varphi_j )_{\Omega(t_n)} \\
                     & \mathbf{b}_{f}^{n+1,i}  = \langle \varphi_i, f^{n+1} 
                     \rangle_{\Omega(t_n)},  \enskip
   \mathbf{b}_{\rm loss}^{n,i}  = \langle h_{\rm loss} 
   (\boldsymbol{\mathrm{U}}^n) \varphi_i, \mathrm{U}_{\rm loss} (t^{n+1}) 
   \rangle_{\partial \Omega^{\rm loss} \cap \partial \Omega(t_n)}.
\end{align*} 

\added{An important characteristic of the physical problem is that, due to 
high heat capacity of metals and small time steps necessary to meet accuracy 
requirements, Eq.~\eqref{eq:semi-implicit} is often a mass-dominated linear 
system. As a result, Jacobi (or diagonal) preconditioning is adopted in the 
numerical examples of Sect.~\ref{sec:results}. Although this preconditioner 
does not alter the asymptotic behaviour of the condition number of the 
conductivity matrix ($\mathcal{O}(h^{-2})$), it corrects relative scales of 
Eq.~\eqref{eq:semi-implicit} arising from the fact that meshes resulting from 
(1) have cells at very different refinement levels, i.e. highly varying sizes.}

\subsection{FE modelling of the moving thermal load} 

As pointed out in Eq.~\eqref{eq:balanceofenergy}, $f$ is a moving thermal load 
that models the action of the laser in the system. But the moving heat source 
also drives the growth of the geometry in time, as the sintering process 
triggered by the laser transforms the metal powder into new solid material.

\begin{figure}[!h]
 \centering
 \scalebox{.70}{

\begin{tikzpicture}

  \begin{scope}

    \fill[fill=blue!20!white] (0,0) rectangle (6,2);
    \fill[fill=blue!20!white] (0,2) rectangle (2,3);
    \fill[fill=blue!70!white] (2,2) rectangle (4,3);
    \fill[pattern=dots] (4,2) rectangle (6,3);
  
    \draw[ultra thin, black!60!gray] (0,0) grid (6,6);
    \draw[scale =  6, black!60!gray] (0,0) grid (1,1);
    
    \draw[ultra thick] (0,0) -- (6,0) -- (6,2) -- (4,2) -- (4,3) -- (0,3) -- cycle;

    \filldraw[fill=yellow,draw=black!60!gray,ultra thin] (6.5,4.7) rectangle (7.0,4.9);
    \filldraw[fill=blue!70!white,draw=black!60!gray,ultra thin] (6.5,3.9) rectangle (7.0,4.1);
    \filldraw[fill=blue!20!white,draw=black!60!gray,ultra thin] (6.5,3.5) rectangle (7.0,3.7);
    \filldraw[pattern=dots,draw=black!60!gray,ultra thin] (6.5,3.1) rectangle (7.0,3.3);
    \draw[draw=black!60!gray,ultra thin] (6.5,2.7) rectangle (7.0,2.9);

    \draw[ultra thick] (6.5,1.9) rectangle (7.0,2.1);
    
    \draw[purple,ultra thick,->] (6.5,1.2) -- (7.0,1.2);
    \draw[teal,ultra thick,->>] (6.5,0.8) -- (7.0,0.8);
    \draw[yellow,ultra thick] (2.25,2.5) -- (3.5,2.5);
    \draw[yellow,thick] (3.5,2.5) -- (3.7,2.75);
    \draw[yellow,thick] (3.5,2.5) -- (3.45,2.8);
    \draw[yellow] (3.5,2.5) -- (3.3,2.6);
    \draw[yellow] (3.5,2.5) -- (3.7,2.6);
    \draw[yellow,thick] (3.5,2.5) -- (3.7,2.25);
    \draw[yellow,thick] (3.5,2.5) -- (3.45,2.2);
    \draw[yellow] (3.5,2.5) -- (3.3,2.4);
    \draw[yellow] (3.5,2.5) -- (3.7,2.4);
    \draw[yellow,thick] (3.5,2.5) -- (3.75,2.5);
    
    \node[font=\large,anchor=west] at (7.2,4.8) {Laser};
    
    \node[font=\large,anchor=west] at (7.2,4.0) {$K_{\rm acd}$: Activated in $\Delta \mathrm{t}$};
	\node[font=\large,anchor=west] at (7.2,3.6) {$\mathcal{T}_{h,j}$: Active at $\mathrm{t}_j$};
	\node[font=\large,anchor=west] at (7.2,3.2) {$K_{\rm in}$: Inactive (powder)};
    \node[font=\large,anchor=west] at (7.2,2.8) {$K_{\rm in}$: Inactive (gas)};
    
    \node[font=\large,anchor=west] at (7.2,2.0) {Contour at $\mathrm{t}^{j+1}$};
    \node[font=\large,anchor=west] at (7.2,1.2) {Heat conduction};
    \node[font=\large,anchor=west] at (7.2,0.8) {Heat conv. \& rad.};
    
    \foreach \i in {0,...,3}
        \draw[teal,ultra thick,->>] (0.5+\i,3) -- (0.5+\i,3.5);
        
    \foreach \i in {0,...,1}
        \draw[purple,ultra thick,->] (4.5+\i,2) -- (4.5+\i,2.35);
        
    \draw[purple,ultra thick,->] (4,2.5) -- (4.35,2.5);    
    
	\foreach \i in {0,...,6}
		\foreach \j in {0,...,2}
			\fill[black] (\i,\j) circle [radius=0.15];
			
	\foreach \i in {0,...,4}	
	    \fill[black] (\i,3) circle [radius=0.15];
	    
	\node[font=\Huge,anchor=west] at (-3,1.5) {$\boldsymbol{\mathcal{T}_{h,j+1}}$};
	\node[font=\Huge,anchor=west,gray] at (-3,5.0) {$\boldsymbol{\mathcal{T}_{h}}$};

  \end{scope}

\end{tikzpicture}}
 \caption{Illustration of the \emph{element-birth} method applied to the 
 thermal simulation of an \ac{am} process. As shown with this 2D \ac{fe} 
 cartesian grid, for any $j = 0, \ldots, N_t$, \acp{dof} are only assigned to 
 the set of active cells $\{K_{\rm ac}\} = \mathcal{T}_{h,j+1}$. A search 
 algorithm is employed to identify the set of activated cells $\{K_{\rm acd}\} 
 = \mathcal{T}_{h,j+1} \setminus \mathcal{T}_{h,j}$, where the laser is 
 focused during $\Delta t$. The computational mesh is then updated, by 
 assigning new \acp{dof} in $K_{\rm acd}$. Afterwards, the energy input, as 
 stated by Eq.~\eqref{eq:heat_source}, is uniformly distributed over $K_{\rm 
 acd}$. Adapted from~\citep{chiumenti_neiva_2017}.}
 \label{fig:feactivation}
\end{figure}
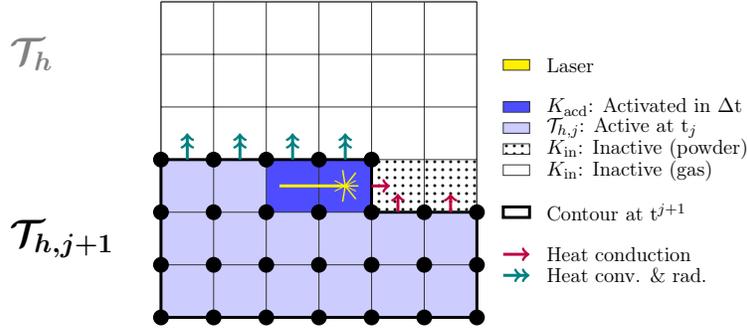

Therefore, the \ac{fe} modelling of the printing process requires a method to 
apply the volumetric heat source $f$ in space and time and track the growing 
$\Omega(t)$. The \ac{ebm} presented in Sect.~\ref{subsec:ebm} can serve both 
purposes, as seen in Fig.~\ref{fig:feactivation}. In this case, the set of 
activated cells $K_{\rm acd}$, representing the incremental growth region, is 
also affected by the laser during the time increment, i.e. the heat source term 
is also integrated in these cells.

Another comment arises on the update of the computational mesh. In general, one 
aims to follow the actual path of the laser in the machine, as faithfully as 
possible. \replaced{To this end, the search algorithm of 
Sect.~\ref{subsec:search} comes into play. The information of the laser path, 
together with other process parameters, such as the laser spot width, defines 
the search volume containing the cells affected by the energy input during the 
time step~\citep{chiumenti_numerical_2017}, subsequently referred to as the 
\ac{hav}.}{Therefore, a search algorithm is needed. It can use the information 
of the laser path, together with other process parameters, such as the width of 
the laser, to find the cells affected by the energy input during the time 
step~\citep{chiumenti_numerical_2017}.}

\replaced{If $f$ is taken as a uniform heat source, the average density 
distribution is }{However, only layer-by-layer \ac{fe} activations are 
considered in Sect.~\ref{sec:results}. In this case, the identification of the 
activated cells is trivial. Moreover, it is not necessary to resort to an 
accurate gaussian heat input model. Hence, $f$ is taken as a uniform heat 
source, with (average) density distribution}computed as 
	\begin{equation}
    f = \frac{\eta \ W}{\mathrm{V}_{\rm acd}},
    \label{eq:heat_source}
	\end{equation}
\noindent{where $W$ is the laser power $\mathrm{[watt]}$, $\eta$ is the heat 
absorption coefficient, a measure of the laser efficiency \added{and 
$\mathrm{V}_{\rm acd}$ is the volume of activated cells, i.e. intersecting the 
\ac{hav}}. \added{Goldak-based or surface Gaussian distributions may also be 
considered. In those cases, the heat source is evaluated with their 
corresponding analytical expressions in $\mathrm{K}_{\rm acd}$.} On the other 
hand, the initial temperature of the new \acp{dof} is set to the same value as 
the initial value at the building platform. More accurate alternatives are 
analysed in the literature~\citep{michaleris_modeling_2014}, but this aspect of 
the model is not relevant to the overall performance of the framework.

\section{\added{Computer implementation.}}
\label{sec:implementation}

\added{This section describes the software design of a parallel \ac{fe} 
framework for growing domains, the so-called \texttt{FEMPAR-AM} module, atop 
the services provided by \texttt{FEMPAR}~\citep{badia-fempar}. \texttt{FEMPAR} 
is an open source, general-purpose, object-oriented scientific software 
framework for the massively parallel \ac{fe} simulation of problems governed by 
\acp{pde}. \texttt{FEMPAR} software architecture is composed by a set of 
mathematically-supported abstractions that let its users break \ac{fe}-based 
simulation pipelines into smaller blocks that can be used and/or extended to 
fit users’ application requirements. Each  abstraction takes charge of a 
different step in a typical \ac{fe} simulation, including, among others, mesh 
generation, adaptation, and partitioning, construction of a global \ac{fe} 
space and \ac{dof} numbering, numerical evaluation of cell and facet integrals 
and linear system assembly, linear system solution or generation of simulation 
output data for later visualization. The reader is referred 
to~\citep{badia-fempar,badia2018on} for a detailed exposition of the main 
software abstractions in \texttt{FEMPAR}. Although \texttt{FEMPAR-AM} exploits 
most of the software abstractions provided by \texttt{FEMPAR}, the discussion 
in this section mainly focuses on those which had to be particularly set up 
and/or customized to support growing domains. Apart from this, the section also 
presents newly introduced software abstractions which are particular to 
\texttt{FEMPAR-AM}. The exposition is intended to help the reader grasp how any 
general-purpose \ac{fe} framework can be customized in order to deal with 
growing domains.}

\added{As \texttt{FEMPAR-AM} relies on the \ac{ebm} (Sect.~\ref{subsec:ebm}), 
the first software requirement that has to be fulfilled by the underlying 
\ac{fe} framework is the ability to build global \ac{fe} spaces that only carry 
out \acp{dof} for cells which are active at the current simulation step. In 
\texttt{FEMPAR}, a global \ac{fe} space is built from two main ingredients: (1) 
\texttt{triangulation\_t}~\citep[Sect. 7]{badia-fempar}, which represents the 
geometrical discretization of the computational domain into cells, and (2) 
\texttt{reference\_fe\_t}~\citep[Sect. 6]{badia-fempar}, which represents an 
\emph{abstract} local \ac{fe} on top of each of the triangulation cells. A 
particular type extension of \texttt{reference\_fe\_t}, referred to as 
\texttt{void\_reference\_fe\_t}~\citep[Sect. 6.5]{badia-fempar}, implements a 
local \ac{fe} with no degrees of freedom. The data structure in charge of 
building the global \ac{fe} space, i.e. \texttt{fe\_space\_t}~\citep[Sect. 
10]{badia-fempar}, is general enough such that one may use a different local 
\ac{fe} on different regions of the computational domain. By mapping 
\emph{active} cells to standard (e.g. Lagrangian) \acp{fe} and \emph{inactive} 
cells to \emph{void} \acp{fe}, \texttt{fe\_space\_t} is constructed such that 
it assigns global \ac{dof} identifiers only to the nodes of \emph{active} 
cells; while  nodes surrounded by \emph{inactive} cells do not receive a 
\ac{dof} identifier and, as a result, they neither assemble any contributions 
from local integration nor form part of the global linear system. On the other 
hand, the \texttt{triangulation\_t} software abstraction lets its users exploit 
the so-called \texttt{set\_id}~\citep[Sect. 7.1]{badia-fempar}, a cell-based 
integer attribute. Each cell of the triangulation can be assigned a 
\texttt{set\_id} number to group the cells into different subsets. In our case, 
this variable is used to store the current status of the cell in the mesh and, 
by \texttt{fe\_space\_t}, to determine which \texttt{reference\_fe\_t} (i.e. 
local \ac{fe} space) to put atop each cell (see discussion above). The update 
of the \texttt{set\_id} in the local cells is carried out within the search 
algorithm (Sect.~\ref{subsec:search}), whereas the update in the off-processor 
ghost cells reuses an existing procedure in \texttt{triangulation\_t} that 
invokes a nearest-neighbour communication. Another readily available method of 
\texttt{triangulation\_t} takes charge of migrating the \texttt{set\_id} values 
when the mesh is redistributed; see Sect.~\ref{subsec:ebm}.}

\added{Apart from the special set up of \texttt{fe\_space\_t} described in the 
previous paragraph, \texttt{FEMPAR-AM} also requires to customize 
\texttt{fe\_space\_t} (as-is in \texttt{FEMPAR}) to support growing domains. In 
particular, one needs to inject \ac{dof} values of any field (e.g. temperature 
in thermal \ac{am}) from $\mathcal{T}_{h,j-1}$ into $\mathcal{T}_{h,j}$ and 
assign initial \ac{dof} values to \emph{activated} cells. For this purpose, 
\texttt{FEMPAR-AM} implements \texttt{growing\_fe\_space\_t}, a data type 
extending the standard \texttt{fe\_space\_t}, that provides an special method, 
referred to as \texttt{increment()}, performing this operation. The 
implementation of this procedure depends on how data structures in charge 
of handling \acp{dof} and \ac{dof} values are laid out and, more importantly, 
on whether each processor stores \ac{dof} values only in its local subdomain 
portion or also at the ghost cells layer. In the first case, an extra 
nearest-neighbour communication is needed to update \ac{dof} values at nodes 
sitting on a subdomain interprocessor interface.}

\added{The following paragraphs introduce the main software abstractions 
exclusive to \texttt{FEMPAR-AM}. They are in charge of supporting the update of the 
computational mesh, tracking the growth of the domain, and driving the main 
top-level AM process simulation loop. The software subsystem is formed by 
(1)~\texttt{activation}, a customizable object enclosing data structures and 
methods necessary to find, at each time step, the subset of \emph{activated} 
cells $K_{\rm acd}$, (2)~\texttt{cli\_laser\_path}, an \ac{am}-specific object 
to handle the geometrical information of the laser path, and (3)~\texttt{discrete\_path}, also \ac{am}-specific, to generate the space-time 
discretization of the laser path. To clarify their structure, contents and 
relationships, an UML class diagram is constructed in 
Fig.~\ref{fig:uml_fempar-am}.}

\tikzstyle{class}=[rectangle, draw=black, align=left, text=black, text 
justified, minimum height=1cm] 
\tikzstyle{inherit}=[open triangle 60-]
\tikzstyle{assoc}=[-angle 60]
\tikzstyle{depen}=[dashed,-angle 60]
\tikzstyle{comp}=[diamond-angle 60]

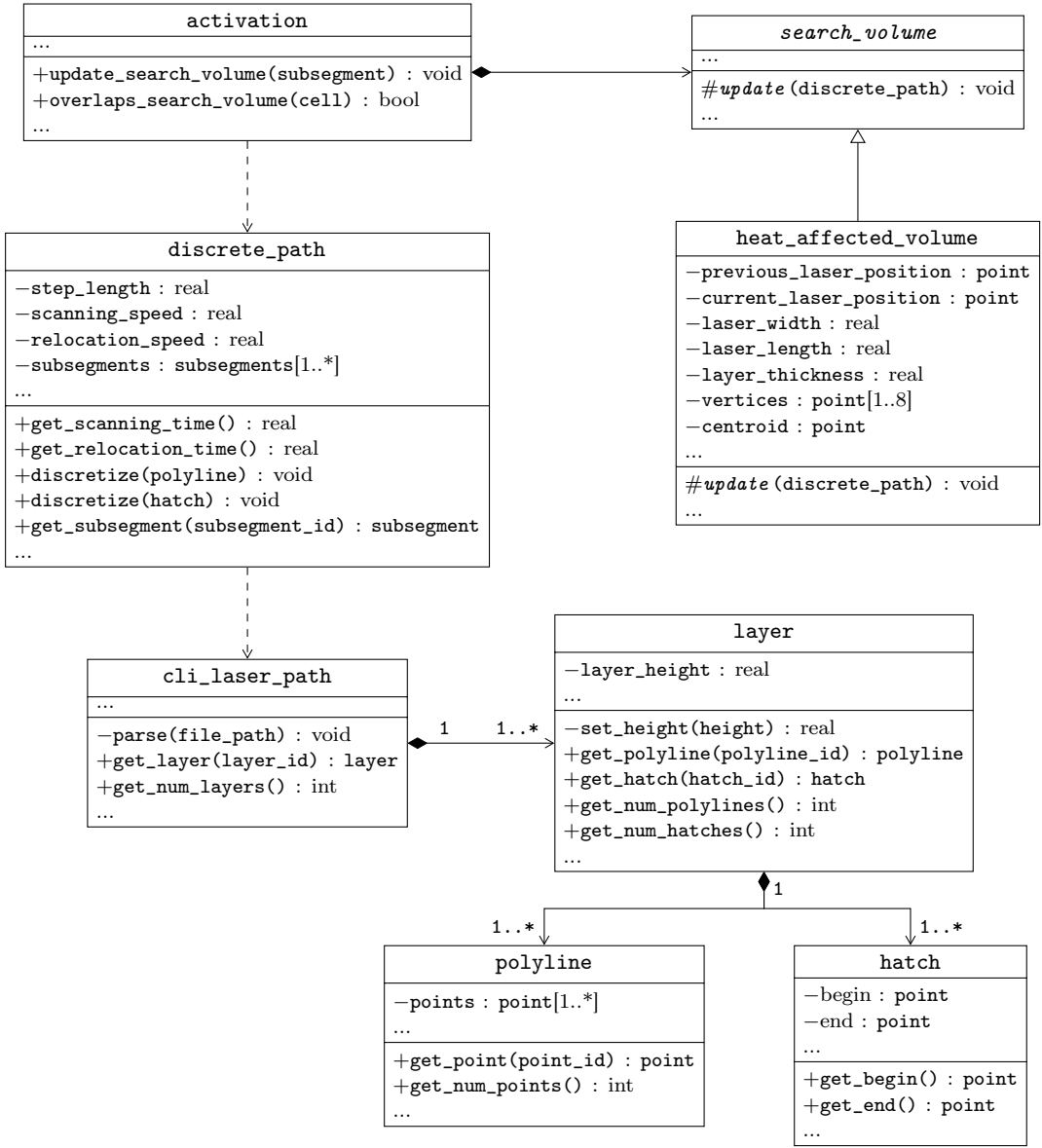
\begin{figure}[htbp]
	\begin{tikzpicture}[node distance=1.cm,auto,font=\footnotesize]

    \node (activation) [class, rectangle split, rectangle split parts=3, 
    rectangle split part align={center,left,left}] { 
          \small \tt activation 
          \nodepart{second}
            ...
          \nodepart{third} 
            $+$\texttt{update\_search\_volume(subsegment)} : void \\
            $+$\texttt{overlaps\_search\_volume(cell)}     : bool \\
            ... };
            
    \node (search) [class, rectangle split, rectangle split parts=3, 
    rectangle split part align={center,left,left}, right=of activation, 
    xshift=2cm] { 
          \small \tt \em search\_volume
          \nodepart{second}
            ...
          \nodepart{third} 
            $\#$\texttt{\emph{update}(discrete\_path)} : void \\
            ... };
            
    \node (hav) [class, rectangle split, rectangle split parts=3, 
    rectangle split part align={center,left,left}, below=of search, 
    yshift=-0.25cm] { 
          \small \tt heat\_affected\_volume
          \nodepart{second}
            $-$\texttt{previous\_laser\_position} : \texttt{point} \\
            $-$\texttt{current\_laser\_position}  : \texttt{point} \\
            $-$\texttt{laser\_width}     : real \\
            $-$\texttt{laser\_length}    : real \\
            $-$\texttt{layer\_thickness} : real \\
            $-$\texttt{vertices}         : \texttt{point}[1..8] \\
            $-$\texttt{centroid}         : \texttt{point} \\
            ...
          \nodepart{third} 
            $\#$\texttt{\emph{update}(discrete\_path)} : void \\
            ... };
            
    \node (discrete) [class, rectangle split, rectangle split parts=3, 
    rectangle split part align={center,left,left}, below=of activation, 
    yshift=-0.25cm] { 
          \small \tt discrete\_path
          \nodepart{second}
            $-$\texttt{step\_length}      : real \\
            $-$\texttt{scanning\_speed}   : real \\
            $-$\texttt{relocation\_speed} : real \\
            $-$\texttt{subsegments}       : \texttt{subsegments}[1..*] \\
            ...
          \nodepart{third} 
            $+$\texttt{get\_scanning\_time()}           : real \\
            $+$\texttt{get\_relocation\_time()}         : real \\
            $+$\texttt{discretize(polyline)}            : void \\
            $+$\texttt{discretize(hatch)}               : void \\
            $+$\texttt{get\_subsegment(subsegment\_id)} : \texttt{subsegment} \\
            ... };
            
    \node (cli) [class, rectangle split, rectangle split parts=3, 
    rectangle split part align={center,left,left}, below=of discrete, 
    yshift=-0.25cm] { 
          \small \tt cli\_laser\_path
          \nodepart{second}
            ...
          \nodepart{third}
            $-$\texttt{parse(file\_path)}     : void \\
            $+$\texttt{get\_layer(layer\_id)} : \texttt{layer} \\
            $+$\texttt{get\_num\_layers()}    : int \\
            ... };
            
    \node (layer) [class, rectangle split, rectangle split parts=3, 
    rectangle split part align={center,left,left}, right=of cli, xshift=1.0cm] 
    { 
          \small \tt layer
          \nodepart{second}
            $-$\texttt{layer\_height}               : real \\
            ...
          \nodepart{third}
            $-$\texttt{set\_height(height)}         : real \\
            $+$\texttt{get\_polyline(polyline\_id)} : \texttt{polyline} \\
            $+$\texttt{get\_hatch(hatch\_id)}       : \texttt{hatch} \\
            $+$\texttt{get\_num\_polylines()}       : int \\
            $+$\texttt{get\_num\_hatches()}         : int \\
            ... };
            
    \node (polyline) [class, rectangle split, rectangle split parts=3, 
    rectangle split part align={center,left,left}, below=of layer, xshift=-3cm] 
    { 
          \small \tt polyline
          \nodepart{second}
            $-$\texttt{points} : \texttt{point}[1..*] \\
            ...
          \nodepart{third}
            $+$\texttt{get\_point(point\_id)} : \texttt{point} \\
            $+$\texttt{get\_num\_points()}    : int \\
            ... };
            
    \node (hatch) [class, rectangle split, rectangle split parts=3, 
    rectangle split part align={center,left,left}, below=of layer, xshift=2cm] 
    { 
          \small \tt hatch
          \nodepart{second}
            $-$begin : \texttt{point} \\
            $-$end   : \texttt{point} \\
            ...
          \nodepart{third}
            $+$\texttt{get\_begin()} : \texttt{point} \\
            $+$\texttt{get\_end()} : \texttt{point} \\
            ... };

		\draw[comp]    (activation.east)  -- (search.west);
		\draw[inherit] (search.south)     -- (hav.north);
		\draw[depen]   (activation.south) -- (discrete.north);
		\draw[depen]   (discrete.south)   -- (cli.north);
		\draw[comp]    (cli.east)         -- node[near start,above]{\tt 1} 
		node[near end,above]{\tt 1..*} (layer.west);
		\draw[comp]    (layer.south)      -- node[near 
		start,right,yshift=-0.1cm]{\tt 1} ++(0,-0.5) -| node[near end,left]{\tt 
		1..*} (polyline.north);
		\draw[comp]    (layer.south)      -- ++(0,-0.5) -| node[near end,right]{\tt 
		1..*} (hatch.north);

\end{tikzpicture}
	\caption{\added{UML class diagram of the software subsystem that 
	\texttt{FEMPAR-AM} uses to support the update of the computational mesh in 
	Alg.~\ref{alg:transform} and to drive the \ac{am} process simulation, 
	tracking the laser path, in Alg.~\ref{alg:main_loop}. \texttt{point} is a 
	simple class encapsulating three real-valued coordinates, whereas 
	\texttt{subsegment} encapsulates two \texttt{point} instances.}}
	\label{fig:uml_fempar-am}
\end{figure}

\added{\texttt{activation} contains the \texttt{search\_volume} abstract 
object, a placeholder for the geometrical description of the region to be 
activated during the time increment. Being a placeholder means that the 
instance is designed to allocate the minimum required memory space to hold a 
\emph{single} search volume at a time. As the definition of the activated 
region depends on the application at hand, the client must specialize the 
behaviour of the object to its needs. In particular, extensions of this data 
type must have all the member variables and implement the methods needed to 
compute and update the dimensions and vertex coordinates defining the search 
cuboid at each time increment. In our context, \texttt{heat\_affected\_volume} 
is the \ac{am}-tailored search volume extended object. \texttt{activation} 
implements two public methods to satisfy the user requirements in the update of 
the computational mesh: (1) \texttt{update\_search\_volume(subsegment)} and (2) 
\texttt{overlaps\_search\_volume(cell)}. The former fills the search volume 
coordinates for a given time increment and the latter is used to test for 
collision between the search cuboid and any cell of the triangulation, using 
the method described in Sect.~\ref{subsec:search}. Alg.~\ref{alg:transform} 
implements the update and increment of the computational domain, i.e. the 
$\mathcal{T}_{h,j-1}$ into $\mathcal{T}_{h,j}$ transformation, described in 
Sect.~\ref{subsec:search} and Fig.~\ref{fig:search}. For simplicity all steps 
regarding the treatment of the FE space and global FE functions are omitted, 
but they must be projected and redistributed. As observed, methods (1) and (2), 
invoked at Lines~\ref{lin:update} and~\ref{lin:overlap}, fulfill important 
steps of the procedure.}

\begin{algorithm}
	\caption{\texttt{update\_and\_increment\_computational\_domain(subsegment)} 
	(see also Fig.~\ref{fig:search}) 
	\label{alg:transform}}
	\SetKwInOut{Input}{input}
	\SetKwInOut{Output}{output}
	\SetKwFunction{overlaps}{overlaps\_search\_volume}
	\SetKwFunction{get}{get\_current\_segment}
	\SetKwFunction{rcf}{set\_refinement\_and\_coarsening\_flags}
	\SetKwFunction{spw}{set\_partition\_weights}
	\SetKwFunction{rac}{refine\_and\_coarsen}
	\SetKwFunction{red}{redistribute}
	\SetKwFunction{mar}{mark\_activated\_cells}
	\SetKwFunction{isl}{is\_local}
	\SetKwFunction{lev}{get\_level}
	\SetKwFunction{forref}{set\_for\_refinement}
	\SetKwFunction{forcor}{set\_for\_coarsening}
	\SetKwFunction{fornot}{set\_for\_do\_nothing}
	\SetKwFunction{increment}{increment}
	\SetKwFunction{upds}{update\_search\_volume}
	\SetKwData{exists}{found\_cell\_for\_refinement}
	\SetKwData{Tria}{triangulation}
	\SetKwData{DPath}{discrete\_path}
	\SetKwData{Subs}{subsegment}
	\SetKwData{Act}{activation}
	\SetKwData{Seg}{segment}
	\SetKwData{Lev}{level}
	\SetKwData{Maxl}{max\_level\_of\_refinement}
	\SetKwData{Minl}{min\_level\_of\_refinement}
	\SetKwData{Cell}{cell}
	\SetKwData{Fes}{growing\_fe\_space}
 	
 	
	$\Act.\upds(\Subs)$\nllabel{lin:update} \tcc*[f]{fill search volume 
	coordinates for current subsegment}
 	
 	\exists $\leftarrow$ \texttt{true}
 	
	\While(\tcc*[f]{repeat until all max level cells intersecting the search 
	volume found}){ \exists }{
		
		\exists $\leftarrow$ \texttt{false}
		
		\For{$\Cell \in \Tria$}{
			
			\If(\tcc*[f]{local = owned by the current MPI task}){$\Cell.\isl()$}{
				
				\uIf(\tcc*[f]{cell intersects the search 
				volume (see 
				Sect.~\ref{subsec:search})}){$\Act.\overlaps(\Cell)$\nllabel{lin:overlap}}{
					
					\uIf{$\Cell.\lev() < \Maxl$}{
						
						$\Cell.\forref()$
						
						\exists $\leftarrow$ \texttt{true}
						
					}
					
					\Else{
						
						$\Cell.\fornot()$
						
					}
					
				}
					
				\Else(\tcc*[f]{cell does not intersect the search volume}){
					
					\uIf{$\Cell.\lev() > \Minl$}{
						
						$\Cell.\forcor()$
						
					}
					
					\Else{
						
						$\Cell.\fornot()$
						
					}
					
				}
				
			}
			
		}
		
		$\Tria.\rac()$ 
		
		... \tcc*[f]{FE space and global FE functions projection/interpolation 
		apply here}
		
		$\spw()$ \tcc*[f]{construct weight function as in Eq.~\eqref{eq:weight}}
		
		$\Tria.\red()$
		
		... \tcc*[f]{FE space and global FE functions redistribution apply here}
		
	}
 	
 	$\mar()$
 	
 	$\Fes.\increment()$ \tcc*[f]{inject \ac{dof} values of $\mathcal{T}_{h,j-1}$ 
 	into $\mathcal{T}_{h,j}$ and initialize \ac{dof} values in $\mathrm{K}_{\rm 
 	acd}$}
 	
\end{algorithm}

\added{Among the application-specific objects, \texttt{cli\_laser\_path} takes 
charge of managing the geometrical information of the laser path. The data 
comes in the same Common Layer Interface (CLI) ASCII file sent to the numerical 
control of the machine. CLI~\citep{CLI20} is a common universal format for the 
input of geometry data to model fabrication systems based on layer 
manufacturing technologies. In the context of \ac{am}, a CLI file describes the 
movement of the laser in the plane of each layer with a complex sequence of 
\emph{polylines}, to define the (smooth) boundary of the component, and 
\emph{hatch} rectiliniar patterns for the inner structures (see \citep{CLI20} 
for several examples). \texttt{cli\_laser\_path} holds a parser for this file 
format (\texttt{parse(file\_path)}) and accommodates its hierarchical structure 
in the following way: First, it aggregates an array of \texttt{layer} entities. 
For each \emph{layer} in the CLI file, a \texttt{layer} instance is created and 
the layer height is stored. Likewise, for each \emph{polyline} or \emph{hatch} 
associated to the layer, an instance of \texttt{polyline} or \texttt{hatch} is 
created and filled with their plane coordinates. The class is supplemented with 
several query methods (e.g. \texttt{get\_layer(layer\_id)}, 
\texttt{get\_hatch(hatch\_id)}, \texttt{get\_point(point\_id)}), such that the 
user can navigate through the laser path subentities and extract their point 
coordinates. Although not implemented, a polymorphic superclass 
\texttt{laser\_path} of \texttt{cli\_laser\_path} may be introduced to consider 
other file formats. In this way, new file formats can be accommodated with new 
child extensions of \texttt{laser\_path}; at the moment, \texttt{FEMPAR-AM} can 
only support CLI files.}

\added{Closely associated to \texttt{cli\_laser\_path} is 
\texttt{discrete\_path}, an entity that generates the space-time discretization 
of the laser path. Given that the printing process is tightly related to the 
movement of the laser, it is more natural to discretize the laser path with a 
\emph{step length} $\Delta \mathrm{x}$, instead of a time step. 
\texttt{discrete\_path} takes the user-prescribed step length and the current 
\emph{polyline} or \emph{hatch} segment and divides it into subsegments of 
$\Delta \mathrm{x}$ size with the method \texttt{discretize(polyline/hatch)}. 
To support the time integration, \texttt{discrete\_path} also computes the time 
increment associated to each subsegment as}
\begin{equation}
	\Delta t = \Delta \mathrm{x} / {\rm V}_{\rm scanning},
\end{equation}
\added{where ${\rm V}_{\rm scanning}$ is the scanning speed. At each time step, 
\texttt{discrete\_path} feeds \texttt{activation} with the current subsegment  
via \texttt{update\_search\_volume(subsegment)}. After \texttt{activation} 
generates the current search volume and drives the mesh transformation, see 
Alg.~\ref{alg:transform}, standard \ac{fe} simulation steps follow until 
solving the linear system modelling the printing during the current time 
increment. The sequence is repeated until all subsegments have been simulated. 
Following this, the system is allowed to cool down, while the laser moves to 
the begin point of the next entity in the laser path or a new layer is spread. 
For the cooling step, \texttt{discrete\_path} also computes recoat and 
relocation time as the quotient of the distance between the end point of the 
current \emph{polyline} or \emph{hatch} and the begin point of the next one 
divided by the relocation speed ${\rm V}_{\rm relocation}$.}

\added{A relevant feature of the design is that \texttt{discrete\_path} is 
another placeholder container object, i.e. it only stores the discretization 
for the current \emph{polyline} or \emph{hatch} being printed and it is updated 
while looping over the laser path entities. Using the \texttt{cli\_laser\_path} 
recursive construction and the design of \texttt{discrete\_path}, this loop, 
which is the one at the top level of \texttt{FEMPAR-AM} simulations, can be 
written in a compact form that reflects the actual printing process, as 
shown in Alg.~\ref{alg:main_loop} and Alg.~\ref{alg:subsegment_loop}. Note that 
the cost of the operations involved in filling and discretizing the laser path 
is negligible w.r.t. other stages of the simulation that concentrate the bulk 
of the computational cost (e.g. linear solver). Given this, and the fact that 
each processor must know the global search cuboid coordinates (see 
Sect.~\ref{subsec:search}), data generated by \texttt{cli\_laser\_path} and 
\texttt{discrete\_path} is not distributed, it is replicated in each processor 
to avoid extra communications.}

\begin{algorithm}
	\caption{Top-level simulation loop of \texttt{FEMPAR-AM} 
	\label{alg:main_loop}}
	\SetKwFunction{init}{initialize\_simulation}
	\SetKwFunction{fina}{finalize\_simulation}
	\SetKwFunction{update}{discretize}
	\SetKwFunction{run}{simulate\_printing\_process}
	\SetKwFunction{get}{get\_relocation\_time}
	\SetKwFunction{rel}{simulate\_cooling}
	\SetKwData{Path}{laser\_path}
	\SetKwData{Layer}{layer}
	\SetKwData{Poly}{polyline}
	\SetKwData{DPath}{discrete\_path}
	\SetKwData{RTime}{relocation\_time}
	\SetKwData{Hatch}{hatch}
 		
 		
 		\For{$\Layer \in \Path$}{
 			
 			\For{$\Poly \in \Layer$}{
 				
 				$\DPath.\update(\Poly)$ \tcc*[f]{divide polyline into user-prescribed 
 				$\Delta \mathrm{x}$-sized subsegments}
 				
 				$\run(\DPath)$ \tcc*[f]{see Alg.~\ref{alg:subsegment_loop}}
 				
 			}
 			
 			\For{$\Hatch \in \Layer$}{
 				
 				$\DPath.\update(\Hatch)$ \tcc*[f]{divide hatch into user-prescribed 
 				$\Delta \mathrm{x}$-sized subsegments}
 				
 				$\run(\DPath)$ \tcc*[f]{see Alg.~\ref{alg:subsegment_loop}}
 				
 			}
 			
 		}
 		

\end{algorithm}

\begin{algorithm}
	\caption{\texttt{simulate\_printing\_process(discrete\_path)}
	\label{alg:subsegment_loop}}
	\SetKwFunction{init}{initialize\_simulation}
	\SetKwFunction{fina}{finalize\_simulation}
	\SetKwFunction{update}{discretize}
	\SetKwFunction{run}{simulate\_printing\_process}
	\SetKwFunction{get}{get\_relocation\_time}
	\SetKwFunction{rel}{simulate\_cooling}
	\SetKwFunction{upd}{update\_and\_increment\_computational\_domain}
	\SetKwData{Path}{laser\_path}
	\SetKwData{Layer}{layer}
	\SetKwData{Poly}{polyline}
	\SetKwData{DPath}{discrete\_path}
	\SetKwData{RTime}{relocation\_time}
	\SetKwData{Hatch}{hatch}
	\SetKwData{Subs}{subsegment}
 		
 		\For{$\Subs\in \DPath$}{
 			
 			$\upd(\Subs)$ \tcc*[f]{see Alg.~\ref{alg:transform}}
 			
 			... \tcc*[f]{for simplicity, remaining simulation steps until linear 
 			system solution are omitted, but follow here}
 			
 		}
 		
		$\RTime \leftarrow \DPath.\get()$
		
		$\rel(\RTime)$ \tcc*[f]{solve Eq.~\eqref{eq:semi-implicit} without heat 
		input (laser relocation or layer recoat)}

\end{algorithm}

\added{Other minor thermal \ac{am} customizations that supplement the design 
are \texttt{heat\_input}, a polymorphic and extensible entity with a suite of 
heat source term descriptions (e.g. uniform, surface Gauss or Goldak double 
ellipsoidal); \texttt{property\_table}, an auxiliary object to allow the 
client to linearly interpolate properties that are known in tabular format, in 
order to evaluate temperature-dependent properties; and  
\texttt{heat\_transfer\_discrete\_integration\_t}, a subclass of 
\texttt{discrete\_integration\_t}~\citep[Sect. 11.2]{badia-fempar}, in charge 
of evaluating the entries of the matrices and vectors defining the linear 
system of Eq.~\eqref{eq:semi-implicit}.}

\section{Numerical experiments and discussion}
\label{sec:results}


\subsection{Verification of the thermal FE model}

First, the thermal \ac{fe} model presented in Sect. \ref{sec:formulation} is 
verified against a 3D benchmark present in the 
literature~\citep{KOLLMANNSBERGER20181483,fachinotti2011analytical} that 
considers a moving single-ellipsoidal heat source on a semi-infinite solid with 
null fluxes at the free surface.

Assuming an Eulerian frame of reference $(x,y,z)$, consider a heat source 
located initially at $z = 0$ that travels at constant velocity $v$ along the 
$x$-axis on top of the semi-infinite solid defined by $z \leq 0$. The heat 
source distribution, derived from Goldak's double-ellipsoidal 
model~\citep{goldak84}, is defined by
\begin{equation}
	q(x,y,z,t) = \frac{6 \sqrt{3}}{\pi \sqrt{\pi}} \frac{Q}{abc} \exp \left[ -3 
	\left( \frac{(x-vt)^2}{a^2} + \frac{y^2}{b^2} + \frac{z^2}{c^2} \right) 
	\right],
\end{equation}
\noindent{where $Q$ is the (effective) rate of energy supplied to the system, 
$v$ is the velocity of the laser and $a,b,c$ are the main dimensions of the 
ellipsoid, as shown in Fig.~\ref{fig:goldak}.}

The problem at hand is linear and admits a semi-analytical solution using 
Green's functions~\cite{carslaw1959conduction,cole2010heat} given by
\begin{equation}
	u(x,y,z,t) = u_0 + \frac{6 \sqrt{3}}{\pi \sqrt{\pi}} \frac{\alpha Q}{k} 
	\int_0^t \frac{\exp \left[ -3 \left( \frac{(x-vt)^2}{a^2 + 12 \alpha 
	(t-\tau)} + \frac{y^2}{b^2 + 12 \alpha (t-\tau)} + \frac{z^2}{c^2 + 12 \alpha 
	(t-\tau)} \right) \right] }{\sqrt{ (a^2 + 12 \alpha (t-\tau)) (b^2 + 12 
	\alpha (t-\tau)) (c^2 + 12 \alpha (t-\tau)) } } \mathrm{d}\tau,
\end{equation}
\noindent{with $u_0$ the initial temperature and $\alpha = k / \rho c$ the 
thermal diffusivity.}

Using the symmetry of the problem, the numerical simulation considers a cuboid 
with coordinates given in Fig.~\ref{fig:verification_setting}. The path of the 
laser follows a segment centred along the edge of the cuboid that sits on the 
$x$-axis. Null fluxes apply at the top surface and the lateral face of 
symmetry, whereas Dirichlet boundary conditions apply at the remaining contour 
surfaces to account for the semi-infinite solid.

\begin{figure}[!h]
 \centering
 \begin{subfigure}[t]{0.45\textwidth}
  	\centering
   \scalebox{1}{\begin{tikzpicture}[scale=0.55]
    \filldraw[fill=brown!20!white,draw=brown!80!white,thick] (0,2) --(4,0) -- (4,2) --(0,4) -- cycle;
    \filldraw[fill=brown!20!white,draw=brown!80!white,thick] (4,0) -- (10,2.5) --(10,4.5) -- (4,2) -- cycle;
    \filldraw[fill=brown!20!white,draw=brown!80!white,thick] (0,4) -- (4,2) --(10,4.5) -- (6,6.5) -- cycle;
    \filldraw[ultra thick,dashed,fill=brown!40!white] (2.69230769231,3.28846153846) {[rotate = 22.61986495] arc [x radius = 2.5cm, y radius=1.25cm, start angle = 180, end angle = 360]};
    \filldraw[ultra thick,fill=brown!40!white] (5,4.25) circle [x radius = 2.5cm, y radius=0.75cm, rotate = 22.61986495];
    \draw[brown!60!black,thick,latex-latex] (2.69230769231,3.28846153846) -- node[pos=0.5,yshift=0.2cm] {\bf a} (5,4.25);
    \draw[brown!60!black,thick,latex-latex] (5,4.25) -- node[pos=0.5,xshift=0.2cm,yshift=0.2cm] {\bf b} (5.8,3.85);
    \draw[brown!60!black,thick,latex-latex] (5,4.25) -- node[pos=0.5,xshift=-0.2cm] {\bf c} (5,2.9);
	\draw[thick,-latex] (2,7) -- node[pos=0.5,xshift=-0.2cm,yshift=0.15cm] {\bf x} (0.95,6.5625);
	\draw[thick,-latex] (2,7) -- node[pos=0.5,xshift=0.2cm,yshift=0.15cm] {\bf y} (2.9,6.55);
    \draw[thick,-latex] (2,7) -- node[pos=0.5,xshift=-0.2cm] {\bf z} (2,8);
\end{tikzpicture}}
   \caption{Single-ellipsoidal heat source model.}
   \label{fig:goldak}
 \end{subfigure}
 \quad
 \begin{subfigure}[t]{0.45\textwidth}
   \centering
   \scalebox{1}{\begin{tikzpicture}[font=\small,scale=0.45]
    \fill[fill=brown!20!white] (0,2) --(4,0) -- (4,4) --(0,6) -- cycle;
    \filldraw[draw=brown!80!white,thick,pattern=north west lines,pattern color=brown!60!white] (0,2) -- (4,0) -- (4,4) --(0,6) -- cycle;
    \filldraw[fill=brown!20!white,draw=brown!80!white,thick] (4,0) -- (12,3) --(12,7) -- (4,4) -- cycle;
    \filldraw[fill=brown!20!white,draw=brown!80!white,thick] (0,6) -- (4,4) --(12,7) -- (8,9) -- cycle;
    \filldraw[ultra thick] (6,4.75) circle (1.5pt) node[yshift=-0.1cm,below] {(2,0,0)} -- (10,6.25) circle (1.5pt) node[yshift=-0.1cm,below] {(0,0,0)};
    \draw[thick, black!80!gray, |-latex] (9,6.5) -- node[pos=0.5,above] {v} (7,5.75);
    \filldraw[brown!80!white] (0,2) circle (3.0pt) node[black,below,yshift=-0.15cm] {(3,-2,-2)} -- (4,0) circle (3.0pt) node[black,below,yshift=-0.15cm] {(3,0,-2)} -- (12,3) circle (3.0pt) node[black,below,yshift=-0.15cm] {(-1,0,-2)} -- (12,7) circle (3.0pt) node[black,above,yshift=0.15cm] {(-1,0,0)};
    \filldraw[brown!80!white] (8,9) circle (3.0pt) node[black,above,yshift=0.15cm] {(-1,-2,0)} -- (0,6) circle (3.0pt) node[black,above,yshift=0.15cm] {(3,-2,0)} -- (4,4) circle (3.0pt) node[black,above,yshift=0.15cm] {(3,0,0)};
\end{tikzpicture}}
   \caption{Domain of analysis, boundary conditions and welding path. 
   Homogeneous Neumann boundary conditions apply at the top surface and the 
   lateral face of symmetry. Dirichlet boundary conditions apply elsewhere.}
   \label{fig:verification_setting}
 \end{subfigure}
 \caption{3D semi-analytical benchmark problem. Heat source distribution, 
 geometry and boundary conditions.}
\end{figure}
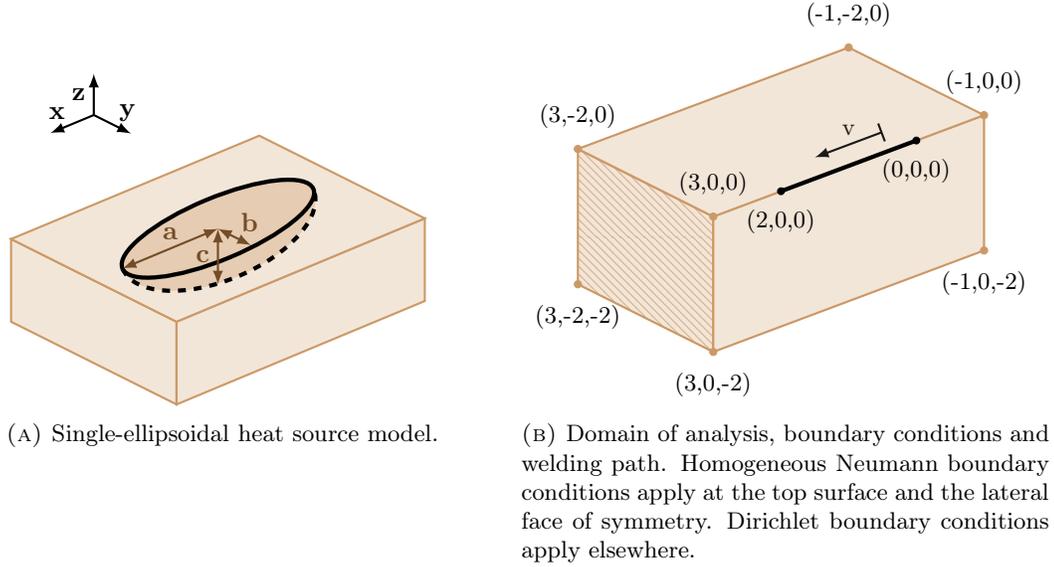

An $h$-adaptive linear \ac{fe} mesh is employed, where the smallest size is 
prescribed around the welding path. Starting with the initial mesh shown in 
Fig.~\ref{fig:initial_mesh} and assigning $u_0 = 20$, $Q = 50$, $v = 1$, 
$\alpha = 0.1$, $k = 1$ and $(a,b,c) = (0.3,0.15,0.25)$, a convergence test is 
carried out.

\begin{figure}[!h]
 \centering
 \begin{subfigure}[t]{0.45\textwidth}
  	\centering
   \includegraphics[width=0.9\textwidth,clip]{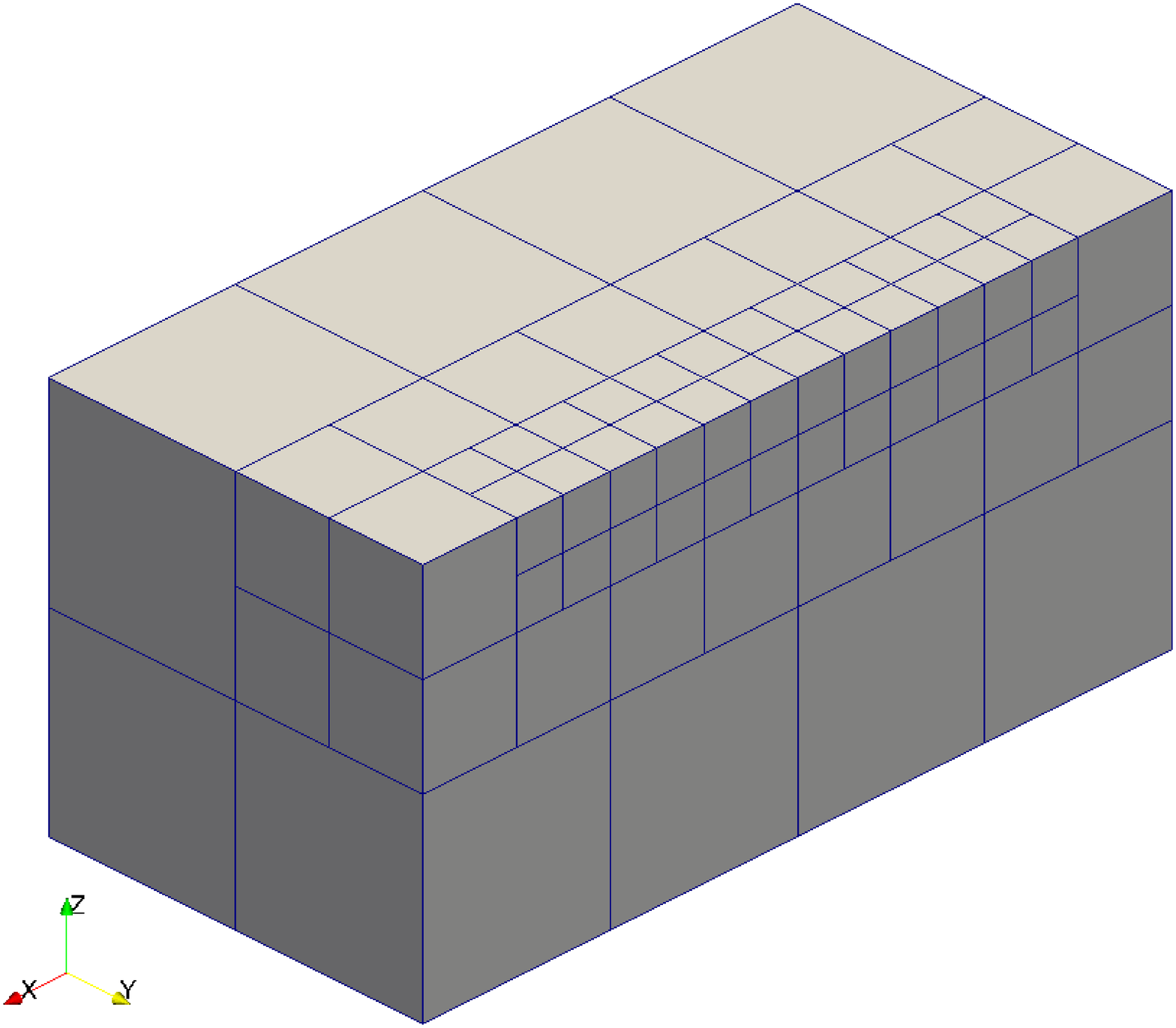}
   \caption{Initial adapted octree mesh.}
   \label{fig:initial_mesh}
 \end{subfigure}
 \quad
 \begin{subfigure}[t]{0.45\textwidth}
   \centering
   \includegraphics[width=0.95\textwidth,clip]{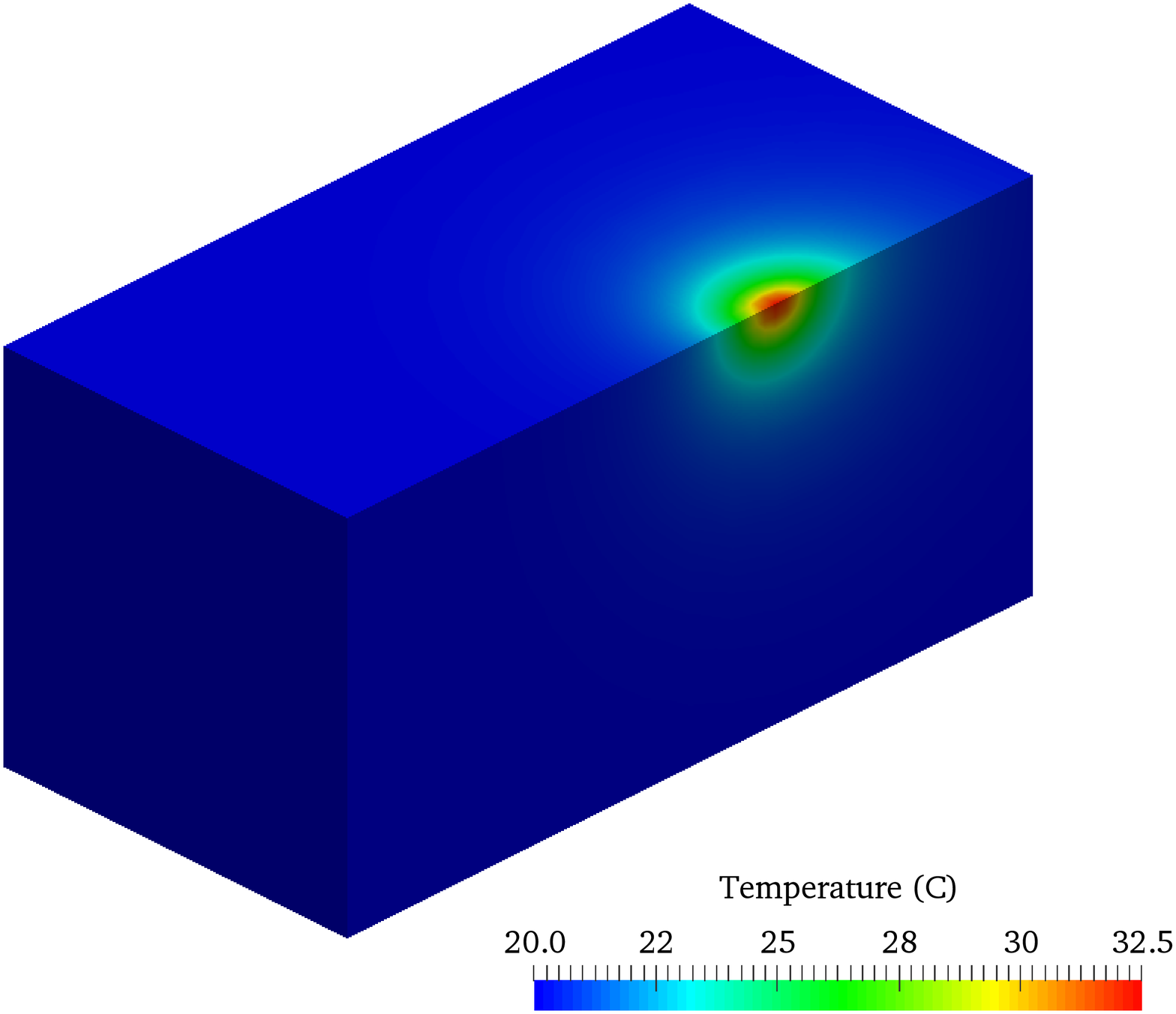}
   \caption{Temperature contour plot at $t = 0.5$ (3\textsuperscript{rd} 
   refinement step).}
   \label{fig:contour_plot}
 \end{subfigure}
 \caption{3D semi-analytical benchmark problem. Initial mesh and contour plot.}
\end{figure}

For a parabolic heat equation with sufficiently smooth solution, the error in 
$L^2( [0,T]; H_0^1(\Omega) )$ with a Backward Euler time integration scheme is 
proportional to $(h^p + \Delta t)$ (see Theorem 6.29 in~\cite{ern2013theory}), 
where $p$ is the order of the \ac{fe}. Since $p = 1$ in this experiment, the 
time discretization should be refined at the same rate of the space 
discretization. 

Taking this into consideration with an initial time step of $\Delta t = 0.008$, 
Fig.~\ref{fig:convergence_test} shows that the numerical error in $L^2( [0,T]; 
H_0^1(\Omega) )$ of the 3D semi-analytical benchmark decreases at the same rate 
as the theoretical one. This indicates a correct implementation of the thermal 
\ac{fe} model.

\begin{figure}[!h]
 \centering
 \scalebox{.70}{\input{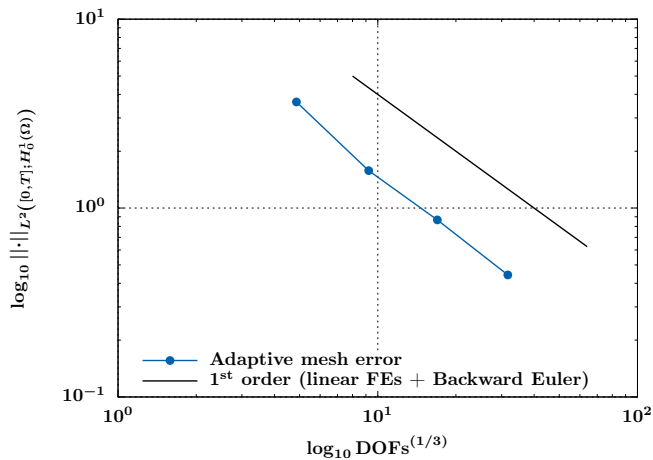}}
 \caption{3D semi-analytical benchmark problem. Convergence test.}
 \label{fig:convergence_test}
\end{figure}

\subsection{Strong-scaling analysis}
\label{sec:ss_example}

Next, the focus is turned to analysing the performance of \texttt{FEMPAR-AM} 
with a strong-scaling analysis\footnote{Strong scalability is the ability of a 
parallel system (i.e. algorithm, software and parallel computer) to efficiently 
exploit increasing computational resources (CPU cores, memory, etc.) in the 
solution of a fixed-size problem. An \emph{ideally} strongly scalable code 
decreases CPU time exactly as $1/P$, where $P$ is the number of processors 
being used. In other words, if the system solves a size $N$ problem in time $t$ 
with a single processor, then it is able to solve the same problem in time 
$t/P$ with $P$ processors.}. The model problem for the subsequent experiments 
is designed to be geometrically simple, but with a computational load 
comparable to a real scenario of an industrial application.

According to this, the object of simulation is now the printing of 48 layers of 
31.25 $[\upmu \mathrm{m}]$ on top of a 32x32x16 $[\mathrm{mm}]$ prism. After 
printing the 48 layers, the prism has dimensions of 32x32x17.5 $[\mathrm{mm}]$, 
as shown in Fig.~\ref{fig:ss_setting}.

Concerning the process parameters, the power of the laser is set to $400$ 
$[\mathrm{W}]$, the volumetric deposition rate during scanning is $d_{\rm p} = 
10.0$ $[\mathrm{mm}^3/\mathrm{s}]$ and the time allowed for lowering the 
platform, recoating and layer relocation between layers is $t_{\rm r} = 10.0$ 
$[\mathrm{s}]$.

Apart from that, the material chosen is the Ti6Al4V Titanium alloy. The 
temperature dependent density, specific heat and thermal conductivity are 
obtained from a handbook and plotted in~\citep{chiumenti_neiva_2017}. 

A constant heat convection boundary condition applies on the boundary of the 
cube with $h_{\rm out} = 50$ $[\mathrm{W}/\mathrm{m^2 K}]$ and $u_{\rm out} = 
35$ $[\mathrm{^\circ C}]$. The initial temperature of both the prism and each 
new layer is $u_{0} = 90$ $[\mathrm{^\circ C}]$.

The root octant of the octree mesh is defined to cover a 32x32x32 
$[\mathrm{mm}]$ cube region. The octree is transformed during the simulation to 
model the layer-by-layer deposition process, by prescribing a maximum 
refinement level of 11, i.e. assigning a mesh size of $h = 32,000/2^{11} = 
15.625$ $[\upmu \mathrm{m}]$, to the elements inside the layer that is 
currently being printed. A mesh gradation is then established according to the 
distribution of thermal gradients (highest at the printing region, lowest at 
the bottom of the cuboid). The mesh size in the $(x,y)$ plane is fixed, whereas 
it decreases in both $z$-directions, until reaching a minimum level of 
refinement of 4 at the top and bottom of the prism. The computational domain is 
then defined by the initial prism and the layers that have been printed up to 
the current time. As seen in Fig.~\ref{fig:ss_mesh}, most elements end up 
concentrating around the current layer, due to the\deleted{aggressive} 
coarsening induced by the 2:1 balance, but this is also the region with the 
highest temperature variations.

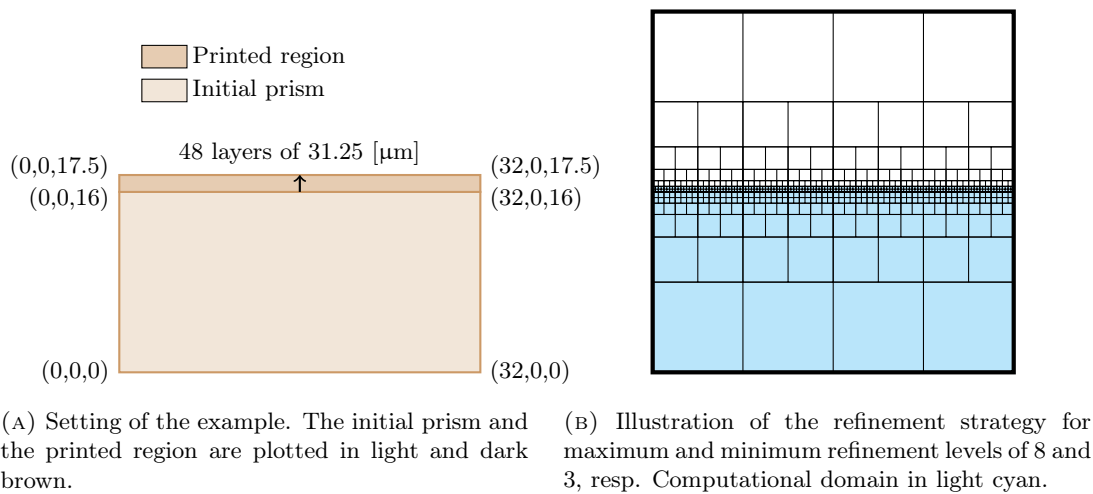
\begin{figure}[!h]
 \centering
 \begin{subfigure}[t]{0.45\textwidth}
  	\centering
   \scalebox{1}{\begin{tikzpicture}[font=\small,scale=0.15]
    \draw[white,ultra thick] (0,0) -- (32,0) -- (32,32) -- (0,32) -- cycle;
    \draw[white] (0,0) circle (1.5pt) node[left,black] {(0,0,0)} -- (32,0) circle (1.5pt) node[right,black] {(32,0,0)};
    \draw[white] (0,16) circle (1.5pt) node[yshift=-0.1cm,left,black] {(0,0,16)} -- (32,16) circle (1.5pt) node[yshift=-0.1cm,right,black] {(32,0,16)};
    \draw[white] (0,17.5) circle (1.5pt) node[yshift=0.1cm,left,black] {(0,0,17.5)} -- (32,17.5) circle (1.5pt) node[yshift=0.1cm,right,black] {(32,0,17.5)};
    \filldraw[fill=brown!20!white,draw=brown!80!white,thick] (0,0) -- (32,0) -- (32,16) -- (0,16) -- cycle;
    \filldraw[fill=brown!40!white,draw=brown!80!white,thick] (0,16) -- (32,16) -- (32,17.5) -- (0,17.5) -- cycle;
    \draw[black,thick,->] (16,16) -- (16,17.5) node[above] {48 layers of 31.25 $[\upmu \mathrm{m}]$};
    \filldraw[fill=brown!20!white,draw=black] (2,24) rectangle (6,26) node[pos=0.5,right,xshift=0.25cm] {Initial prism};
	\filldraw[fill=brown!40!white,draw=black] (2,27) rectangle (6,29) node[pos=0.5,right,xshift=0.25cm] {Printed region};
\end{tikzpicture}}
   \caption{Setting of the example. The initial prism and the printed region 
   are plotted in light and dark brown.}
   \label{fig:ss_setting}
 \end{subfigure}
 \quad
 \begin{subfigure}[t]{0.45\textwidth}
   \centering
   \scalebox{1}{\begin{tikzpicture}[font=\small,scale=0.15]
	\fill[cyan!25!white] (0,0) rectangle (32,16.5);
    \draw[scale=8]    (0,0)  grid    (4,4);
    \draw[scale=4]    (0,2)  grid    (8,6);
    \draw[scale=2]    (0,6)  grid  (16,10);
    \draw[scale=1]    (0,14) grid  (32,18);
    \draw[scale=0.5]  (0,30) grid  (64,34);
    \draw[scale=0.25] (0,64) grid (128,66);
    \draw[black,ultra thick] (0,0) node[left,white] {(0,0,0)} -- (32,0) node[right,white] {(0,0,0)} -- (32,32) -- (0,32) -- cycle;
\end{tikzpicture}}
   \caption{Illustration of the refinement strategy for maximum and minimum 
   refinement levels of 8 and 3, resp. Computational domain in light cyan.}
   \label{fig:ss_mesh}
 \end{subfigure}
 \caption{Strong-scaling problem set up. Plane XZ view of the setting 
 and the mesh.}
\end{figure}

This refinement strategy leads to a simulation workflow that, for each layer, 
comprises the following steps:
\begin{enumerate}
	\item \textbf{Remeshing:} The previous mesh is refined and coarsened to 
	accommodate the current layer.
	\item \textbf{Redistribution:} The new mesh is partitioned and redistributed 
	among all processors to maintain load balance.
	\item \textbf{Activation:} New \acp{dof} are distributed and initialized over 
	the cells within the current layer.
	\item \textbf{Printing:} The problem is solved for the printing step. This 
	step consists in the application of the heat needed to fuse the powder of the 
	current layer and the time increment is calculated as $\Delta t = t_{\rm p} = 
	{\rm V}_{\rm layer} / d_{\rm p}$ $[\mathrm{s}]$. 
	\item \textbf{Cooling:} The problem is solved for the cooling step, 
	accounting for lowering of building platform, recoat time and laser 
	relocation with a time increment of $\Delta t = t_{\rm r}$ $[\mathrm{s}]$. 
	During the cooling step, the laser is off and the prism is allowed to cool 
	down. 
\end{enumerate}
\noindent{According to this workflow, the simulation of each layer is carried 
out in two time steps, printing and cooling, so the total number of time steps 
is $48 \cdot 2 = 96$. However, with the exception of the first layer, each new 
layer is meshed with a single refine and coarsen step. Hence, the simulation 
has about half as many \ac{amr} events as time steps. Besides, the linear 
system in Eq.~\eqref{eq:semi-implicit}, arising at each time step, is solved 
with the \replaced{Jacobi}{\ac{bddc}(f)}-\ac{pcg}\deleted{\ac{dd}} method 
\added{with unit relaxation parameter.}}

The numerical experiments for this example were run at the 
Marenostrum-IV~\citep{marenostrum} (MN-IV) supercomputer, hosted by the 
Barcelona Supercomputing Center (BSC). It is equipped with 3,456 compute nodes 
connected together with the Intel OPA HPC network. Each node has 2x Intel Xeon 
Platinum 8160 multi-core CPUs, with 24 cores each (i.e. 48 cores per node) and 
96 GBytes of RAM. 

Apart from that, \texttt{FEMPAR-AM} was compiled with Intel Fortran 18.0.1 
using system recommended optimization flags and linked against the Intel 
\ac{mpi} Library (v2018.1.163) for message-passing and the BLAS/LAPACK 
\deleted{and PARDISO available on the Intel MKL }library for optimized dense 
linear algebra kernels.\deleted{and sparse direct solvers, respectively.} All 
floating-point operations were performed in IEEE double precision.

\deleted{The parallel algorithm is set up such that each subdomain is 
associated to a different \ac{mpi} task, with a one-to-one mapping among 
\ac{mpi} tasks and CPU cores. In the \ac{bddc} framework, these are referred to 
as the \emph{fine} tasks. Following the two-level algorithm 
in~\cite{badia2014highly}, the \emph{coarse} duties are carried out by an 
additional \ac{mpi} task, which is mapped to a full node, i.e. it has access to 
all its memory (96 GBytes) and computational resources (48 cores). The 
multithreaded sparse direct solvers in PARDISO are then employed for the 
solution of the coarse problem, by mapping as many threads to cores available 
in the node.}

\deleted{An effort was done in order to tune, with performance and scalability 
in mind, the values of the relevant parameters affecting the behaviour of the 
two-level \ac{bddc}(f) customized for growing domains. The resulting 
algorithm-parameter values combination is referred to as \ac{bddc}-ref. In 
particular, regarding dynamic load balancing, the partition weights are set to 
$w_a = 10$ for active cells and $w_i = 1$ for inactive cells. As for the coarse 
subgraph extraction procedure of Sect.~\ref{sec:redundant}, the maximum allowed 
degree $\Delta^{\max}$ is 20, while the maximum allowed distance 
$\mathrm{d}^{\max}$ is 3.}

\deleted{Likewise, the example was designed such that it filled, as much as 
possible, the available memory per node for the smallest number of subdomains 
considered in the analysis (2 nodes, 96 cores / fine tasks).} \added{The 
parallel framework is set up such that each subdomain is associated to a 
different \ac{mpi} task, with a one-to-one mapping among \ac{mpi} tasks and CPU 
cores. Regarding dynamic load balancing, the partition weights are set to 
$w_a = 10$ for active cells and $w_i = 1$ for inactive cells.} 
\replaced{Using}{Following this criterion and using} linear \acp{fe}, the 
average total number of cells $N_{\rm cells}$ and global \acp{dof} $N_{\rm 
dofs}$ (excluding hanging \acp{dof}) across all time steps are 12,585,216 and 
10,273,920. Note that, if a fixed uniform mesh was used, specifying the maximum 
refinement level of 11 all over the cube, the number of cells would be 
$(2^{11})^3 = 8.59 \cdot 10^9$ and the number of \acp{dof} would grow from 
$(2^{11}+1)^2 (2^{11}/2+1) = 4.30 \cdot 10^9$, initially, to $(2^{11}+1)^3 = 
8.60 \cdot 10^9$, at the end of the simulation. Hence, it is readily exposed 
how $h$-adaptivity drastically reduces (almost by three orders of magnitude) 
the size of the problem and the required computational resources, while 
preserving accuracy around the growing printing region.

\deleted{Fig.~\ref{fig:ss_bddc_cg} and Tab.~\ref{tab:ss_bddc_cg} represent the 
speed-up and the total simulation wall time $[\mathrm{s}]$ of \ac{bddc}-ref 
against a non-preconditioned parallel implementation of the Conjugate Gradient 
(CG) method, as the number of subdomains is increased. As observed, 
\ac{bddc}-ref scales up to 3,072 fine tasks with a peak speed-up of 8.8. Above 
3,072 cores, time-to-solution increases due to parallelism related overheads; 
more computationally intensive simulations (i.e. larger loads per processor) 
would be required to exploit additional computational resources efficiently. 
However, CG saturates about six times earlier at 576 processors because 
interprocessor communication dominates the time devoted to solve the linear 
system. Moreover, CG is about 35\% slower than \ac{bddc}-ref. With 
PCG-\ac{bddc} the total wall time reduces with the number of subdomains to 
approximately 6 minutes, whereas CG is not able to simulate the 48 layers in 
less than 20 minutes. The superiority of PCG-\ac{bddc} with respect to CG will 
even be stronger as larger total problem sizes are considered (i.e. $h \to 0$), 
due to the fact that only PCG-\ac{bddc} is weakly scalable.}

\added{Fig.~\ref{fig:ss_jacobi} and Tab.~\ref{tab:ss_jacobi} report speed-up 
and total simulation wall time $[\mathrm{s}]$ of \texttt{FEMPAR-AM} using the 
Jacobi-\ac{pcg} method, as the number of subdomains is increased. As observed, 
\texttt{FEMPAR-AM} scales up to 6,144 fine tasks with a peak speed-up of 19.2. 
Above 6,144 cores, time-to-solution increases due to parallelism related 
overheads (e.g. interprocessor communication); more computationally intensive 
simulations (i.e. larger loads per processor) would be required to exploit 
additional computational resources efficiently. As observed, the total wall 
time reduces with the number of subdomains to approximately two minutes. This 
means that it takes 2.5 seconds in average to simulate the printing and cooling 
of a single layer (in two time steps). However, at larger scales of simulation 
and/or different problem physics, \emph{weakly scalable}}~\footnote{Weak 
scalability is the ability of a parallel system to efficiently exploit 
increasing computational resources in the solution of a problem \emph{with 
fixed local size per processor}. An \emph{ideally} weakly scalable code does 
not vary the time-to-solution with the number of processors and fixed local 
size per processor. In other words, if the system solves a problem in time $t$ 
with a given amount of processors, then it is able to solve also in time $t$ an 
$X$ times larger problem with $X$ times the number of processors. The Jacobi 
method is not weakly scalable because the number of iterations grows with the 
global problem size.}\added{ methods, such as \ac{amg} or \ac{bddc}, may have 
superior performance.}

\begin{figure}[!h]
\centering
\begin{subfigure}[t]{0.49\textwidth}
  	\centering
  \scalebox{0.62}{\input{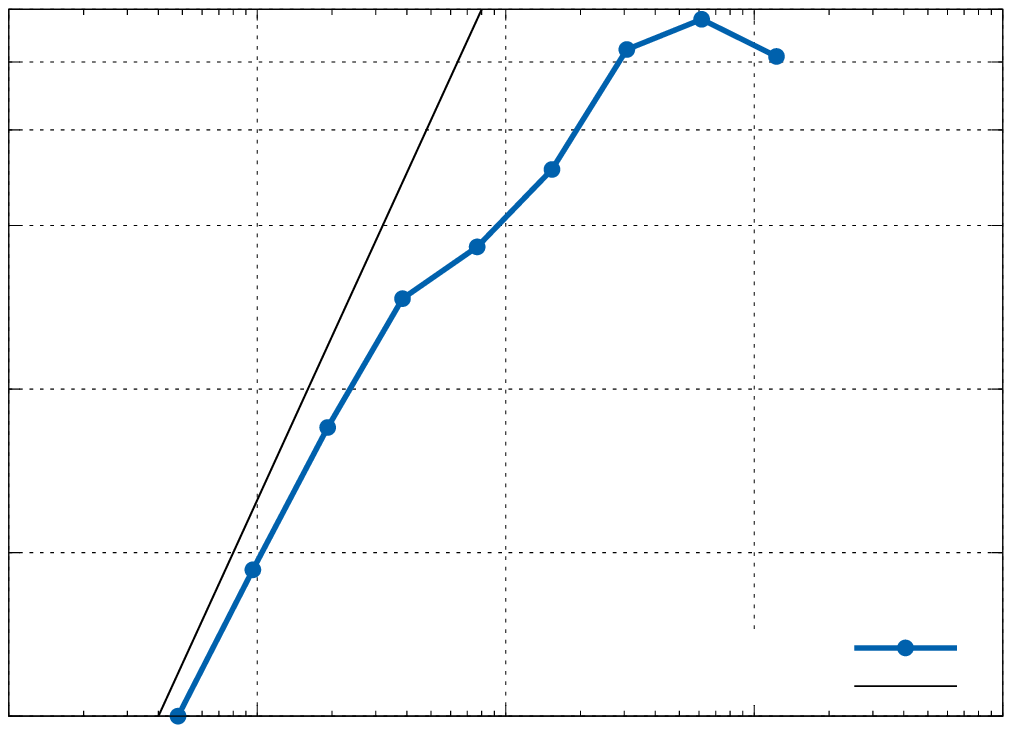}}
  \label{fig:ss_bddc_cg_speedup}
\end{subfigure}
\begin{subfigure}[t]{0.49\textwidth}
  \centering
  \scalebox{0.62}{\input{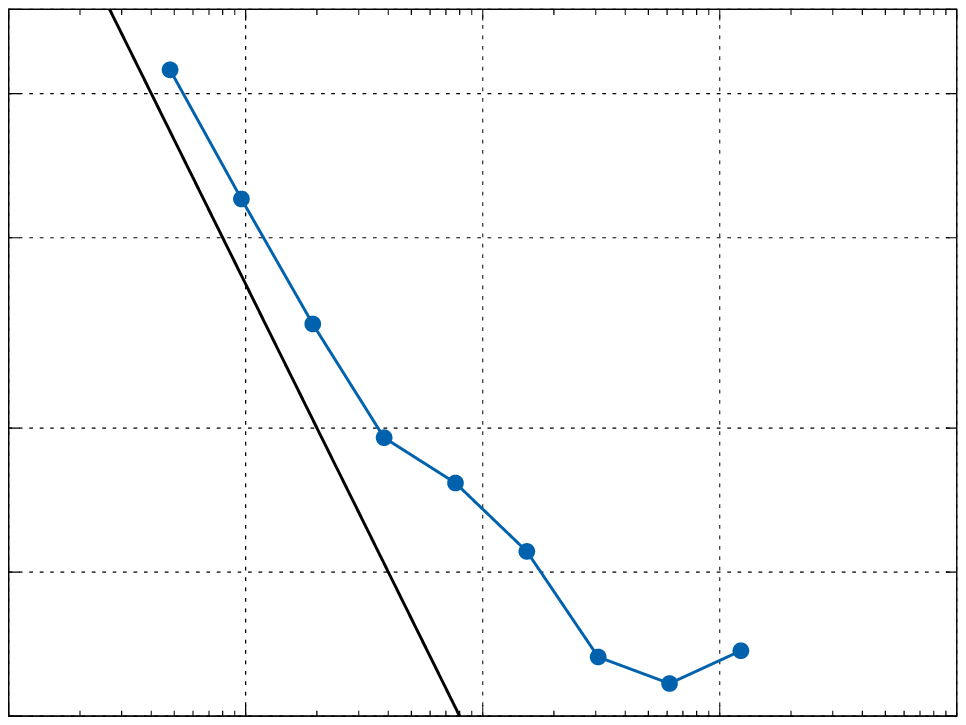}}
  \label{fig:ss_bddc_cg_times}
\end{subfigure}
\caption{\replaced{Strong-scaling example: Results of \texttt{FEMPAR-AM} for 
$(w_a,w_i) = (10,1)$. Maximum speed-up of 19.2 and minimum wall time of 117 
$[\mathrm{s}]$ is attained at 6,144 processors.}{Strong scaling analysis: 
Results of \ac{bddc}-ref vs CG. \ac{bddc}-ref takes $(w_a,w_i) = (10,1)$, 
$\deg^{\max} = 20$ and $\mathrm{d}^{\max} = 3$. Compared to CG, \ac{bddc}-ref 
achieves higher scalability and faster runtimes.}}
\label{fig:ss_jacobi}
\end{figure}

\begin{table}[!h]
    \centering
    \renewcommand{\arraystretch}{1.5}
    \begin{tabular}{ | c | c | c | c | c | c | }
        \hline
        \multirow{2}{*}{$\mathbf{P}$} & \textbf{Total wall}  & 
        \multirow{2}{*}{$\mathbf{S_P} = \frac{t_{48}}{t_P}$} & 
        \multirow{2}{*}{$\mathbf{E_P} = \frac{S_P}{P/48}$}   & 
        \multirow{2}{*}{$\mathbf{\bar{n}_{dofs}^{local}}$}   & 
        \multirow{2}{*}{$\mathbf{\bar{n}^{iters}}$} \\ 
               & \textbf{time [s]} & & & & \\ \hline
            48 & 2,244 &  1.00 & 1.00 & 222,355 & \multirow{9}{*}{68} \\ 
            96 & 1,205 &  1.86 & 0.93 & 111,812 & \\ 
				   192 &   660 &  3.40 & 0.85 &  56,390 & \\ 
				   384 &   382 &  5.87 & 0.73 &  28,533 & \\ 
				   768 &   307 &  7.31 & 0.46 &  14,508 & \\ 
				 1,536 &   221 & 10.15 & 0.32 &   7,418 & \\ 
				 3,072 &   133 & 16.87 & 0.26 &   3,827 & \\
				 6,144 &   117 & 19.18 & 0.15 &   1,993 & \\
				12,288 &   137 & 16.38 & 0.06 &   1,052 & \\ \hline
    \end{tabular} \\
    \vspace{0.2cm}
    \caption{\replaced{Strong-scaling analysis results of \texttt{FEMPAR-AM} 
    for $(w_a,w_i) = (10,1)$. Total wall time accounts for the computational 
    time of all simulation stages. $S_P$ is the speed-up, $E_P$ is the parallel 
    efficiency, $\bar{n}_{\rm dofs}^{\rm local}$ is the average size of the 
    local fine problem across processors and time steps and $\bar{n}^{\rm 
    iters}$ is the average number of iterations of the Jacobi-\ac{pcg} solver 
    across time steps.}{Strong scaling analysis results of \ac{bddc}-ref vs CG. 
    \ac{bddc}-ref takes $(w_a,w_i) = (10,1)$, $\Delta^{\max} = 20$ and 
    $\mathrm{d}^{\max} = 3$. Total wall time accounts for the computational 
    time of all simulation stages. $S_P$ is the speed-up, $E_P$ is the parallel 
    efficiency, $\bar{n}_{\rm dofs}^{\rm local}$ is the average size of the 
    local fine problem (size of $K^i$) across processors and time steps, 
    $\bar{n}_{\rm dofs}^{\rm coarse}$ is the average size of the coarse problem 
    (size of $A_{\rm C}$) and $\bar{n}^{\rm iters}$ is the average number of 
    iterations of the CG solver, both among time steps. At $P = 3,072$ maximum 
    speed up with \ac{bddc} is achieved.}}
    \label{tab:ss_jacobi}
\end{table}

\added{Another point of interest is to analyse the fraction of total wall time  
spent in different phases of the simulation and their scalability, shown in 
Fig.~\ref{fig:ss_bddc_phases}. As observed, the assembly phase dominates at low 
number of tasks, followed by the triangulation one. However, while assembly is 
the most scalable simulation phase, the triangulation is the least one. That is 
why the latter gradually dominates with increasing number of tasks and also 
leads the degradation of parallel efficiency. It is interesting to see that the 
solver phase is not relevant, with the exception of a (reproducible) spike in 
computation time at $786$ and $1536$ tasks. While the low-importance is caused 
by solving a relatively simple problem that can be efficiently preconditioned 
with the Jacobi method (the average number of iterations is merely 68), the 
abnormal deviation could be explained by the irregularity of the partition, 
although the authors were not able to find clear correlation. Another 
particularity related to the construction of the problem is that, when 
following a z-ordering, active and inactive cells are generally mixed. Hence, a 
standard $(w_a,w_i) = (1,1)$ partition more or less equally distributes the 
active cells. It follows that there is a rather low sensitivity to the 
partition weights.}

\deleted{Another point of interest is to analyse the fraction of total wall 
time of \ac{bddc}-ref dedicated to different phases of the simulation and their 
scalability. As observed in Fig.~\ref{fig:ss_bddc_phases}, almost two thirds of 
the computational time is devoted to the setup of the \ac{bddc} and the 
solution of the linear system at each printing and cooling step. This is partly 
related to the fact that there are twice as many linear system solutions as 
\ac{amr} steps. This evidence reinforces the convenience of selecting partition 
weights in the search of equal distribution of \acp{dof} among processors, and 
thus higher parallel efficiency in this stage (see Sect.~\ref{subsec:dynamic}). 
In addition, the speed-up of the total simulation is also governed by the 
solver. Even if the triangulation phase is the least scalable, the solver phase 
is the one determining when \ac{bddc}-ref loses parallel efficiency. In fact, 
this is a typical behaviour of two-level preconditioners: Although average 
iteration counts increase mildly with the number of processors, the size of the 
coarse problem keeps growing, while the local problems become smaller. 
Therefore, at some point, the coarse solver, which only exploits a bounded 
number of cores (i.e. those in a single compute node in this experiment), 
dominates computing times, loosing parallel efficiency. Fortunately, there is a 
large room for improvement in this direction, e.g. a multilevel version of the 
preconditioner is expected to push forward the limits of the presented strong 
scalability results (see, e.g. \cite{badia2016multilevel}).}

\begin{figure}[!h]
 \centering
 \begin{subfigure}[t]{0.49\textwidth}
  	\centering
   \scalebox{0.62}{\input{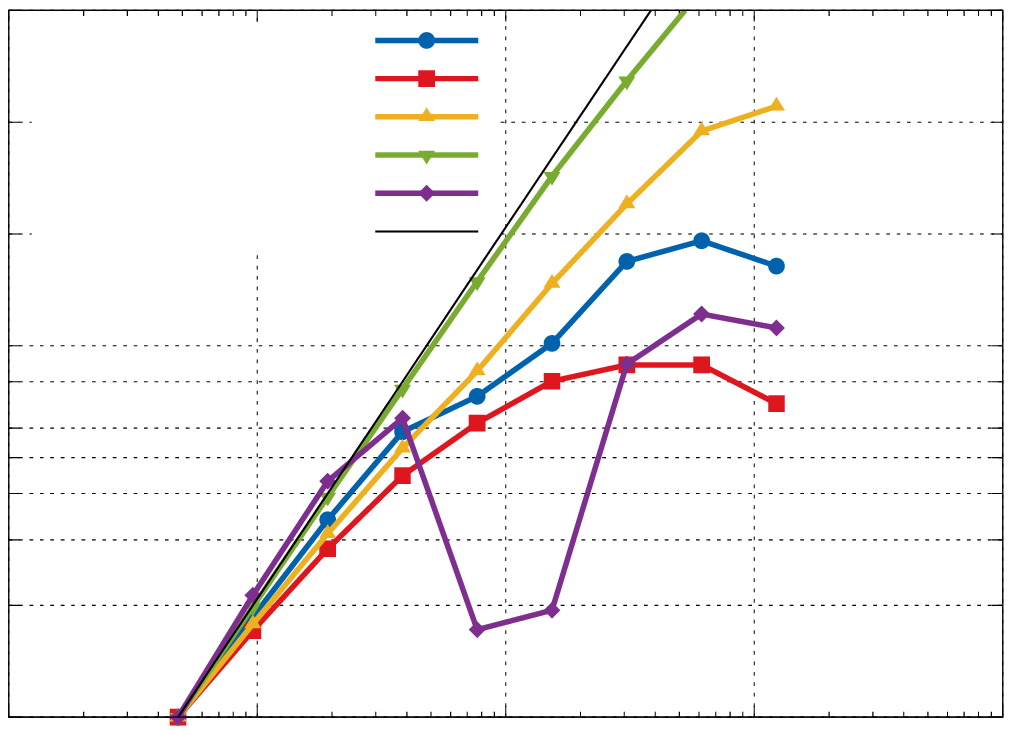}}
 \end{subfigure}
 \begin{subfigure}[t]{0.49\textwidth}
   \centering
   \scalebox{0.62}{\input{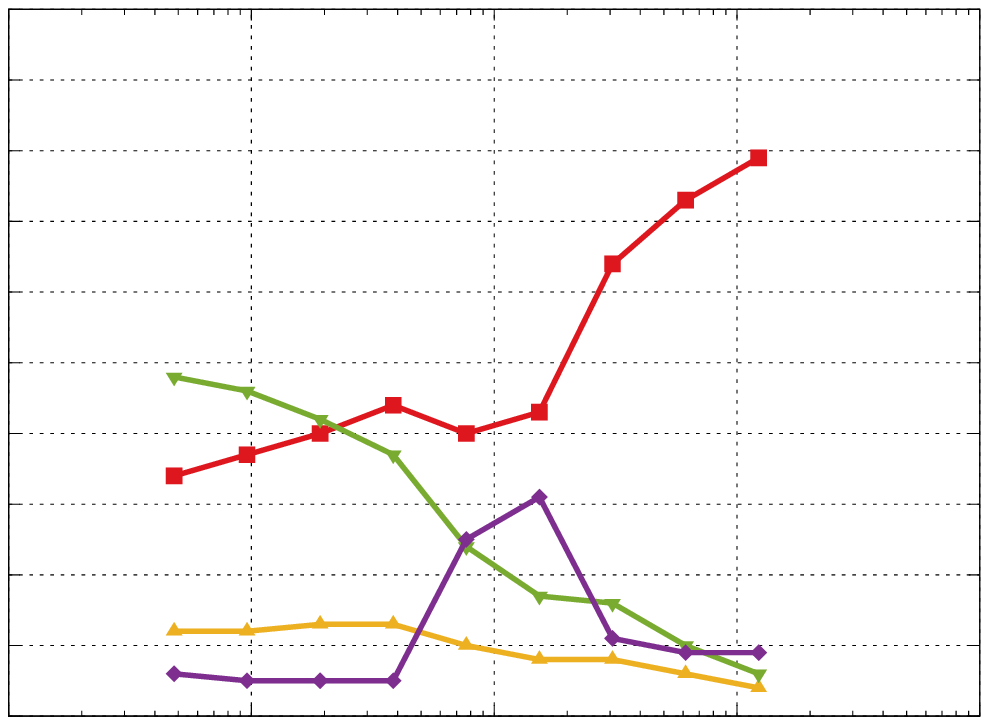}}
 \end{subfigure}
 \caption{Strong-scaling example: Results of \replaced{\texttt{FEMPAR-AM} for 
 $(w_a,w_i) = (10,1)$}{\texttt{\ac{bddc}-ref}} per simulation phases. The 
 \emph{triangulation} phase accounts for the remesh  and redistribute steps of 
 the simulation workflow, including projections and  redistributions of the FE 
 solution, whenever the mesh is transformed or  redistributed. The 
 \emph{activation} phase accounts for the search of activated cells and the 
 generation of the  FE space with new \acp{dof} assigned within the current 
 layer. The \emph{assembly} phase consists of local integration of the weak 
 form and construction of the global linear system, during printing and cooling 
 steps. Finally, the \emph{solver} phase includes \deleted{the setup of the 
 \ac{bddc} and }the solution of the linear system, also during printing and 
 cooling steps. Although the assembly phase is initially dominant, 
 computational time and scalability are dominated by the triangulation phase. 
 \deleted{Although the triangulation phase is the least scalable, computational 
 time and scalability in this example are dominated by the solver phase.}}
 \label{fig:ss_bddc_phases}
\end{figure}

Further insights are drawn by studying the local distribution of cells and 
\acp{dof} in space (among processors) and in time (among layers) up to 3,072 
processors. As the number of cells and \acp{dof} vary for each layer, all 
geometrical quantities are studied in terms of the mean $\mu$ and the 
coefficient of variation $c_v$, also known as relative standard deviation, 
which is the ratio of the standard deviation $\sigma$ to the mean $\mu$. $c_v$ 
measures the extent of variability in relation to $\mu$. Thus, it can be used 
to compare the variability among different quantities.

Given a quantity $x(t,p)$ that \added{can} depend\deleted{s} on the time step 
and processor, the time average at every processor is represented as 
$\mu^t(x)$, the average among processors at every time step as $\mu^p(x)$ and 
the mean value across processors and time steps as $\mu(x)$. \deleted{For 
quantities that only depend on time or processors, the superscripts are 
omitted.}\added{In what follows, the magnitude of $x$ is studied with 
$\mu(x)$, whereas dispersion is analysed with the coefficient of variation 
among processors, i.e. $c_v^p(x) = \sigma^p(x) / \mu^p(x)$. This statistic 
informs about possible computational load unbalances, due to an uneven 
distribution of $x$ among processors. As $c_v^p(x)$ depends on the time step, 
for the sake of simplicity, the average across time steps $\mu^t(c_v^p(x))$ is 
reported instead.}


Tab.~\ref{tab:ss_cells} gathers the local distribution of cells and degrees of 
freedom (excluding the hanging ones). The values of $\mu^t(c_v^p(x))$ show that 
the local number of cells is slightly unbalanced, but the local weighted number 
of cells, i.e. the sum of the cell weights at each processor for $(w_a,w_i) = 
(10,1)$, is perfectly balanced. Apart from that, Fig.~\ref{fig:total_cells} 
shows that the number of cells oscillates with the height of the layer, but it 
does not grow in time. This behaviour propagates to other quantities such as 
the number of active cells or \acp{dof} and it is caused by 
the\deleted{aggressive} 2:1 balance: The cells concentrate at the current layer 
and immediately below. Even if the domain grows in time, the number of cells 
away from the layer is much smaller than the number of those close or at the 
layer.

\begin{table}[!h]
   \centering
   \renewcommand{\arraystretch}{1.5}
   \begin{tabular}{ | c | c c | c c | c c | c c | }
       \cline{2-9}
	 \multicolumn{1}{c|}{} & \multicolumn{2}{c|}{$\mathbf{n_{cells}}$} & 
	 \multicolumn{2}{c|}{$\mathbf{n_{cells}^{weighted}}$} &
	 \multicolumn{2}{c|}{$\mathbf{n_{cells}^{active}}$} & 
	 \multicolumn{2}{c|}{$\mathbf{n_{dofs}}$} \\ \hline
       $\mathbf{P}$ & $\boldsymbol{\mu}$ & $\boldsymbol{\mu^t(c_v^p)}$ & 
       $\boldsymbol{\mu}$ & $\boldsymbol{\mu^t(c_v^p)}$ & $\boldsymbol{\mu}$ & 
       $\boldsymbol{\mu^t(c_v^p)}$ & $\boldsymbol{\mu}$ & 
       $\boldsymbol{\mu^t(c_v^p)}$ \\ \hline
       48 & 262.2k & 0.61 & 2,228k & 0.00 & 218.5k & 0.08 & 222.4k & 0.26 \\
       96 & 131.0k & 0.75 & 1114k & 0.00 & 109.2k & 0.10 & 111.8k & 0.37 \\
       192 & 65.5k & 1.02 & 557k & 0.01 & 54.6k & 0.14 & 56.4k & 0.45 \\
       384 & 32.8k & 1.44 & 279k & 0.01 & 27.3k & 0.20 & 28.5k & 0.63 \\
       768 & 16.4k & 2.28 & 139k & 0.02 & 13,7k & 0.31 & 14.5k & 0.73 \\
       1,536 & 8.2k & 3.46 & 69.6k & 0.05 & 6,8k & 0.48 & 7.4k & 1.10 \\
       3,072 & 4.1k & 5.46 & 35.8k & 0.09 & 3,4k & 0.76 & 3.8k & 1.39 \\ \hline
   \end{tabular}
   \vspace{0.2cm}
   \caption{Strong-scaling example. \replaced{\texttt{FEMPAR-AM} for $(w_a,w_i) 
   = (10,1)$}{Case PCG-\ac{bddc}-ref}. Local distribution of cells and degrees 
   of freedom (excluding hanging). $\mu^t(c_v^p)$ expressed in \%.}
   \label{tab:ss_cells}
\end{table}

\begin{figure}[!h]
 \centering
 \scalebox{0.7}{\input{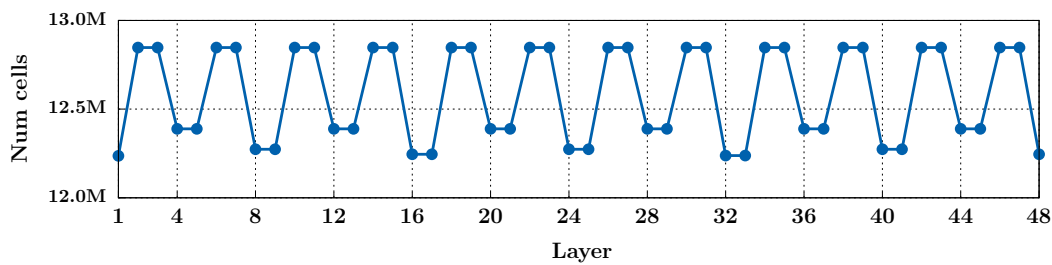}}
 \caption{Evolution of global number of cells with the height of the layer.}
 \label{fig:total_cells}
\end{figure}

It is also apparent that the pair $(w_a,w_i) = (10,1)$ effectively equilibrates 
the number of active cells among processors. Hence,\deleted{the} degrees of 
freedom are also evenly distributed, though with a slightly higher dispersion. 
This is especially beneficial for the integration and assembly phases, as they 
are implemented in \texttt{FEMPAR}, such that the bulk of the computational 
load is concentrated on the active cells set, and also the linear solver phase, 
as the size of the local systems depend on the number of \acp{dof} that the 
processor owns. On the other hand,\deleted{the} mesh generation, refinement, 
coarsening and redistribution phases suffer from an uneven distribution of 
total cells. \deleted{However, as discussed in Sect.~\ref{subsec:dynamic} and 
assessed here in Fig.~\ref{fig:ss_bddc_phases}, these tend to be secondary 
phases in terms of computational time.} 

\deleted{Concerning the local number of interface \acp{dof} in 
Tab.~\ref{tab:ss_cells}, high variations in space expose the frequent 
irregularity of the Z-curve partitions. Even if the partition becomes 
increasingly regular with $P$, the imbalance of interface \acp{dof} must be a 
source of \ac{mpi} wait time in important operations of the non-preconditioned 
CG such as the matrix-vector product. As for \ac{bddc}, the negative effects of 
this imbalance are possibly relevant in the constrained Neumann solver for 
large P, when the number of interior \acp{dof} and interface \acp{dof} are of 
the same order.}

\added{Concerning the local number of subdomain neighbours and interface 
\acp{dof}, in Tab.~\ref{tab:ss_interface}, high variations in space expose the 
extreme irregularity of Z-curve partitions. Due to the refinement strategy in 
Fig.~\ref{fig:ss_mesh} and the Z-ordering, most subdomains are embedded in the 
current layer, while only few are made of bottom coarser cells. This explains 
why, as seen in the fourth column, listing the maximum number of neighbours 
across processors and time steps, there can be subdomains touching all the 
remaining ones. Even if the partition becomes increasingly regular with $P$, 
the imbalance of neighbours and interface \acp{dof} increases synchronization 
times in important operations of the  Jacobi-\ac{pcg}, such as the 
matrix-vector product. It could also be the main cause for the deviation 
observed in the solver times of Fig.~\ref{fig:ss_bddc_phases}.}

\begin{table}[!h]
   \centering
   \renewcommand{\arraystretch}{1.5}
   \begin{tabular}{ | c | c c c | c c | }
       \cline{2-6}
       \multicolumn{1}{c|}{} & \multicolumn{3}{c|}{$\mathbf{n_{neighbours}}$} & 
       \multicolumn{2}{c|}{$\mathbf{n_{dofs}^{interface}}$} \\ \hline
       $\mathbf{P}$ & $\boldsymbol{\mu}$ & $\boldsymbol{\mu^t(c_v^p)}$ & 
       $\boldsymbol{\max}$ & $\boldsymbol{\mu}$ & $\boldsymbol{\mu^t(c_v^p)}$ 
       \\ \hline
       48    & 14 & 47.9  & 48    & 7.1k & 35.1 \\
       96    & 16 & 62.5  & 95    & 4.8k & 34.4 \\
       192   & 17 & 79.4  & 191   & 3.3k & 28.3 \\
       384   & 18 & 91.1  & 383   & 2.3k & 25.8 \\
       768   & 19 & 103.6 & 767   & 1.6k & 20.5 \\
       1,536 & 20 & 92.0  & 1,024 & 1.1k & 20.7 \\
       3,072 & 21 & 80.1  & 1,048 & 0.8k & 16.6 \\ \hline
   \end{tabular}
   \vspace{0.2cm}
   \caption{Strong scaling example. Local distribution of subdomain neighbours 
   and interface degrees of freedom. $\mu^t(c_v^p)$ and $c_v$ expressed in \%.}
   \label{tab:ss_interface}
\end{table}

\subsection{Extruded wiggle}
\label{subsec:curved}

\added{If the previous example focused in performance, the one in this section 
aims to show the capabilities of the framework in a more realistic scenario. 
The setting now is the printing of a 3D curved geometry \emph{following the 
actual scan path}, instead of a layer-averaging approach. In this way, the 
geometry is no longer rectangular, the temperature field is captured with 
higher detail and, more importantly, the parallel search algorithm introduced 
in Sect.~\ref{subsec:search} is more intensively stressed.}

\added{The simulation spans the build of eight 60 $[\upmu \mathrm{m}]$ layers 
on top of a 7.68 $[\mathrm{mm}]$ height extruded wiggle. The geometry, 
represented in Fig.~\ref{fig:wiggle-res}, is obtained in the following way: (1) 
a one-dimensional wiggle is defined by joining two parabolic functions, the 
distance between its edges is 30.72 $[\mathrm{mm}]$; (2) extrusion along the 
$\mathrm{x}$-axis yields a 15.36 $[\mathrm{mm}]$ thick plane wiggle; (3) 
extrusion perpendicular to the $\mathrm{xy}$-plane completes the 3D shape.}

\begin{figure}[!h]
 \centering
 \begin{subfigure}[t]{0.45\textwidth}
   \centering
   \includegraphics[width=0.97\textwidth]{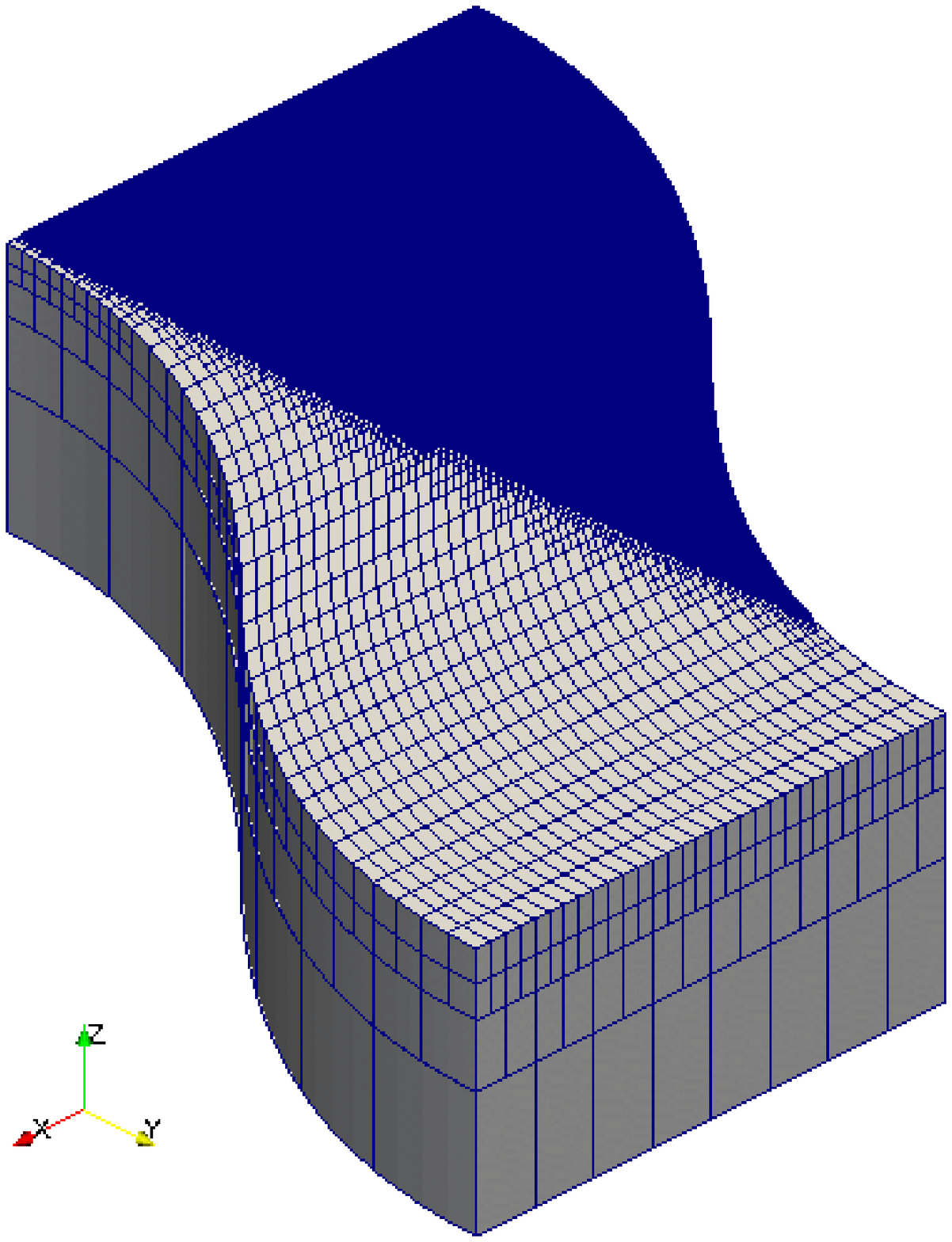}
   \caption{The absolute minimum level of refinement at the bottom of the 
   wiggle is three, while the search one below the current layer is at least 
   five. A blue shaded area indicates the highly-refined cell concentration at 
   the current layer.}
   \label{fig:wiggle-res-a}
 \end{subfigure}
 \quad
 \begin{subfigure}[t]{0.45\textwidth}
   \centering
   \includegraphics[width=0.97\textwidth]{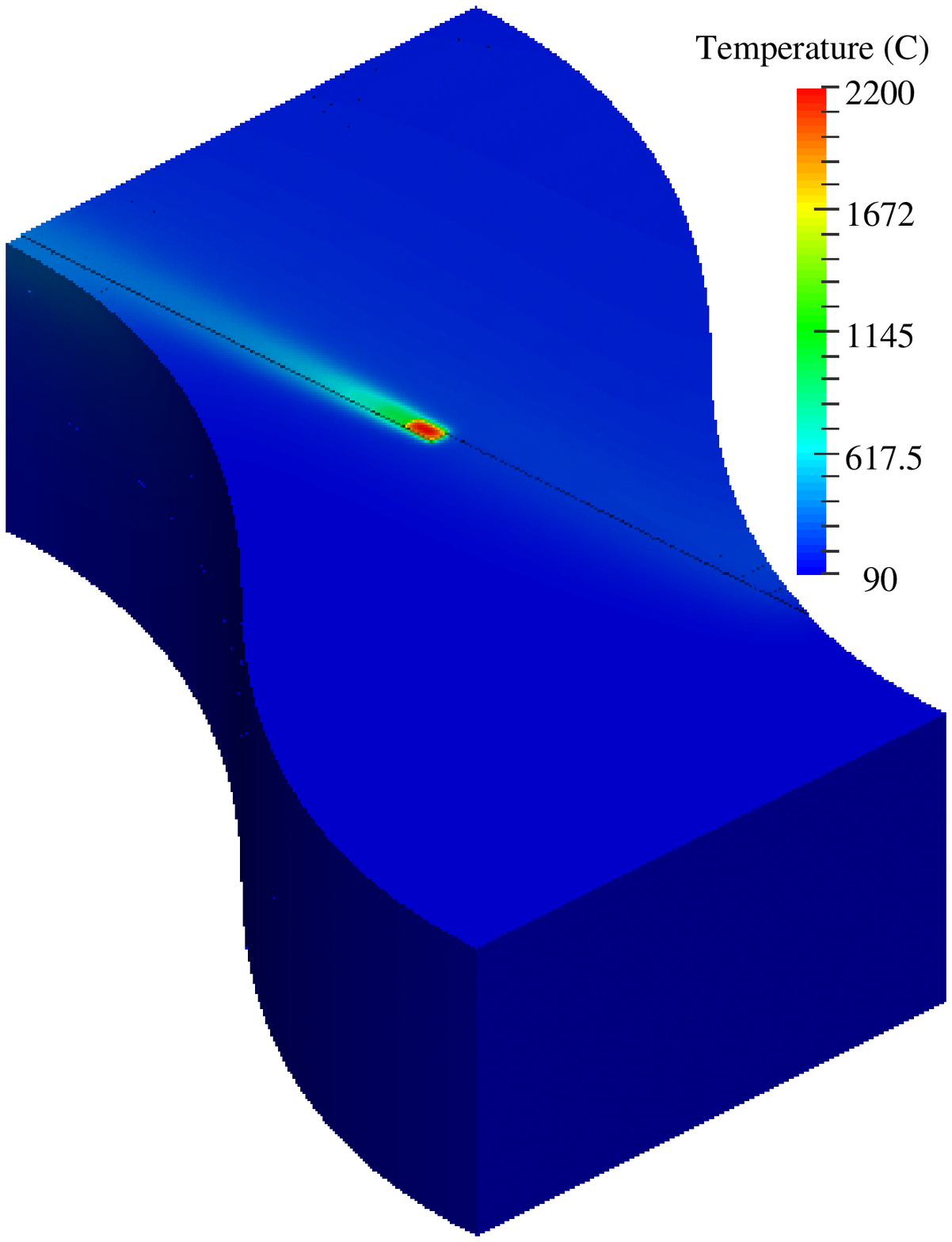}
   \caption{Contour plot of temperatures for the given time step. The relevant 
   thermal features are localized around the \ac{hav} and its tail.}
   \label{fig:wiggle-res-b}
 \end{subfigure} \\ \vspace{0.2cm}
 \begin{subfigure}[t]{0.45\textwidth}
   \centering
   \includegraphics[width=0.97\textwidth]{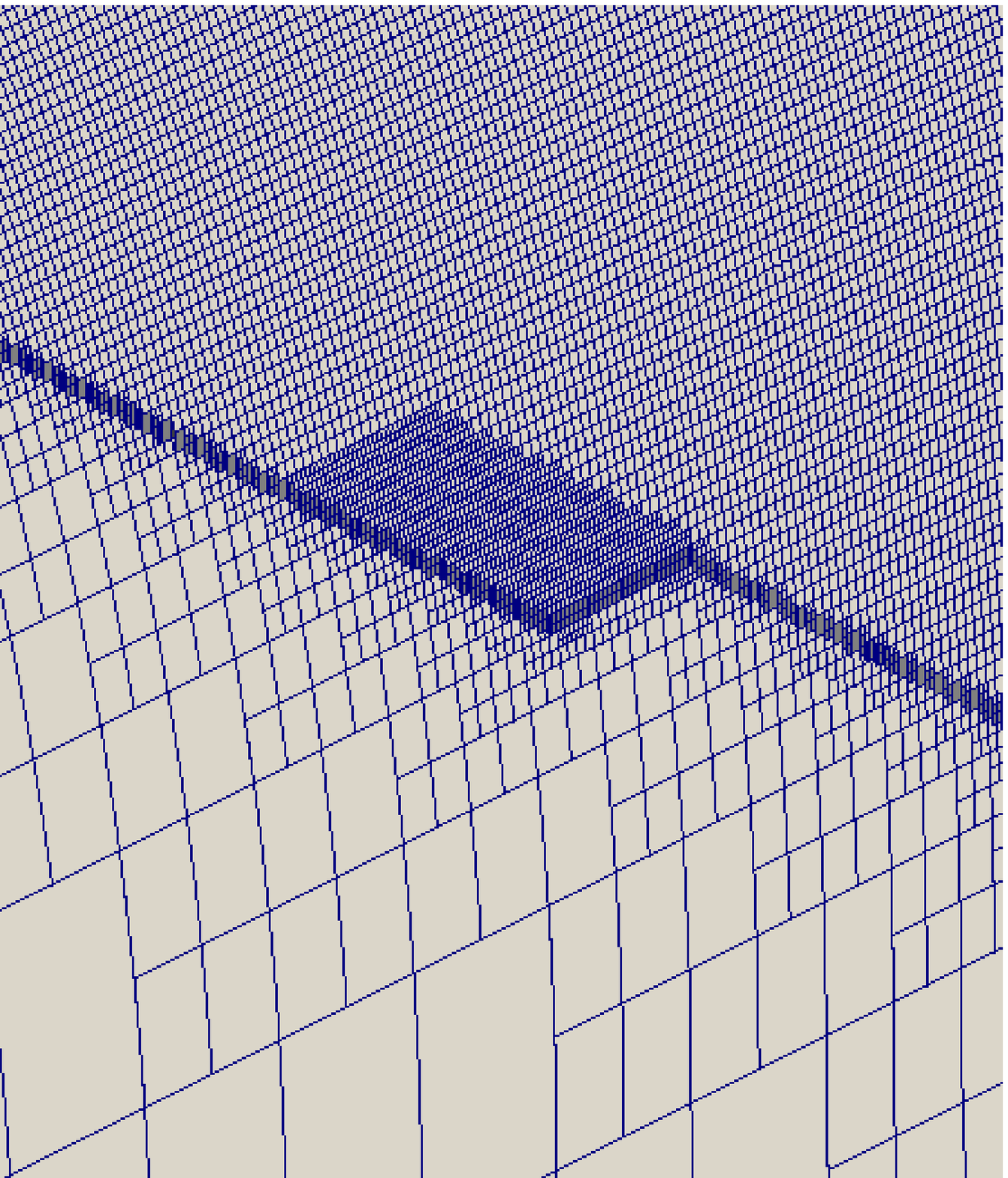}
   \caption{Close-up of the mesh around the \ac{hav}. The search algorithm 
   finds the cells inside the \ac{hav} on top of a quasi-rectangular region and 
   assigns to them the maximum level of refinement (10).}
   \label{fig:wiggle-res-c}
 \end{subfigure}
 \quad
 \begin{subfigure}[t]{0.45\textwidth}
   \centering
   \includegraphics[width=0.97\textwidth]{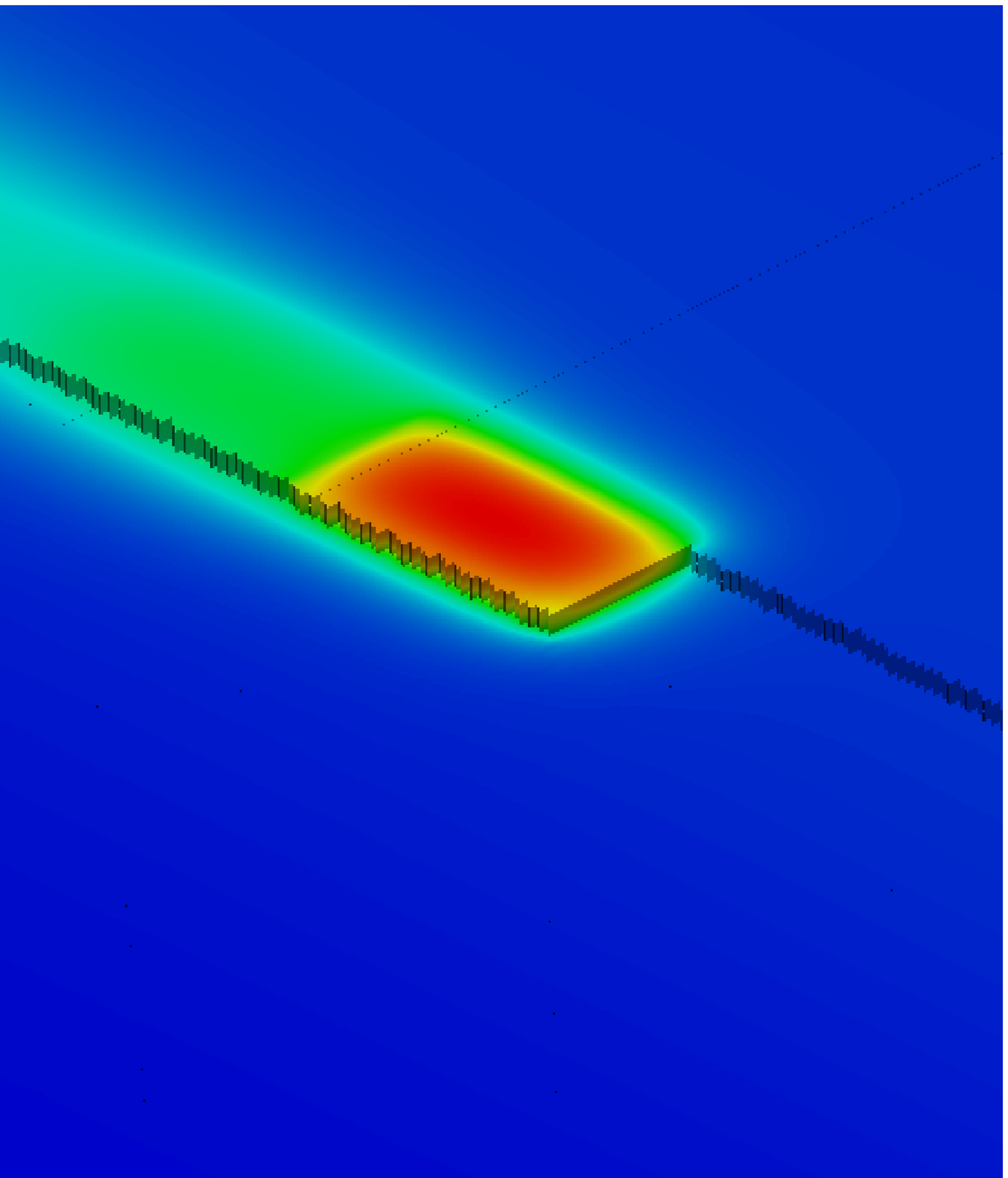}
   \caption{Close-up of the contour plot around the \ac{hav}. Away from the 
   \ac{hav}, highly-refined layer cells are only necessary to capture the 
   growing geometry; they do not increase the resolution.}
   \label{fig:wiggle-res-d}
 \end{subfigure}
 \caption{\added{Mesh and temperature contour plot of the 3D wiggle at an 
 intermediate time step of the simulation of the first layer for $(w_a,w_i) = 
 (10,1)$.}}
 \label{fig:wiggle-res}
\end{figure}

\added{One feature of the simulation is the use of \emph{second order} 
lagrangian \acp{fe} both for geometry and continuous \ac{fe} space 
approximations. In particular, given that the wiggle is parametrized by 
(piecewise) polynomials of order less or equal than two, an exact 
discretization can be easily defined. Although this choice clashes with the 
rectangularity assumption of the search algorithm in Sect.~\ref{subsec:search}, 
if the \ac{hav} overlaps a refined enough region, the mesh cells can be 
regarded as quasi-rectangular. As a result, the meshing approach considers two 
different user-prescribed minimum levels of refinement: the (1) \emph{absolute} 
one, a lower bound of the level of any cell in the mesh, and the (2) 
\emph{search} one, i.e. the lowest level a cell below the layer being printed 
must have to ensure quasi-rectangularity. It is apparent that (1) is always 
lower or equal than (2).}

\added{The octree mesh is constructed by working on two different configuration 
spaces; \texttt{p4est} generates and transforms the mesh in a \emph{reference} 
unit cube, whereas \texttt{FEMPAR} maps the unit cube mesh coordinates into the 
\emph{physical} wiggle. The root octant of the octree mesh is mapped into a 
30.72 $[\mathrm{mm}]$ height wiggle. The absolute minimum level of refinement 
is set to three, enough to have an exact discrete geometry; the search one is 
set to five, after several trial and error experiments; and the maximum one at 
the \ac{hav} is ten, i.e. there are two cells along the thickness of the layer. 
This setting is apparent in Figs.~\ref{fig:wiggle-res-a} 
and~\ref{fig:wiggle-res-c}. The initial computational domain has a height of 
7.68 $[\mathrm{mm}]$, cells located above this threshold are inactive. After 
printing the eight layers, the active region reaches 8.16 $[\mathrm{mm}]$ 
height.}

\added{Regarding the laser path, odd layers are built with nonoverlapping 
hatches along the $\mathrm{x}$-axis, whereas even ones along the 
$\mathrm{y}$-axis. Even though the scanning path is orthogonal, the mesh cells 
are skewed due to the cube-to-wiggle mapping. Therefore, during the simulation, 
the search algorithm generally tests the intersection of non-aligned cuboids. A 
uniform heat input of 200 $[\mathrm{W}]$ and heat absorption $\eta$ of 0.5 is 
distributed in a 0.96x0.48x0.06 $[\mathrm{mm}]$ \ac{hav}. Note that the 
\ac{hav} has been scaled 1\% extra in the $\mathrm{xy}$-plane for safety, a 
practice that is advisable even for linear \acp{fe} to avoid failures of the 
search algorithm in situations that depend on arithmetic precision. The 
scanning speed is 100 $[\mathrm{mm}/\mathrm{s}]$ and the relocation speed is 
200 $[\mathrm{mm}/\mathrm{s}]$. This means that the time step during printing 
is $0.96/100 = 9.6 \cdot 10^{-3} \ [\mathrm{s}]$.}

\added{The material is, once again, Ti6Al4V, and the initial and boundary 
conditions are also the same of the previous example, i.e. $u_{0} = 90$ 
$[\mathrm{^\circ C}]$ and convection on the boundary ($h_{\rm out} = 50$ 
$[\mathrm{W}/\mathrm{m^2 K}]$ and $u_{\rm out} = 35$ $[\mathrm{^\circ C}]$). 
An important simplification is that the powder is not simulated. Although 
modelling the powder bed has already been object of study by the 
authors~\citep{neiva2018numerical}, it is left out of this example, because it 
does not contribute to the main purpose, i.e. the study of the search 
algorithm.}

\added{The simulation workflow is the same as the one in 
Sect.~\ref{sec:ss_example}, but the time discretization is set up such that, at 
each time step, 0.96 $[\mathrm{mm}]$ along the laser path are advanced, not a 
whole layer. Simulation steps (1) to (4), i.e. mesh transformation, activation 
and printing step, are carried out at every 0.96 $[\mathrm{mm}]$ step, while 
cooling (5) steps only apply between hatches and layers. In this strategy, the 
mesh explicitly tracks the laser path, but there are many more \ac{amr} events 
than time steps, because the next \ac{hav} is not accommodated in a single 
refine and coarsen step, but (at most) $11-5=6$ steps. Although not covered in 
this work, other simulation strategies could be analysed. For instance, one 
could mesh the whole layer as in Sect.~\ref{sec:ss_example}, then simulate 
printing following the laser path point to point. Compared to the 
laser-tracking mesh approach, there would be many more time steps than mesh 
transformations, but the problem solved at each time step would also be bigger.}


\added{Numerical experiments were run with 96 \ac{mpi} tasks one-to-one 
assigned to 96 cores, distributed in six computing nodes. Each node has 2x 
Intel Xeon E5-2670 multi-core CPUs, with 8 cores each and 64 GBytes of RAM. The 
computing nodes were located at the Acuario~\citep{acuario} computer cluster, 
hosted by the International Centre of Numerical Methods in Engineering (CIMNE) 
in Barcelona, Spain. Linear systems were solved with a nonrelaxed 
Jacobi-\ac{pcg} method. Contour plots were generated in ParaView 
5.1.0~\citep{ayachit2015paraview} with second order visualization cells written 
in VTK format~\citep{schroeder2004visualization}.}

\added{Simulation results at several time steps are gathered in 
Figs.~\ref{fig:wiggle-res} and~\ref{fig:wiggle-res-ii} for $(w_a,w_i) = (10,1)$.
Average global problem size is 2.0M (nonhanging) \acp{dof} and total number of 
time steps is 8,727. For 96 \ac{mpi} tasks, the average number of local 
\acp{dof} is 21.1k and $\mu^t ( c_v^p(x) ) = 3.3 \%$, with $c_v^p(x) = 
\sigma^p(x) / \mu^p(x)$. In front of the strong-scaling example, dispersion of 
local \acp{dof} is higher and the total number of \acp{dof} notably oscillates 
in time (Fig.~\ref{fig:wiggle_dofs}) due to the laser-tracking refinement 
strategy.}

\begin{figure}[!h]
 \centering
 \scalebox{0.7}{\input{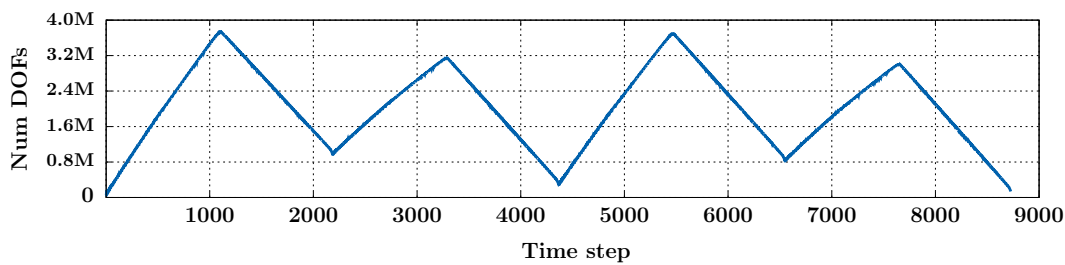}}
 \caption{\added{Total number of (nonhanging) dofs heavily oscillates with the 
 laser-tracking refinement approach. Increasing/decreasing trend reverses 
 at the start of a new layer.}}
 \label{fig:wiggle_dofs}
\end{figure}

\begin{figure}[!h]
 \centering
 \begin{subfigure}[t]{0.95\textwidth}
   \centering
   \includegraphics[width=0.45\textwidth]{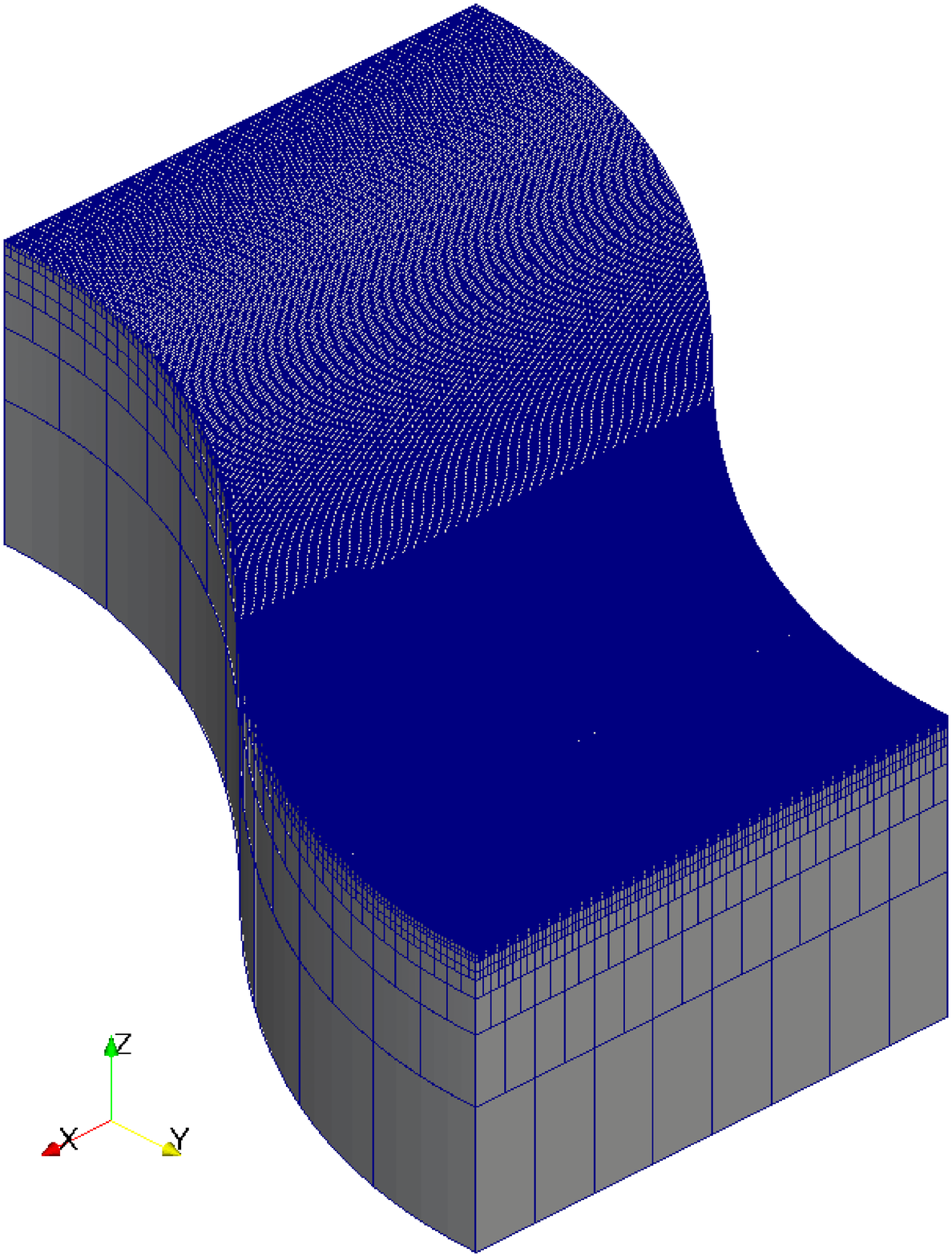}
   \hfill
   \includegraphics[width=0.45\textwidth]{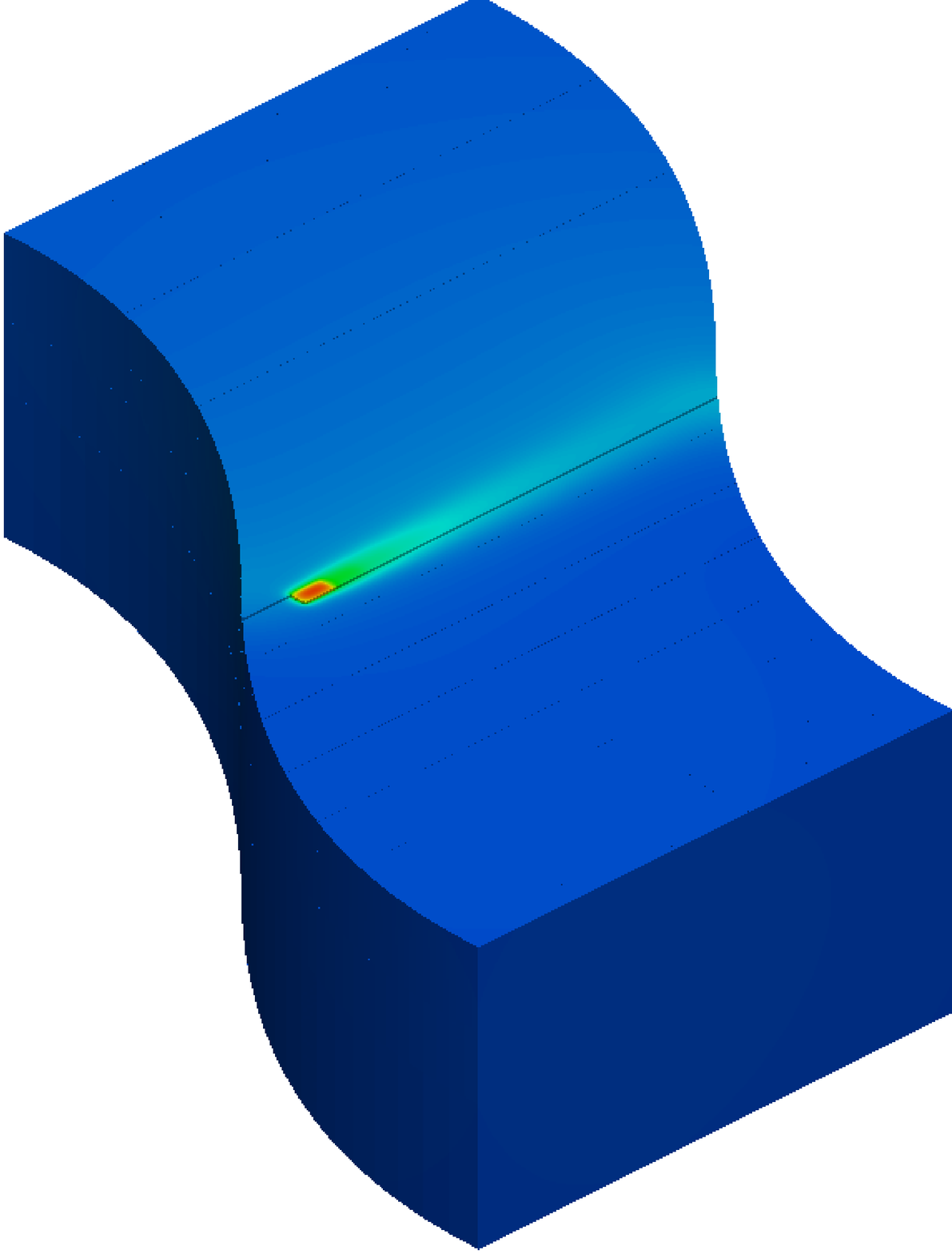}
   \caption{As the mesh must be layer-conforming, many cells concentrate at the 
   last printed layers, even though they do not contribute to capture better 
   the thermal field.}
   \label{fig:wiggle-res-e}
 \end{subfigure} \\ \vspace{0.2cm}
 \begin{subfigure}[t]{0.95\textwidth}
   \centering
   \includegraphics[width=0.45\textwidth]{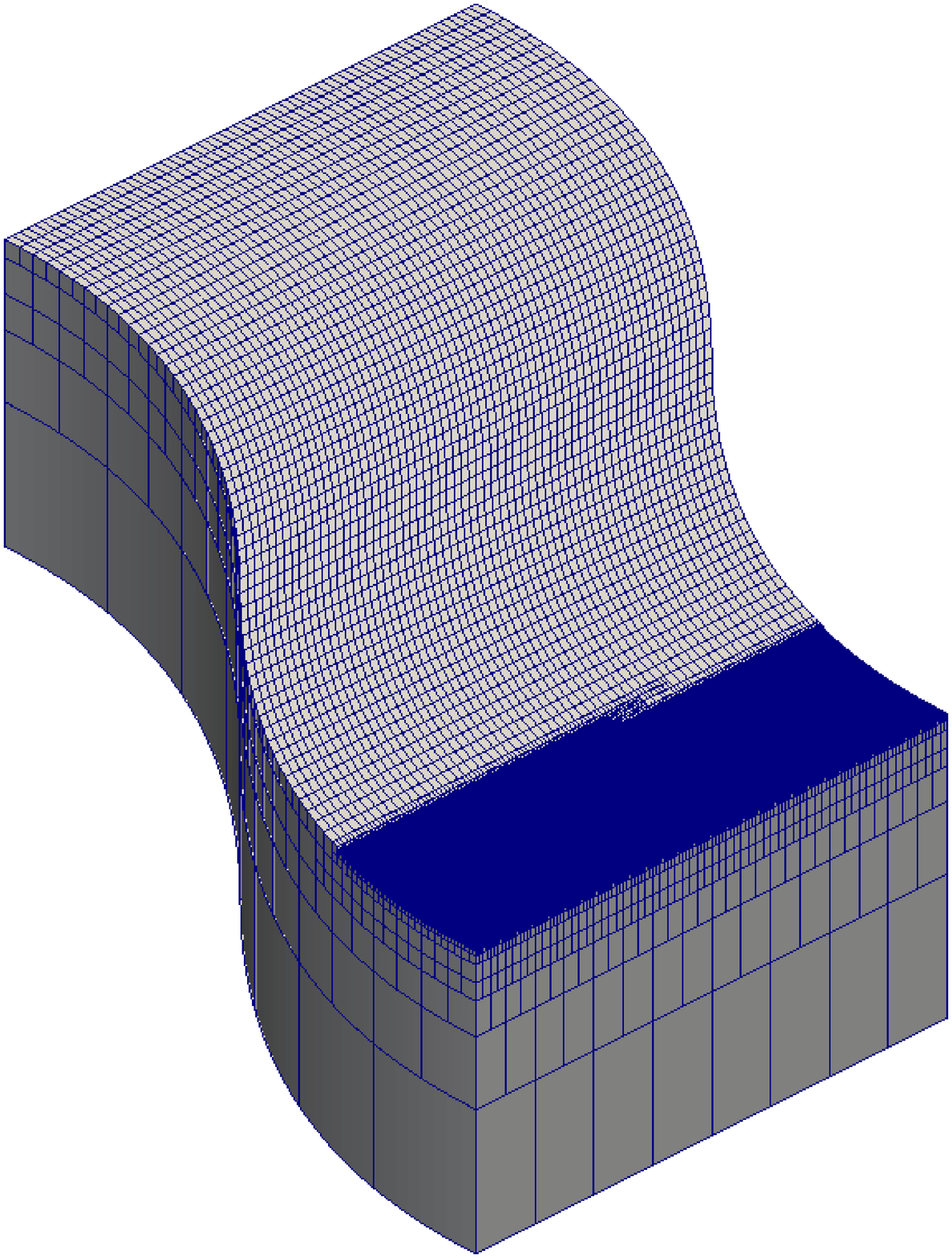}
   \hfill
   \includegraphics[width=0.45\textwidth]{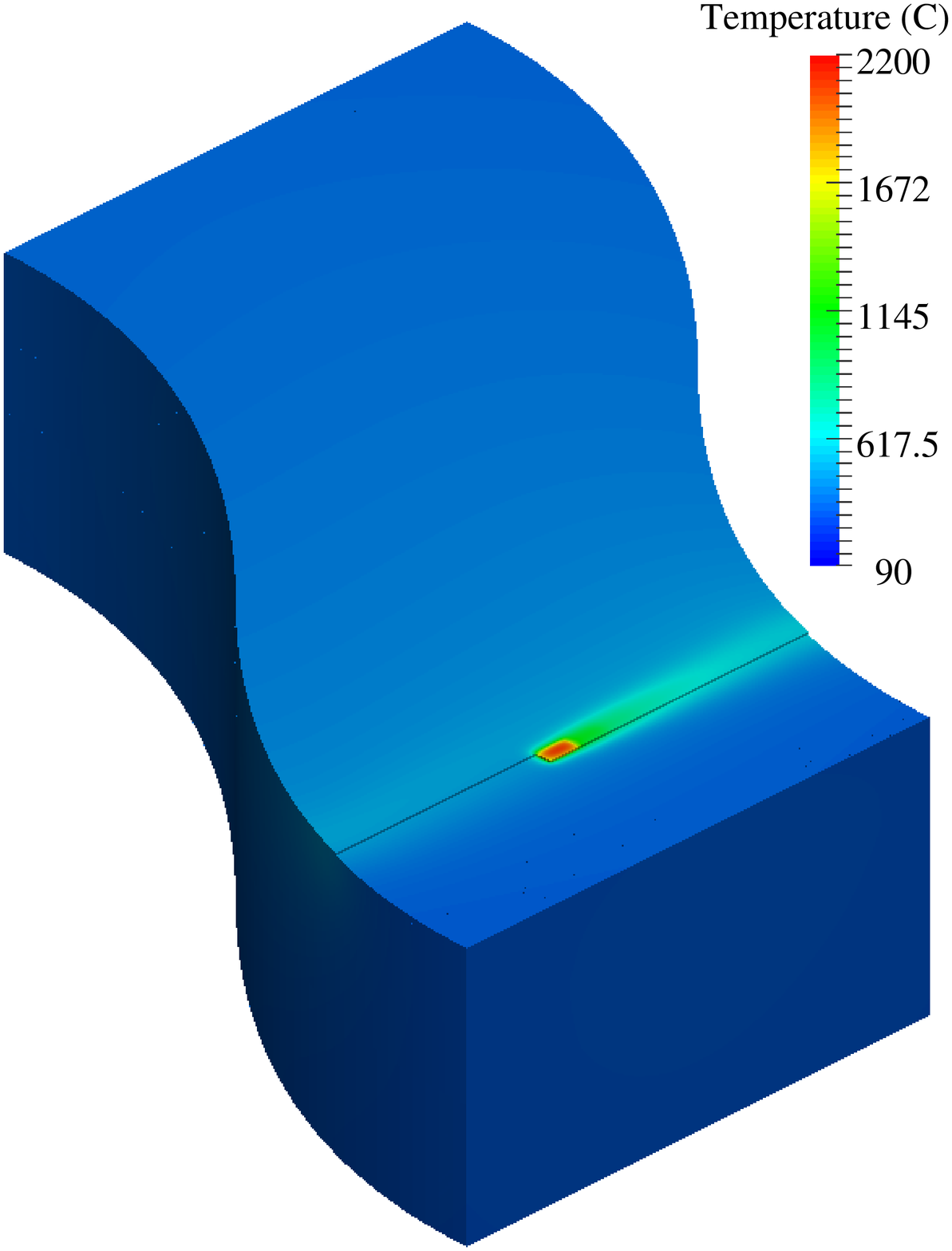}
   \caption{Homogeneous (same level) coarsening at the last layer indicates 
   that there are no interior or boundary holes, i.e. cells that the search 
   algorithm has failed to activate.}
   \label{fig:wiggle-res-f}
 \end{subfigure}
 \caption{\added{Mesh and temperature field at intermediate time steps of the 
 sixth~\ref{fig:wiggle-res-e} and eighth~\ref{fig:wiggle-res-f} layers.}}
 \label{fig:wiggle-res-ii}
\end{figure}

\added{The duration of the simulation is 13 $[\mathrm{h}]$; each layer is 
printed in 97 $[\mathrm{min}]$ average, but the individual time varies 
significantly, due to the oscillation of total number of \acp{dof}. On the 
other hand, average number of Jacobi-\ac{pcg} iterations is 381. In spite of 
being numerous, mesh generation, adaptation and redistribution roughly amounts 
to 52\% of the simulation time. Next phases by runtime are local integration 
and assembly (22\%) and linear solver (24\%). An important outcome is that 
activation with the \ac{hst} nonitersection test and initialization of new 
\acp{dof} only occupies 3\% of the simulation. Moreover, mesh and contour plots 
(e.g. Fig.~\ref{fig:wiggle-res-f}) do not expose any holes during the filling 
of the layers. Therefore, the parallel search algorithm is both efficient and 
robust in this example.}

\added{Apart from that, there is a pronounced sensitivity to the partition 
weights. Indeed, Tab.~\ref{tab:wiggle} gathers distribution of \acp{dof} and 
runtimes for several pairs of $(w_a,w_i)$. A weighted partition decreases up to 
34\% the computation time with respect to a nonweighted one. Runtime decrease 
affects especially the assembly and solver phases, whose complexity depends on 
the number of active cells or \acp{dof}, respectively. Reductions could be more 
significant for larger number of subdomains $P$, because the disequilibrium of 
local \acp{dof} grows with $P$. Hence, partition weights not only dynamically 
equilibrate the \acp{dof} among processors, they also have the potential to 
significantly shorten simulation times. Besides, in relation to the discussion 
at the end of Sect.~\ref{subsec:dynamic}, performance of pairs $(w_a,w_i)$ 
satisfying $w_a \gg w_i$ is barely distinguished from $(w_a,w_i) = (1,0)$, i.e. 
they appear capable of equilibrating the computational load, while precluding 
extreme imbalance of cells.}

\begin{table}[!h]
   \centering
   \renewcommand{\arraystretch}{1.5}
   \begin{tabular}{ | c | c c | c c | c c c c | }
       \cline{2-9}
			 \multicolumn{1}{c|}{} & \multicolumn{2}{c|}{$\mathbf{n_{dofs}}$} & 
			 \multicolumn{2}{c|}{Runtime} &
			 \multicolumn{4}{c|}{Phase percentage [\%]} \\ \hline
       $\mathbf{(w_a,w_i)}$ & $\boldsymbol{\mu}$ & $\boldsymbol{\mu^t(c_v^p)}$ 
       [\%] & $\mathrm{T} \ [\mathrm{h}]$ & $\mathrm{T}/\mathrm{T}_{(1,1)}$ 
       [\%] & 
       Triang. & Activation & Assembly & Solver \\ \hline
       (1,1)   & \multirow{4}{*}{21.1k} & 25.6 & 19.7 & 100 & 48 & 3 & 26 & 23 
       \\
       (2,1)   & & 15.4 & 15.2 & 77 & 51 & 2 & 23 & 24 \\
       (10,1)  & & 3.3  & 13.0 & 66 & 52 & 3 & 22 & 24 \\
       (100,1) & & 3.0  & 13.0 & 66 & 52 & 2 & 23 & 23 \\
       (1,0)   & & 2.4  & 13.0 & 66 & 51 & 2 & 23 & 24 \\ \hline
   \end{tabular}
   \vspace{0.2cm}
   \caption{\added{Sensitivity of \ac{dof} distribution and computation times 
   to the partition weights. Simulation phases correspond to the same ones 
   described in Fig.~\ref{fig:ss_bddc_phases}. Compared to the nonweighted 
   case, partition weights equilibrate the \ac{dof} distribution and reduce up 
   to 34\% the total simulation runtime.}}
   \label{tab:wiggle}
\end{table}

\added{A final comment arises on two computational bottlenecks identified in 
this example that guide how the framework could be further improved, towards 
dealing with real industrial scenarios. The first and most obvious one is the 
layer-conforming mesh constraint that precludes coarsening at cells touching 
the active/inactive interface (e.g. Fig.~\ref{fig:wiggle-res-e}). Unfitted 
\ac{fe} methods could likely get rid of this requirement and dramatically drop 
the mesh size. On the other hand, even if concurrency in space is efficiently 
exploited, the framework is still fully sequential in time. Very small time 
steps must be prescribed to follow the laser with high precision, leading to 
extremely long transient simulations. This motivates exploring adaptive 
space-time approximations and 
solvers~\citep{badia2017space,falgout2014parallel,gander201550}. The idea is to 
provide the solution at all time steps in one shot using the vast computational 
resources available nowadays.}

\section{Conclusions}
\label{sec:conclusions}

This work has introduced a novel HPC numerical framework for growing geometries 
that has been applied to the thermal \ac{fe} analysis of metal additive 
manufacturing by powder-bed fusion at the part scale. The abstract framework is 
constructed from three main blocks (1) hierarchical \ac{amr} with octree 
meshes, distributed across processors with space-filling curves; (2) the 
extension of the element-birth method to \replaced{(1)}{a parallel 
environment}; to track the growth of the geometry \added{with an embarrassingly 
parallel search algorithm, based on the hyperplane separation theorem}; and (3) 
\replaced{state-of-the-art parallel iterative linear solvers.}{a customization 
of \ac{bddc}, a highly scalable two-level preconditioner founded on DD methods, 
for the parallel linear solver.}

\added{After implementation in \texttt{FEMPAR}, an advanced object-oriented 
\ac{fe} code for large scale computational science and engineering, a 
strong-scaling analysis, of a problem with 10M unknowns, evaluated the 
performance of the code up to 12,288 CPU cores. Timing of simulation 
phases and statistical treatment of cell- and \ac{dof}-related quantities 
revealed uneven \ac{dof} distribution and partition irregularity as the main 
sources of load imbalance, thus increasing parallel overhead.} 

\added{A second numerical example of 2.0M unknowns in a curved geometry 
examined the search algorithm, driving the update of the computational domain. 
The results not only verified efficiency and robustness of the search routine, 
but also showed that the mesh rectangularity assumption of the algorithm can be 
slightly relaxed. A further study exposed the potential of partition weights to 
tune dynamic load balance and bring down simulation times; the weight function 
has been defined to seek a compromise between equally distributing among 
processors the number of cells or \acp{dof}.}

\added{Average simulation times \emph{per time step} were cut down to 1.2 
$\mathrm{[s]}$ and 5.4 $\mathrm{[s]}$ in the first and second examples. Even 
at the rate of 1 $\mathrm{[s]}$ per time step, fully resolved high-fidelity 
\ac{am} simulations, involving, e.g. other physics than heat transfer, stronger 
nonlinearities, historical variables, among others, can still take long hours 
in massively parallel systems. Higher efficiency and parallelism could be 
attained by resorting to (1) unfitted \ac{fe} 
methods~\citep{badia2018aggregated} to eliminate the mesh body- \emph{and} 
layer-fitting requirement and (2) weakly scalable adaptive and nonlinear  
space-time solvers that also exploit concurrency in time.}

\added{In spite of the limitations, \texttt{FEMPAR-AM} is, to the authors 
knowledge, the first \emph{fully} parallel and distributed \ac{fe} framework 
for part-scale metal \ac{am} and the numerical experiments have shown the 
potential to efficiently address levels of complexity and accuracy unseen in 
the literature of metal \ac{am}. Apart from turning to other part-scale physics 
in metal or polymer \ac{am}, this HPC framework could also be useful for other 
growing-geometry problems, as long as the growth is modelled by adding new 
elements into the computational mesh.}

\deleted{After implementation in \texttt{FEMPAR}, an advanced object-oriented 
\ac{fe} code for large scale computational science and engineering, a 
strong-scaling analysis considered the printing of 48 layers in a problem with 
10 million unknowns. It exposed the two main bottlenecks for efficiency and 
scalability: the load imbalance of (1) \acp{dof}, as the computational domain 
is a subset of the partitioned octree mesh, and (2) coarse \acp{dof}, due to 
the irregularity of the mesh partition.}

\deleted{A weighted partition can effectively deal with (1); the weight 
function has been defined to seek a compromise between equally distributing 
among processors the number of cells or \acp{dof}. As for (2), a coarse 
subgraph extraction algorithm has been proposed to identify and eliminate 
redundant coarse \acp{dof}. This procedure does not only equilibrate the coarse 
\ac{dof} distribution, it also reduces considerably the linear solver time, 
without disturbing the robustness and optimality of the preconditioner. After 
all these enhancements, the total wall time of the simulation was brought down 
to 377 $[\mathrm{s}]$ for 3,072 processors, i.e. the simulation of each layer 
(printing + cooling) lasted in average 8 $[\mathrm{s}]$.}

\deleted{Even if the adaptively growing nature of the problem hinders 
significantly the possibility to recover optimal scalability, in the sense of 
an ideal setting on a structured mesh and partition, it clearly compensates to 
adapt a high-end preconditioner, as it results in faster runtimes and higher 
scalability. In the case of \ac{bddc}, further scaling could be achieved 
through a multilevel extension~\citep{badia2016multilevel}. On the other hand, 
the heuristic for reduction of coarse \acp{dof} could also be of interest for 
other problems with highly-localized features treated with \ac{amr} and 
extended to a distributed multilevel version.}

\deleted{The numerical experiments with \texttt{FEMPAR-AM} have shown how this 
HPC framework could open the path to efficiently address levels of complexity 
and accuracy unseen in the literature of metal \ac{am}. Apart from turning to 
other part-scale physics in metal or polymer \ac{am}, it can also be exploited 
for other growing-geometry problems, as long as the growth is modelled by 
adding new elements to the computational mesh. Finally, with regards to 
possible industrial applications in \ac{am}, another interesting line of work 
is to blend the framework with unfitted \ac{fe} methods to eliminate the mesh 
generation step for complex \ac{am} geometries and permit a coarsening of the 
mesh without affecting much the geometrical error.}

\section{Acknowledgements}
\label{sec:acknowledgements}

Financial support from the EC - International Cooperation in Aeronautics with 
China (Horizon 2020) under the \emph{EMUSIC} project (\emph{Efficient 
Manufacturing for Aerospace Components USing Additive Manufacturing, Net Shape 
HIP and Investment Casting}), the EC - Factories of the Future (FoF) Programme 
under the \emph{CA}$\times $\emph{Man} Project (\emph{Computer Aided 
Technologies for Additive Manufacturing}) within \emph{Horizon 2020} Framework 
Programme and the Spanish Government-MINECO-Proyectos de I+D 
(Excelencia)-DPI2017-85998-P-ADaMANT-Computational Framework for Additive 
Manufacturing of Titanium Alloy are gratefully acknowledged. The authors 
thankfully acknowledge the computer resources at Titani (Camins-UPC), Acuario 
(CIMNE), Marenostrum-IV (RES-ActivityIDs: FI-2018-1-0014, FI-2018-2-0009, 
FI-2018-3-0029 and FI-2019-1-0007) and the technical support provided by the 
Barcelona Supercomputing Center. E. Neiva gratefully acknowledges the support 
received from the Catalan Government through a FI fellowship (2018 FI-B1-00095; 
2017 FI-B-00219). S. Badia gratefully acknowledges the support received from 
the Catalan Government through the ICREA Acad\`emia Research Program. Financial 
support to CIMNE via the CERCA Programme of the Generalitat de Catalunya is 
also acknowledged.

\bibliographystyle{abbrvnat}
\bibliography{art035}

\end{document}